\documentstyle[12pt]{article}
\begin{document}
\newcommand{\openone}{1\!\!1}
\title{Equivalence of Light-Front and Covariant Field Theory }
\author{N.E. Ligterink and B.L.G. Bakker
\\ Department of Physics and Astronomy            \\
       Vrije Universiteit, Amsterdam, The Netherlands   }
\date{}
\bibliographystyle{unsrt}
\begin{titlepage}
\pagestyle{empty}
\maketitle
\begin{abstract}
In this paper we discuss the relation between the standard covariant
quantum field theory and light-front field theory. We define covariant
theory by its Feynman diagrams, whereas light-front field theory is defined
in terms of light-cone time-ordered diagrams. A general algorithm is proposed
that produces the latter from any  Feynman diagram. The procedure is
illustrated in several cases. Technical problems that occur in the light-front
formulation and have no counterpart in the covariant formulation are
identified and solved.
\end{abstract}
\end{titlepage}
\section{Introduction}
Dirac's paper \cite{D49} on forms of relativistic dynamics opened up a whole
field of investigation: the study of different ways of quantizing and the
relationship between different forms of dynamics. Three forms were identified,
the instant form, that corresponds to ordinary time ordered theories, the point
form, that will be of no concern to us here, and the front form, where the
dynamical variables refer to physical conditions on a plane advancing with the
velocity of light. (Such surfaces are called null planes or light fronts.)
The latter form has the advantage that it requires only three dynamical
operators, "Hamiltonians", the other seven  (kinematical) generators of the
Poincar\'{e}
group containing no interaction. The other advantage noted by Dirac was that
there is no square root in the Hamiltonians, thus avoiding the degeneracy
of the free field solutions. (Particles and anti-particles cannot have the same
kinematical momenta.)

Dirac published his
paper right in the middle of the period in the development of quantum
electrodynamics, when old-fashioned perturbation theory was replaced by the
covariant formalisms of Feynman, Schwinger and Tomonaga \cite{Sch58}.
Only much later were his ideas taken up again\footnote{Leutwyler and Stern
\cite{LS78} extended
Dirac's analysis to include two more forms of dynamics. A review on the
present situation can be found in \cite{L93}.}.
Quantum electrodynamics held center stage for a long time, not only for its
unparalleled phenomenological success, but also because it functioned as a
role model for  many new theories, notably  the gauge theories of the weak and
the strong interactions.

A new development occurred when the infinite-momentum frame, which had appeared
in connection with current
algebra (see {\it e.g.} De Alfaro {\it et al.} \cite{A+73}), was proposed
by Weinberg \cite{W66} as a tool in the study of scalar theories, because it
simplifies the vacuum structure in those theories. Not long after Weinberg's
paper was published, it was suggested \cite{CM69} that the use of a new set of
variables, {\it viz},
\begin{equation}
 x^+ = \frac{x^0 + x^3}{\sqrt{2}},\;  x^- = \frac{x^0 - x^3}{\sqrt{2}}, \;
 x^1, \; x^2
 \label{eq1}
\end{equation}
would provide the advantages which are present in the use of the
infinite-momentum frame. In such a description, $x^+$ plays the role of
"time", {\it i.e.}, the evolution parameter, and the connection to
Dirac's front-form of dynamics seems immediate.
(We will use the terminology light-cone time (l.c.t.) for this variable.)
However, the connection between
the rest frame and the infinite-momentum frame involves a limiting procedure
of Lorentz transformations.
Therefore, the equivalence of descriptions in those frames cannot be derived
using arguments based on Lorentz invariance alone.

So the question concerning the relationship between different forms of dynamics
remains difficult to answer. In particular, the connection between the
manifestly covariant formulations and the front form is not yet fully clear.
The main reason is that quantization using planes $x^+ = \tau$ as surfaces on
which the initial conditions are specified--initial surfaces--is beset with
difficulties, that occur already at the classical level in scalar theories. It
is a well established result from the theory of partial-differential equations
\cite{Hormander}
that the Cauchy problem with an initial surface that contains a light-like
direction, is ill-posed.

This disturbing fact might hinder the development of a Hamiltonian formulation
of front-form field theory, if it could not be circumvented. A possible way out
was shown by Chang and Ma \cite{CM69} and later by Kogut and Soper \cite{KS70}:
one may attack the problem at the
level of Feynman diagrams. If one follows that line, one must show that the
usual Feynman rules can be reformulated in terms of the new variables,
eq.~(\ref{eq1}), or their conjugate momenta
\begin{equation}
  p^+ = \frac{p^0 + p^3}{\sqrt{2}}, \; p^- = \frac{p^0 - p^3}{\sqrt{2}}, \;
  p^1, \; p^2 .
 \label{eq2}
\end{equation}

In this paper we use this approach and show that it is successful for theories
describing spinless particles.
We show in sect.~\ref{Chapter2} how to derive light-cone time-ordered
(l.c.t.-ordered)
diagrams from a given Feynman diagram by integrating over the l.c. energy
$p^-$. The general algorithm is illustrated there by applying it to the
box diagram.
On the way to the proof of equivalence we encounter questions of
regularization\footnote{The problem of renormalization is not discussed in this
paper.}.
For scalar theories they are not more difficult to answer than in the
manifestly
covariant formulation.

The true difficulty lies in theories containing spinning particles. In the case
of spin-1/2 particles one encounters the following expression for the
free propagator \cite{CY73}
\begin{equation}
 \frac{i (p\!\!\!/+m)}{p^2 - m^2 + i\epsilon} =
 \frac{i (p\!\!\!/{}_{on}+m)}{p^2 - m^2 + i\epsilon} - \frac{i \gamma^+}{2 p^+}
  = \sum_\alpha \frac{i u^{(\alpha )} \otimes \bar u^{(\alpha )}}{p^2 - m^2 +
i\epsilon} - \frac{i \gamma^+}{2 p^+}
 \label{eq3}
\end{equation}
where $p_{on}$ is the on-shell value of the four momentum of the spin-1/2
particle with mass $m$, if its components $p^+$, $p^1$ and $p^2$ are given.
The component $p^-_{on}$ is computed from
\begin{equation}
 p^2 = 2 p^+ p^- - p_{\perp}^2 = m^2
 \label{eq4}
\end{equation}
and so
\begin{equation}
 p^-_{on} = \frac{ p_{\perp}^2 + m^2}{2 p^+}.
 \label{eq5}
\end{equation}

The occurrence of the non-propagating part $i \gamma^+ / 2 p^+$ makes the
treatment of fermions in front-form field theories much more
difficult, as it gives rise to integrals that are much more singular than the
corresponding integrals in time-ordered or manifestly covariant formulations.
In sect.~\ref{Chapter3} the general case of spin-1/2 particles is discussed.
As an illustration a box diagram, describing two fermions exchanging scalar
bosons, is reduced to a set of l.c.t.-ordered diagrams.

The algorithm we propose demonstrates the equivalence of Feynman diagrams to
sets of $x^+$-ordered diagrams in the case of scalar  and spin-1/2 fields.
(In a way, this is the reverse of Wick's theorem for time-ordered perturbation
theory. The fact that the Cauchy problem with a null-plane as initial surface
is ill posed makes the Wick theorem in front-form dynamics a strictly formal
result. Different interpretations have lead to different perturbative
expansions \cite{KS70,CY73}.) The popular belief that massive fields
do not have these problems is a misconception. The leading behavior of the
fields near the light front is independent of the mass.

A brief discussion on the extension of our treatment to diagrams with several
loops is given in sect.~\ref{Chapter4}.

In the course of our investigation we encountered several technical
difficulties. They are discussed in sect.~\ref{Chapter5}, where solutions are
given too. In particular we argue that there is no problem concerning zero
modes, if the $p^-$-integrals are regularized properly, {\it viz}, using a
regularization that preserves covariance. But it shows that Feynman diagrams
give rise to terms in the perturbative expansion of the $S$-matrix that act
on $p^+=0$-states.

The next section  (sect. \ref{Chapter6}) is concerned with the many
mathematical details that were
left out from the preceding sections, lest the main line of argument be
blurred.

We close with a discussion of our results and compare them to some of the
literature on light-front field theory.
\section{Equivalence}
\label{Chapter2}
Before we proceed with the equivalence proof, we define what we mean by
equivalence. By application of the Feynman rules as ordinarily understood, one
obtains manifestly-covariant expressions\footnote{Except for non-covariant
gauge terms in a non-covariant gauge like the light-cone gauge.} for terms in
the perturbative
expansion of $S$-matrix elements, expressed in terms of four-momenta, masses,
spins, and dynamical ingredients: coupling constants. Wick's theorem can be
understood as asserting that the S-matrix elements could be calculated as well
in time-dependent perturbation theory, and as giving us an algorithm to
combine the terms found in the latter case into manifestly-covariant
expressions. Thus Wick's theorem establishes the equivalence of time-dependent
("old-fashioned") and covariant perturbation theory.

In this paper we use the word equivalence in a similar way: each term in
covariant perturbation theory--Feynman diagram--can be written as the sum of
amplitudes that can be {\it interpreted} as terms in a l.c.t.-ordered
perturbation series. (In the interest of brevity, we will use the terminology
l.c.t.-ordered diagrams.)
In fact, those amplitudes are expressed in momentum-energy-space quantities,
but it is a straightforward matter to translate them into space-time language,
thus justifying our terminology.

By taking Feynman diagrams as our point of departure, we avoid the problems of
front-form quantization mentioned before. Besides, we also side-step the
problem of identifying the independent degrees of freedom and the determination
of commutation relations between them for a constrained system.

The splitting of a Feynman diagram into l.c.t.-ordered ones results in
amplitudes of the form
\begin{equation}
 V_\alpha {1 \over P^- -H_0} V_\beta {1 \over P^- - H_0} \cdots
{1 \over P^- -H_0} V_\sigma {1 \over P^- - H_0} V_\omega .
 \label{eq8}
\end{equation}
Here, $P^-$ plays the role of the "energy" variable, conjugate to the
light-cone time (l.c.t.) $x^+$. $H_0$  is again the energy, but now expressed
in terms of the kinematical components $p^+$ and $p_\perp$ of the momenta of
the
particles in the intermediate state between two interactions. The objects
$V$ are the vertices that correspond to the local interactions. As we will
show, expressions of this form arise naturally upon integration of a Feynman
diagram over the  the minus-component of the integration variable. In general,
a number of l.c.t.-ordered diagrams are derived from a single Feynman diagram.
This is directly analogous to old-fashioned perturbation theory, where $n!$
time-ordered
diagrams sum up to one Feynman diagram with $n$ vertices. However, there exists
an important difference: in l.c.t.-ordered theory there are less diagrams
owing to the linearity of the denominator of the single-particle propagator in
the $p^-$ variable
\begin{equation}
 {1 \over p^2 - m^2 + i \epsilon} =
 {1 \over 2 p^+ [ p^- - {p_\perp^2+m^2 - i \epsilon \over 2 p^+}]} .
 \label{eq9}
\end{equation}
Consequently, to every propagator there corresponds only one pole in $p^-$ and
its location in the complex $p^-$-plane depends on the sign as well as the
magnitude of $p^+$. This property, already alluded to by Dirac \cite{D49},
allows us to pose the condition that in any state $p^+ > 0$. In the past the
status of this so-called "spectrum condition" has remained somewhat unclear. We
shall demonstrate that it follows directly from our splitting procedure and
from a natural reinterpretation of amplitudes, quite similar to the
reinterpretation of negative-energy states as states of positive energy of
anti-particles in time-ordered perturbation theory.
In fact, our reduction algorithm shows that the l.c.t.-ordered diagrams have
the property that the internal lines carry positive $p^+$-momentum only.
Therefore a better terminology might be {\it spectrum property}, but we stick
to the term spectrum condition, because it is commonly used.
In case one would like to formulate diagram rules in l.c.t.-ordered
perturbation
theory, one could use the spectrum condition as a limitation of all
intermediate
states to states where every particle has positive $p^+$-momentum.

The spectrum condition is intimately related to causality. When the causal
single-particle propagator eq.~(\ref{eq9}) is Fourier transformed, one finds
that
the sign of $p^+$ determines whether one can extend the integral over the
$p^-$-axis to an integral along a closed contour in the complex $p^-$-plane
by adding a semi-circle at infinity for positive or negative $\Im p^-$:
$p^+ > 0$ corresponds to positive $x^+$-evolution. States with positive energy
go forward in time and states with negative energy go backward in time.

One can argue formally that the spectrum condition holds for all intermediate
states. As a result of the completeness of the physical Hilbert-space each
state is a superposition of free states and thus has a positive $p^+$ momentum.
Any particle in a free state with positive $p^+$ has positive energy and goes
forward in time.
The conservation of kinematical momentum ($p^+, p_\perp$) restricts the
creation
of particles in a Hamiltonian formulation, which is not the case in the
equal-time formulation.  We will show that this property
indeed holds for the l.c.t.-ordered perturbative expansion.

This result seems rather obvious for spin-0 bosons,
but, to our knowledge, its proof has never before been given. For spin-1/2
particles, there are
complications due to the non-propagating part $i \gamma^+ / 2 p^+$ of the
fermion propagator. These render the equivalence proof in this case more
difficult.

There is an important point worth mentioning: the pole moves in the
complex $p^-$-plane as a function of $p^+$, and even crosses the real axis at
infinity for $p^+=0$. This makes the propagator
undefined as it stands. The crossing at infinity give rise to so-called
"zero-modes" which will be dealt with later  (see sect.~\ref{zeromodes}).

\subsection{Examples}
\label{boxdiagram}
As a pedagogical example we reduce the box diagram
and the crossed-box diagram in $\phi^3$-theory\footnote{The type of theory
is not essential for the arguments, the presence of a loop is.}
to the associated l.c.t.-ordered diagrams.
This gives us the opportunity to show the working of the reduction
algorithm in great detail. Later on we will give the general form of
the algorithm.

\subsubsection{Box diagram}

The box diagram consists of four propagators:

\begin{eqnarray}
 FD_\Box & = & \int \frac{{\rm d} k^+ {\rm d}^2 k_\perp}{(2\pi)^3}
  D_\Box , \nonumber \\
 D_\Box \rule{4mm}{0mm} & = & \int  \frac{ {\rm d} k^-}{2\pi} \frac{1} { (k^2_1
- m^2 + i \epsilon)(k^2_2 -m^2 +i \epsilon)
 (k^2_3 - m^2 + i \epsilon)(k^2_4 -m^2 +i \epsilon )} ,
\label{eq2.1}
\end{eqnarray}
with incoming momenta $p$ and $q$ and outgoing momenta $l$ and $p+q-l$.
The four momenta in the loop are: $k_1 = k$, $k_2 = k-l$, $k_3 = k-p-q$ and
$k_4 = k-p$.
\begin{figure}
\begin{center}
\setlength{\unitlength}{0.012500in}%
\begin{picture}(470,169)(100,580)
\thicklines
\put(120,735){\line( 1,-1){ 40}}
\put(160,695){\line( 1, 0){ 60}}
\put(220,695){\line( 1, 1){ 40}}
\put(220,695){\line( 0,-1){ 60}}
\put(220,635){\line(-1, 0){ 60}}
\put(160,635){\line( 0, 1){ 60}}
\put(160,635){\line(-1,-1){ 40}}
\put(220,635){\line( 1,-1){ 40}}
\put(100,580){\makebox(0,0)[lb]{\raisebox{0pt}[0pt][0pt]{\twlrm $p$}}}
\put(100,740){\makebox(0,0)[lb]{\raisebox{0pt}[0pt][0pt]{\twlrm $q$}}}
\put(270,580){\makebox(0,0)[lb]{\raisebox{0pt}[0pt][0pt]{\twlrm $l$}}}
\put(185,610){\makebox(0,0)[lb]{\raisebox{0pt}[0pt][0pt]{\frtnrm $k_1$}}}
\put(140,665){\makebox(0,0)[lb]{\raisebox{0pt}[0pt][0pt]{\frtnrm $k_4$}}}
\put(185,705){\makebox(0,0)[lb]{\raisebox{0pt}[0pt][0pt]{\frtnrm $k_3$}}}
\put(225,655){\makebox(0,0)[lb]{\raisebox{0pt}[0pt][0pt]{\frtnrm $k_2$}}}
\put(270,730){\makebox(0,0)[lb]{\raisebox{0pt}[0pt][0pt]{\twlrm $p+q-l$}}}
\put(345,735){\line( 1,-1){ 40}}
\put(385,695){\line( 1, 0){100}}
\put(485,695){\line( 1, 1){ 40}}
\put(485,695){\line(-5,-3){100}}
\put(385,635){\line(-1,-1){ 40}}
\put(385,635){\line( 1, 0){100}}
\put(485,635){\line(-5, 3){100}}
\put(485,635){\line( 1,-1){ 40}}
\put(425,615){\makebox(0,0)[lb]{\raisebox{0pt}[0pt][0pt]{\twlrm $k_1$}}}
\put(380,650){\makebox(0,0)[lb]{\raisebox{0pt}[0pt][0pt]{\twlrm $k_4$}}}
\put(335,600){\makebox(0,0)[lb]{\raisebox{0pt}[0pt][0pt]{\twlrm $p$}}}
\put(525,595){\makebox(0,0)[lb]{\raisebox{0pt}[0pt][0pt]{\twlrm $l$}}}
\put(325,740){\makebox(0,0)[lb]{\raisebox{0pt}[0pt][0pt]{\twlrm $q$}}}
\put(415,700){\makebox(0,0)[lb]{\raisebox{0pt}[0pt][0pt]{\twlrm $k_3$}}}
\put(540,735){\makebox(0,0)[lb]{\raisebox{0pt}[0pt][0pt]{\twlrm $p+q-l$}}}
\put(465,650){\makebox(0,0)[lb]{\raisebox{0pt}[0pt][0pt]{\twlrm $k_2$}}}
\end{picture}
\caption{Box  and crossed-box diagrams. \label{box}}
\end{center}
\end{figure}
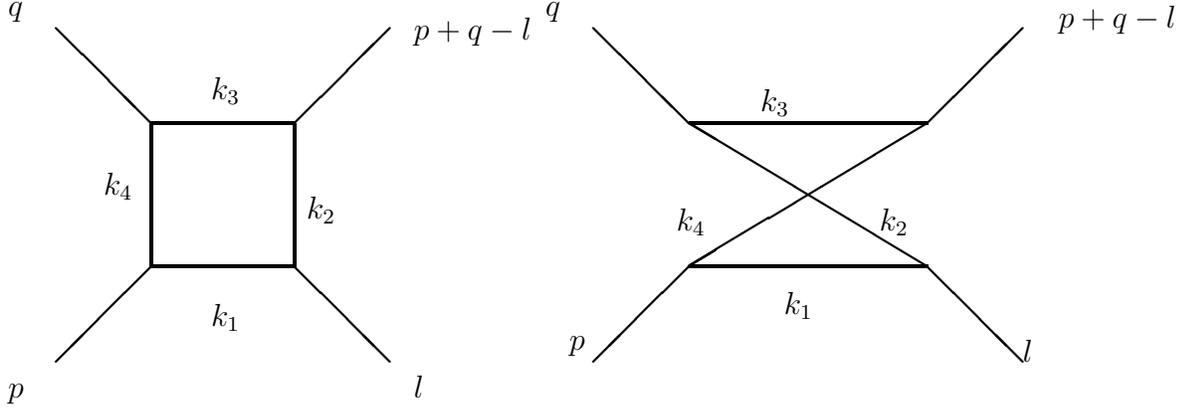
We can rewrite the integral in terms of energy denominators and phase
space factors.
We define the quantities $\{ H_i \}^{4}_{i=1}$
as follows:
\begin{eqnarray}
 H_1  & = & \frac{k_\perp^2+m^2-i \epsilon }{2 k^+} , \nonumber  \\
 H_2  & = & l^- + \frac{(k_\perp - l_\perp)^2+m^2-i \epsilon}{2(k^+ - l^+)}
 , \nonumber \\
 H_3  & = & p^- + q^- + \frac{(k_\perp - p_\perp - q_\perp)^2 + m^2-i \epsilon}
 {2 (k^+ -p^+-q^+)} , \nonumber  \\
 H_4  & = & p^- + \frac{(k_\perp - p_\perp)^2 + m^2-i \epsilon}{2 (k^+-p^+)} .
 \label{eq2.2}
\end{eqnarray}
Then the $k^-$-integral can be written as follows:
\begin{equation}
 D_\Box = \int \frac{{\rm d} k^-}{2\pi \phi }
 \frac{1}{( k^- - H_1) ( k^- - H_2) ( k^- - H_3)
 ( k^- - H_4)} .
 \label{eq2.4}
\end{equation}

\noindent The phase-space factor is given by
$\phi= 16k^+ (k^+-l^+) (k^+-p^+-q^+) (k^+-p^+)$.

\begin{figure}
\begin{center}
\setlength{\unitlength}{0.012500in}%
\begin{picture}(440,144)(60,570)
\put(155,610){\makebox(0,0)[lb]{\raisebox{0pt}[0pt][0pt]{\twlrm $k_1^+$}}}
\put(110,610){\makebox(0,0)[lb]{\raisebox{0pt}[0pt][0pt]{\twlrm $k_2^+$}}}
\put(195,610){\makebox(0,0)[lb]{\raisebox{0pt}[0pt][0pt]{\twlrm $k_4^+$}}}
\put( 70,610){\makebox(0,0)[lb]{\raisebox{0pt}[0pt][0pt]{\twlrm $k_3^+$}}}
\put(415,610){\makebox(0,0)[lb]{\raisebox{0pt}[0pt][0pt]{\twlrm $k_2^+$}}}
\put(370,610){\makebox(0,0)[lb]{\raisebox{0pt}[0pt][0pt]{\twlrm $k_1^+$}}}
\put(455,610){\makebox(0,0)[lb]{\raisebox{0pt}[0pt][0pt]{\twlrm $k_3^+$}}}
\put(330,610){\makebox(0,0)[lb]{\raisebox{0pt}[0pt][0pt]{\twlrm $k_4^+$}}}
\thicklines
\multiput( 60,710)(9.00000,0.00000){21}{\makebox(0.4444,0.6667){\tenrm .}}
\multiput( 60,690)(9.00000,0.00000){21}{\makebox(0.4444,0.6667){\tenrm .}}
\multiput( 60,670)(9.00000,0.00000){21}{\makebox(0.4444,0.6667){\tenrm .}}
\multiput( 60,650)(9.00000,0.00000){21}{\makebox(0.4444,0.6667){\tenrm .}}
\multiput( 60,630)(9.00000,0.00000){21}{\makebox(0.4444,0.6667){\tenrm .}}
\multiput(320,710)(9.00000,0.00000){21}{\makebox(0.4444,0.6667){\tenrm .}}
\multiput(320,690)(9.00000,0.00000){21}{\makebox(0.4444,0.6667){\tenrm .}}
\multiput(320,670)(9.00000,0.00000){21}{\makebox(0.4444,0.6667){\tenrm .}}
\multiput(320,650)(9.00000,0.00000){21}{\makebox(0.4444,0.6667){\tenrm .}}
\multiput(320,630)(9.00000,0.00000){21}{\makebox(0.4444,0.6667){\tenrm .}}
\put( 60,640){\line( 1, 0){ 40}}
\put(100,640){\line( 0, 1){ 40}}
\put(100,680){\line( 1, 0){ 40}}
\put(140,680){\line( 0, 1){ 20}}
\put(140,700){\line( 1, 0){ 40}}
\put(180,700){\line( 0,-1){ 40}}
\put(180,660){\line( 1, 0){ 40}}
\put(220,660){\line( 0,-1){ 20}}
\put(220,640){\line( 1, 0){ 20}}
\put(320,640){\line( 1, 0){ 40}}
\put(360,640){\line( 0, 1){ 60}}
\put(360,700){\line( 1, 0){ 40}}
\put(400,700){\line( 0,-1){ 40}}
\put(400,660){\line( 1, 0){ 40}}
\put(440,660){\line( 0, 1){ 20}}
\put(440,680){\line( 1, 0){ 40}}
\put(480,680){\line( 0,-1){ 40}}
\put(480,640){\line( 1, 0){ 20}}
\put(245,705){\makebox(0,0)[lb]{\raisebox{0pt}[0pt][0pt]{\twlrm 1}}}
\put(245,685){\makebox(0,0)[lb]{\raisebox{0pt}[0pt][0pt]{\twlrm 2}}}
\put(245,665){\makebox(0,0)[lb]{\raisebox{0pt}[0pt][0pt]{\twlrm 3}}}
\put(245,645){\makebox(0,0)[lb]{\raisebox{0pt}[0pt][0pt]{\twlrm 4}}}
\put(245,625){\makebox(0,0)[lb]{\raisebox{0pt}[0pt][0pt]{\twlrm 5}}}
\put(100,570){\makebox(0,0)[lb]{\raisebox{0pt}[0pt][0pt]{\twlrm (a)}}}
\put(360,570){\makebox(0,0)[lb]{\raisebox{0pt}[0pt][0pt]{\twlrm (b)}}}
\end{picture}
\caption{The relative values of the longitudinal momenta, for the box diagram
(a), and the crossed box diagram (b). All other box diagrams follow from these
two through discrete symmetries. The absolute scale for the $k^+_i$'s is set
by the value of $k^+$, the loop momentum.} \label{sign}
\end{center}
\end{figure}
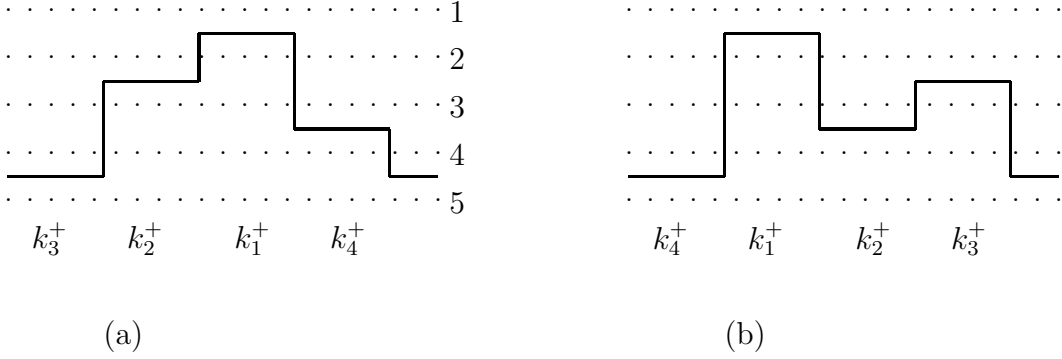

The positions of the poles in the complex $k^-$ plane depend on the values
of the external momenta and the value of $k^+$.
To calculate the time ordered diagram we must set the values of the external
momenta $p^+,\ q^+$ and $l^+$. For specific values of $k^+$, the poles cross
the axis
and end up in the opposite half plane. When that happens, the integral changes
discontinuously. The order in which the different poles cross the real axis
depends on the values of the external momenta.
In order to make our example definite, and without loss of generality,
we assume $p^+>l^+$. Then we have five regions on the $k^+$-axis:
\begin{enumerate}
 \item $k^+<0;\; \Im H_i > 0$, $\;$ (i=1,2,3,4);

 \item $0<k^+<l^+; \; \Im H_1 < 0, \Im H_i > 0$, $\;$ (i=2,3,4);

 \item $l^+<k^+<p^+; \;\Im H_1, \Im H_2 <0, \Im H_4,
                    \Im H_3 > 0$;

 \item $p^+<k^+<q^++p^+; \; \Im H_3 > 0, \Im H_i < 0$, $\;$
 (i=1,2,4);

 \item $p^++q^+<k^+ ; \; \Im H_i <  0$, $\;$ (i=1,2,3,4).

\end{enumerate}
(see fig.~\ref{sign})
with respectively zero, one, two, three, and four poles below the real axis.
Each region corresponds with a different process in light-cone time.
This observation leads naturally to the definition of a {\it skeleton graph}.
Each physical region in $k^+$ corresponds to one skeleton graph.
It is a graph that is topologically equivalent to the original Feyman diagram,
but has its internal lines graded $+$ or $-$ corresponding to the signs of
the $\Im H_i$, associated with the internal momenta $k^{\mu}_i$.

In the first  and the last case all the poles are at the same side of
the real axis so the integral over $k^-$ vanishes. Cases 2 and 4 are similar to
each other.  The integrals are calculated by closing the
contour in the  upper (lower) half-plane of complex $k^-$-values.
The application of the residue theorem gives for case 2

\begin{equation}
D^{2}_\Box =
 \frac{i}{\phi} \frac{1} { (H_4-H_1)(H_3-H_1) (H_2-H_1)} ,
 \label{eq2.5}
\end{equation}
and for case 4
\begin{equation}
D^{4}_\Box =
 \frac{i}{\phi} \frac{1} { (H_3-H_1)(H_3-H_4) (H_3-H_2)} .
 \label{eq2.6}
\end{equation}

Case 3 is the most interesting one. Straightforward application of the residue
theorem gives the result
\begin{equation}
D^{3}_\Box =
 \frac{-i} {\phi} \frac{(H_1 -H_2) ( H_4 - H_3) (H_4 + H_3 - H_1 - H_2)}
 {(H_4 - H_1) (H_3 - H_1) (H_1-H_2)(H_4-H_3)(H_4-H_2)(H_3-H_2)} ,
 \label{eq2.7}
\end{equation}
which can be split into two parts
\begin{equation}
 D^{3}_\Box  =
 -{i \over \phi} {1 \over (H_4-H_1)(H_4-H_2)(H_3-H_2)}
 -{i \over \phi}{1 \over (H_4-H_1)(H_3-H_1)(H_3-H_2)}  .
 \label{eq2.8}
\end{equation}

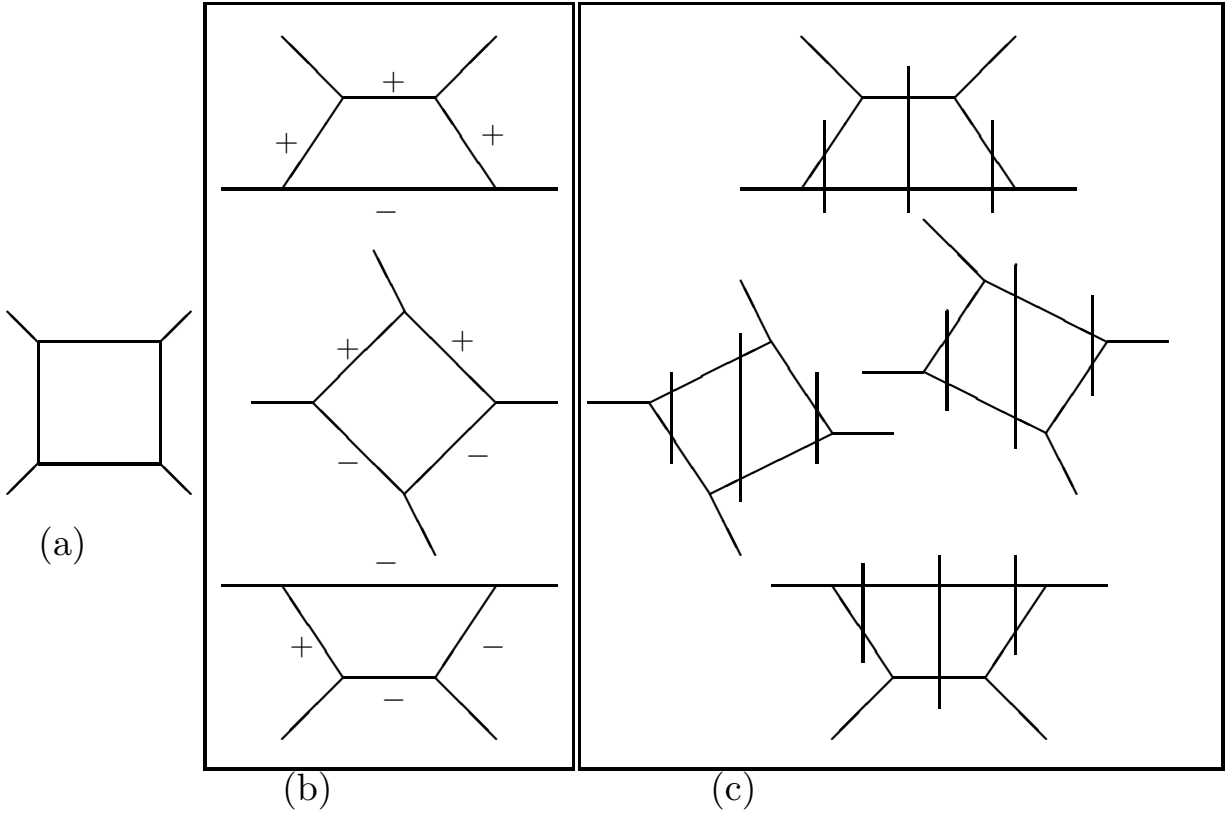
\begin{figure}
\begin{center}
\setlength{\unitlength}{0.00800in}%
\begin{picture}(795,520)(105,195)
\thicklines
\put(285,595){\line( 1, 0){140}}
\put(425,595){\line( 0, 1){  0}}
\put(425,595){\line(-2, 3){ 40}}
\put(385,655){\line(-1, 0){ 60}}
\put(325,655){\line(-2,-3){ 40}}
\put(285,595){\line(-1, 0){ 20}}
\put(265,595){\line(-1, 0){ 20}}
\put(325,655){\line(-1, 1){ 40}}
\put(385,655){\line( 1, 1){ 40}}
\put(425,595){\line( 1, 0){ 40}}
\put(625,595){\line( 1, 0){140}}
\put(765,595){\line( 0, 1){  0}}
\put(765,595){\line(-2, 3){ 40}}
\put(725,655){\line(-1, 0){ 60}}
\put(665,655){\line(-2,-3){ 40}}
\put(625,595){\line(-1, 0){ 20}}
\put(605,595){\line(-1, 0){ 20}}
\put(665,655){\line(-1, 1){ 40}}
\put(725,655){\line( 1, 1){ 40}}
\put(765,595){\line( 1, 0){ 40}}
\put(285,335){\line( 1, 0){140}}
\put(425,335){\line(-2,-3){ 40}}
\put(385,275){\line(-1, 0){ 60}}
\put(325,275){\line(-2, 3){ 40}}
\put(285,335){\line( 0, 1){  0}}
\put(325,275){\line(-1,-1){ 40}}
\put(385,275){\line( 1,-1){ 40}}
\put(425,335){\line( 1, 0){ 40}}
\put(285,335){\line(-1, 0){ 40}}
\put(645,335){\line( 1, 0){140}}
\put(785,335){\line(-2,-3){ 40}}
\put(745,275){\line(-1, 0){ 60}}
\put(685,275){\line(-2, 3){ 40}}
\put(645,335){\line( 0, 1){  0}}
\put(685,275){\line(-1,-1){ 40}}
\put(745,275){\line( 1,-1){ 40}}
\put(785,335){\line( 1, 0){ 40}}
\put(645,335){\line(-1, 0){ 40}}
\put(125,495){\line( 0,-1){ 80}}
\put(125,415){\line( 1, 0){ 80}}
\put(205,415){\line( 0, 1){ 80}}
\put(205,495){\line(-1, 0){ 80}}
\put(305,455){\line( 1, 1){ 60}}
\put(365,515){\line( 1,-1){ 60}}
\put(425,455){\line(-1,-1){ 60}}
\put(365,395){\line(-1, 1){ 20}}
\put(345,415){\line(-1, 1){ 40}}
\put(525,455){\line( 2, 1){ 80}}
\put(605,495){\line( 2,-3){ 40}}
\put(645,435){\line(-2,-1){ 80}}
\put(565,395){\line(-2, 3){ 40}}
\put(525,455){\line( 0, 1){  0}}
\put(205,415){\line( 1,-1){ 20}}
\put(205,495){\line( 1, 1){ 20}}
\put(305,455){\line(-1, 0){ 40}}
\put(365,515){\line(-1, 2){ 20}}
\put(365,395){\line( 1,-2){ 20}}
\put(425,455){\line( 1, 0){ 40}}
\put(485,455){\line( 1, 0){ 40}}
\put(525,455){\makebox(0.4444,0.6667){\tenrm .}}
\put(565,395){\line( 1,-2){ 20}}
\put(645,435){\line( 1, 0){ 40}}
\put(605,495){\line(-1, 2){ 20}}
\put(235,215){\framebox(240,500){}}
\put(480,215){\framebox(420,500){}}
\put(665,475){\line( 1, 0){ 40}}
\put(705,475){\line( 2, 3){ 40}}
\put(745,535){\line( 2,-1){ 80}}
\put(825,495){\line(-2,-3){ 40}}
\put(705,475){\line( 2,-1){ 80}}
\put(705,575){\line( 1,-1){ 40}}
\put(825,495){\line( 1, 0){ 40}}
\put(785,435){\line( 1,-2){ 20}}
\put(125,355){\makebox(0,0)[lb]{\raisebox{0pt}[0pt][0pt]{\frtnrm (a)}}}
\put(285,195){\makebox(0,0)[lb]{\raisebox{0pt}[0pt][0pt]{\frtnrm (b)}}}
\put(565,195){\makebox(0,0)[lb]{\raisebox{0pt}[0pt][0pt]{\frtnrm (c)}}}
\put(125,495){\line(-1, 1){ 20}}
\put(125,415){\line(-1,-1){ 20}}
\put(695,675){\line( 0,-1){ 95}}
\put(640,640){\line( 0,-1){ 60}}
\put(750,640){\line( 0,-1){ 60}}
\put(540,475){\line( 0,-1){ 60}}
\put(585,500){\line( 0,-1){110}}
\put(635,475){\line( 0,-1){ 60}}
\put(720,515){\makebox(0.4444,0.6667){\tenrm .}}
\put(720,515){\line( 0,-1){ 65}}
\put(765,545){\makebox(0.4444,0.6667){\tenrm .}}
\put(765,545){\line( 0,-1){120}}
\put(815,525){\line( 0,-1){ 65}}
\put(665,350){\line( 0,-1){ 65}}
\put(715,355){\line( 0,-1){100}}
\put(765,355){\line( 0,-1){ 65}}
\put(345,575){\makebox(0,0)[lb]{\raisebox{0pt}[0pt][0pt]{\frtnrm $-$}}}
\put(415,625){\makebox(0,0)[lb]{\raisebox{0pt}[0pt][0pt]{\frtnrm $+$}}}
\put(350,660){\makebox(0,0)[lb]{\raisebox{0pt}[0pt][0pt]{\frtnrm $+$}}}
\put(280,620){\makebox(0,0)[lb]{\raisebox{0pt}[0pt][0pt]{\frtnrm $+$}}}
\put(320,410){\makebox(0,0)[lb]{\raisebox{0pt}[0pt][0pt]{\frtnrm $-$}}}
\put(405,415){\makebox(0,0)[lb]{\raisebox{0pt}[0pt][0pt]{\frtnrm $-$}}}
\put(395,490){\makebox(0,0)[lb]{\raisebox{0pt}[0pt][0pt]{\frtnrm $+$}}}
\put(320,485){\makebox(0,0)[lb]{\raisebox{0pt}[0pt][0pt]{\frtnrm $+$}}}
\put(290,290){\makebox(0,0)[lb]{\raisebox{0pt}[0pt][0pt]{\frtnrm $+$}}}
\put(350,255){\makebox(0,0)[lb]{\raisebox{0pt}[0pt][0pt]{\frtnrm $-$}}}
\put(415,290){\makebox(0,0)[lb]{\raisebox{0pt}[0pt][0pt]{\frtnrm $-$}}}
\put(345,345){\makebox(0,0)[lb]{\raisebox{0pt}[0pt][0pt]{\frtnrm $-$}}}
\end{picture}
\caption{ The plane box. (a) Feynman diagram, (b) skeleton graphs, and (c)
l.c.t.-ordered diagrams.}
\label{fourpoint}
\end{center}
\end{figure}
A point to be clarified is the meaning of the denominators $(H_i - H_j)$.
We choose the sign such that in these denominators
$\Im H_i >0$, corresponding to backward moving particles,
while  $H_j $ has a negative imaginary part and refers therefore to forward
moving particles.

Four-momentum conservation gives $k_4 = k_1 - p$. We set $k \equiv k_1$ and
consider the case $\Im H_1 < 0, \, \Im H_4 > 0$, which means that
$0 < k^+ < p^+$. Then, according to the residue theorem, we
have a factor $H_4 - H_1$ in the denominator corresponding to this diagram.
This factor can be written as follows
\begin{equation}
 H_4 - H_1 = p^- - \frac{(p_\perp - k_{\perp})^2 +m^2} {2 (p^+ - k^+)}
 -\frac{k^2_\perp + m^2}{2 k^+}
 \label{eq2.9}
\end{equation}
The interpretation of the factor  $H_4 - H_1$ is facilitated by
cutting the diagram. We cut it first by a line cutting the legs of the loop
with momenta $k_1$ and $k_4$ and all incoming lines except $p$.
For every internal line, we define an on-shell value of the corresponding
minus-component as
\begin{equation}
 k^-_{i,on} = \frac{k^2_{i \perp} + m^2}{2 k^+_i}.
 \label{eq2.10}
\end{equation}
We define $H_0$ as the sum of the on-shell minus-momenta on the lines cut.
In our example we find for the internal lines 1, 4 and the external line $q$
\begin{equation}
 H_0(1,4) = q^- + \frac{k^2_\perp + m^2}{2 k^+} +
       \frac{(p_{1 \perp} - k_\perp)^2 + m^2}{2 (p^+ - k^+)} .
 \label{eq2.11}
\end{equation}

The cutting line defines an intermediate state with total minus momentum $P^- =
p^- + q^-$ and total on-shell minus momentum $H_0(1,4)$. The difference
between these two is just $H_4 - H_1$, eq. (\ref{eq2.9}). Up till now, the
direction of the internal four momenta is determined by the direction in which
the loop is passed. If we reverse the direction of $k_4$, the momentum with
negative plus component, {\it viz}, $k^+-p^+$
eqs.~(\ref{eq2.9}, \ref{eq2.10}), is replaced by $p^+ - k^+$, which is then
correctly interpreted as the plus component of the momentum of the particle
corresponding to line 4.

If we consider a cut through $k_1, \, k_3$ and $p, q$, we will
find $k_3 = k - p - q$. The same algebra that led to the result
eq.~(\ref{eq2.11}), will now give
\begin{eqnarray}
P^- - H_0(1,3) & = & p^-+ q^-  -
 \left( \frac{k^2_{\perp} + m^2}{2 k^+} +
        \frac{(p+q-k)^2_\perp + m^2}
         {2 (p^+ + q^+ - k^+)}\right) \nonumber \\
        & = & H_3 - H_1
\label{eq2.13}
\end{eqnarray}
It is clear that this procedure can be followed until the cut considered is
cutting outgoing external lines only. There it stops. So, we conclude that we
have the general result that any factor $(H_i - H_j)$
is equal to the difference of the minus-component of the total momentum
$P^-=p^- + q^-$, and the on-shell minus-component of the momentum carried by
the lines cut. Generally this holds for all combinations $i$ and $j$ such that
$\Im H_i>0$ and $\Im H_j <0$ since
the imaginary part is related to the sign of the on-shell momentum.

The l.c.t.-ordered box-diagrams can be interpreted as fourth order diagrams in
l.c.-perturbation theory, having the form
\begin{equation}
 V_4 \frac{1}{P^- - H_0} V_3 \frac{1}{P^- - H_0} V_2 \frac{1}{P^- - H_0} V_1
 \label{eq2.14}
\end{equation}

In order to systematize the reduction of $D_\Box$ to a sum of residues that
correspond to l.c.t.-ordered diagrams, it is appropriate to
consider both the algebraic structure and the connection of residues with
diagrams.

First, we demonstrate the use of some concepts that will be of crucial
importance for the proof of equivalence of the Feynman-diagram approach and
l.c.t.-ordered perturbation theory in the simple case of the box diagram.
(The general proof is to be found in sect.~\ref{Chapter6}).

The first object of interest is the Vandermonde determinant of order $k$:

\begin{equation}
\Delta(H_1, \ldots , H_k) =
 \left| \begin{array}{ccccc}
        H^{k-1}_1 & \cdots & H^2_1 & H_1 & 1 \\
        \vdots    &        & \vdots& \vdots& \vdots \\
        H^{k-1}_k & \cdots & H^2_k & H_k & 1
        \end{array}
 \right|
 \label{eq2.15}
\end{equation}

The second one is $W_{n,m}(H_1, \ldots , H_n | H_{n+1}, \ldots
 , H_{n+m})$ defined as
\begin{equation}
W_{n,m}(H_1, \ldots , H_n | H_{n+1}, \ldots , H_{n+m}) = (-1)^m
 \left| \begin{array}{cccccc}
        H^{n+m-2}_1 & \cdots & H^2_1 & H_1 & 0 & 1\\
        \vdots    &        & \vdots& \vdots& \vdots & \vdots\\
        H^{n+m-2}_n & \cdots & H^2_n & H_n & 0 & 1\\
        H^{n+m-2}_{n+1} & \cdots & H^2_{n+1} & H_{n+1} & 1 & 0 \\
        \vdots    &        & \vdots& \vdots& \vdots & \vdots \\
        H^{n+m-2}_{n+m} & \cdots & H^2_{n+m} & H_{n+m} & 1 & 0
        \end{array}
 \right|
 \label{eq2.16}
\end{equation}
By direct computation one verifies easily the following statements:
\begin{equation}
\frac{W_{1,n} (y|x_1, \ldots , x_n)}{\Delta(y, x_1, \ldots , x_n)} =
 \frac{(-1)^{n}}{\prod^{n}_{i=1} (y-x_i)} ,
 \label{eq2.17}
\end{equation}
\begin{equation}
\frac{W_{n,1} (x_1, \ldots , x_n|y)}{\Delta(x_1, \ldots , x_n, y)} =
 \frac{-1}{\prod^{n}_{i=1} (x_i-y)} .
 \label{eq2.18}
\end{equation}
Straightforward application of our rules gives for the skeleton graphs the
following corresponding amplitudes:
\begin{equation}
 D^{2} =  - \frac{i}{\phi}
          \frac{W_{3,1}( H_2, H_3, H_4| H_1)}{\Delta( H_2, H_3, H_4, H_1)} ,
 \label{eq2.19}
\end{equation}
\begin{equation}
 D^{4} =  - \frac{i}{\phi}
          \frac{W_{1,3}(H_3 | H_1, H_2, H_4)}{\Delta(H_3, H_1, H_2, H_4)} .
 \label{eq2.20}
\end{equation}
In case 3 we have
\begin{equation}
 D^{3} =  - \frac{i}{\phi}
          \frac{W_{2,2}(H_4, H_3 | H_1, H_2)}{\Delta(H_4, H_3, H_1, H_2)} .
 \label{eq2.21}
\end{equation}
$D^3$ needs to be rewritten such that energy denominators appear. The energy
denominator $(H_4-H_1)^{-1}$ should appear in all l.c.t.-ordered diagrams
associated with $D^3$. It can be extracted as follows:
\begin{equation}
{W_{2,2}(H_4,H_3|H_1,H_2) \over \Delta (H_4,H_3,H_1,H_2)}= {1 \over (H_4-H_1)}
\left( {W_{1,2}(H_3|H_1,H_2) \over \Delta (H_3,H_1,H_2)} +{
W_{2,1}(H_4,H_3|H_2) \over \Delta( H_4,H_3,H_2)} \right)
 \label{eq2.22}
\end{equation}
and upon using eq.~(\ref{eq2.17}) we recover the final expression
(\ref{eq2.8}).
The proof in sect.~\ref{Chapter6} demonstrates how this type of reduction can
be carried out in the general case.

Secondly, we describe the relation of this algebraic procedure with
l.c.t.-ordered
diagrams. We begin with the Feynman-diagram and enumerate the possible
configurations of poles in the complex $k^-$-plane. (Cases $1,\ldots , 5$.)
A pictorial representation
of those cases where the contour integral over $k^-$ does not vanish (Cases 2,
3 and 4) is given by diagrams where the sign of $\Im H$ is indicated.
(See fig.~\ref{fourpoint}.)
The box diagram is relatively easy to reduce to l.c.t.-ordered diagrams because
in any associated skeleton graph there are at most two vertices that need to be
ordered
with respect to each other.
There are only two internal lines which connect
an incoming line with an outgoing line. Generally we call this kind of diagrams
{\it flat}.
The box diagram being flat, it does not show all the possible complications.
Therefore, we also discuss the crossed box.

\subsubsection{Crossed box diagram}
\label{CrossedBox}
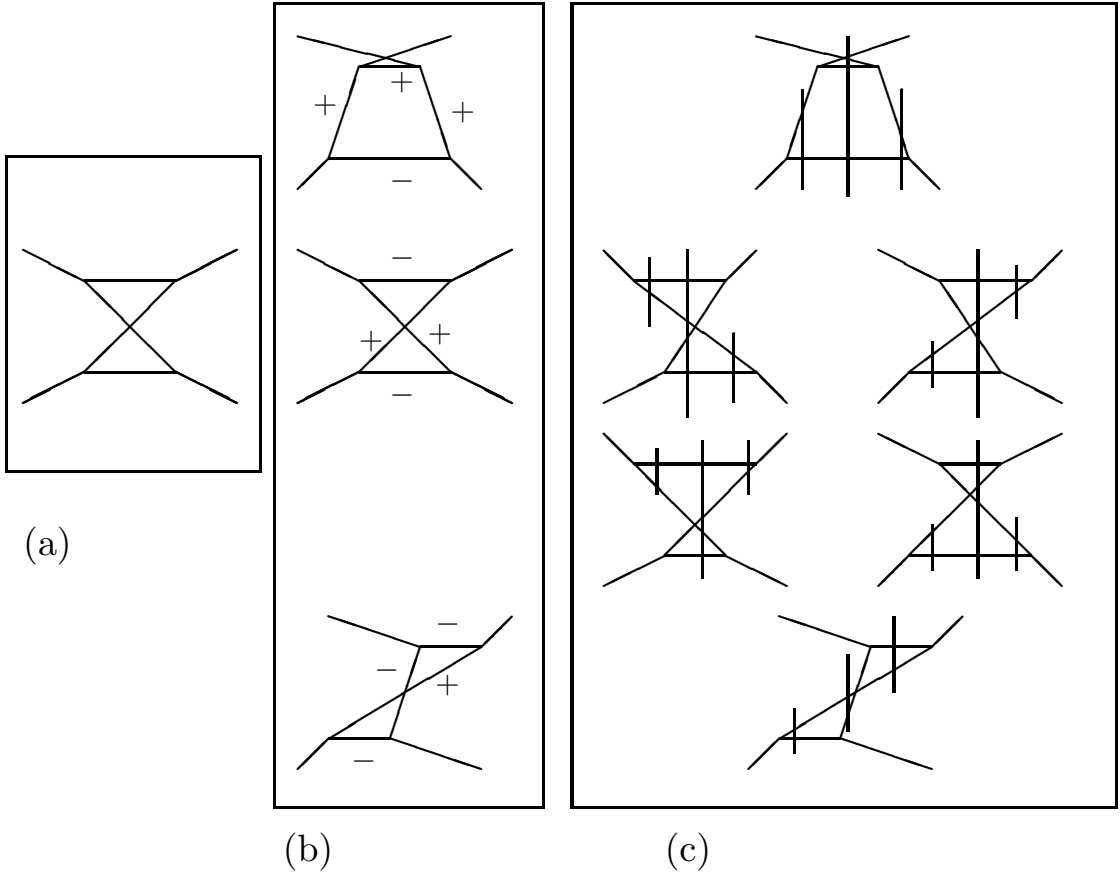
\begin{figure}
\begin{center}
\setlength{\unitlength}{0.00800in}%
\begin{picture}(845,560)(70,200)
\thicklines
\put(170,520){\line(-2,-1){ 40}}
\put(230,520){\line( 2,-1){ 40}}
\put(170,520){\line(-2,-1){ 40}}
\put(230,520){\line( 2,-1){ 40}}
\put(130,600){\line( 2,-1){ 40}}
\put(170,580){\line( 1, 0){ 60}}
\put(230,580){\line( 2, 1){ 40}}
\put(270,600){\line(-2,-1){ 40}}
\put(230,580){\line(-1,-1){ 60}}
\put(170,520){\line( 1, 0){ 60}}
\put(230,520){\line(-1, 1){ 60}}
\put(350,520){\line(-2,-1){ 40}}
\put(410,520){\line( 2,-1){ 40}}
\put(350,520){\line(-2,-1){ 40}}
\put(410,520){\line( 2,-1){ 40}}
\put(310,600){\line( 2,-1){ 40}}
\put(350,580){\line( 1, 0){ 60}}
\put(410,580){\line( 2, 1){ 40}}
\put(450,600){\line(-2,-1){ 40}}
\put(410,580){\line(-1,-1){ 60}}
\put(350,520){\line( 1, 0){ 60}}
\put(410,520){\line(-1, 1){ 60}}
\put(310,640){\line( 1, 1){ 20}}
\put(330,660){\line( 1, 0){ 80}}
\put(410,660){\line( 1,-1){ 20}}
\put(390,720){\line(-4, 1){ 80}}
\put(350,720){\line( 3, 1){ 60}}
\put(610,640){\line( 1, 1){ 20}}
\put(630,660){\line( 1, 0){ 80}}
\put(710,660){\line( 1,-1){ 20}}
\put(690,720){\line(-4, 1){ 80}}
\put(650,720){\line( 3, 1){ 60}}
\put(605,260){\line( 1, 1){ 20}}
\put(625,280){\line( 1, 0){ 40}}
\put(665,280){\line( 1, 3){ 20}}
\put(685,340){\line( 1, 0){ 40}}
\put(725,340){\line(-5,-3){100}}
\put(665,280){\line( 3,-1){ 60}}
\put(685,340){\line(-3, 1){ 60}}
\put(725,340){\line( 1, 1){ 20}}
\put(310,260){\line( 1, 1){ 20}}
\put(330,280){\line( 1, 0){ 40}}
\put(370,280){\line( 1, 3){ 20}}
\put(390,340){\line( 1, 0){ 40}}
\put(430,340){\line(-5,-3){100}}
\put(370,280){\line( 3,-1){ 60}}
\put(390,340){\line(-3, 1){ 60}}
\put(430,340){\line( 1, 1){ 20}}
\put(330,660){\line( 1, 3){ 20}}
\put(350,720){\line( 1, 0){ 40}}
\put(390,720){\line( 1,-3){ 20}}
\put(510,600){\line( 1,-1){ 20}}
\put(530,580){\line( 1, 0){ 60}}
\put(590,580){\line( 1, 1){ 20}}
\put(510,500){\line( 2, 1){ 40}}
\put(550,520){\line( 1, 0){ 60}}
\put(610,520){\line( 1,-1){ 20}}
\put(610,520){\line(-4, 3){ 80}}
\put(590,580){\line(-2,-3){ 40}}
\put(690,600){\line( 2,-1){ 40}}
\put(730,580){\line( 1, 0){ 60}}
\put(790,580){\line( 1, 1){ 20}}
\put(690,500){\line( 1, 1){ 20}}
\put(790,580){\line(-4,-3){ 80}}
\put(710,520){\line( 1, 0){ 60}}
\put(770,520){\line(-2, 3){ 40}}
\put(770,520){\line( 2,-1){ 40}}
\put(510,480){\line( 1,-1){ 20}}
\put(530,460){\line( 1, 0){ 80}}
\put(610,460){\line( 1, 1){ 20}}
\put(610,460){\line(-1,-1){ 60}}
\put(550,400){\line( 1, 0){ 40}}
\put(590,400){\line(-1, 1){ 60}}
\put(550,400){\line(-2,-1){ 40}}
\put(590,400){\line( 2,-1){ 40}}
\put(730,460){\line( 1, 0){ 40}}
\put(770,460){\line(-1,-1){ 60}}
\put(710,400){\line( 1, 0){ 80}}
\put(790,400){\line(-1, 1){ 60}}
\put(730,460){\line(-2, 1){ 40}}
\put(770,460){\line( 2, 1){ 40}}
\put(790,400){\line( 1,-1){ 20}}
\put(710,400){\line(-1,-1){ 20}}
\put(650,720){\line(-1,-3){ 20}}
\put(650,720){\line( 1, 0){ 40}}
\put(690,720){\line( 1,-3){ 20}}
\put(120,455){\framebox(165,205){}}
\put(295,235){\framebox(175,525){}}
\put(490,235){\framebox(355,525){}}
\put(130,400){\makebox(0,0)[lb]{\raisebox{0pt}[0pt][0pt]{\frtnrm (a)}}}
\put(550,200){\makebox(0,0)[lb]{\raisebox{0pt}[0pt][0pt]{\frtnrm (c)}}}
\put(300,200){\makebox(0,0)[lb]{\raisebox{0pt}[0pt][0pt]{\frtnrm (b)}}}
\put(540,595){\line( 0,-1){ 45}}
\put(565,600){\line( 0,-1){110}}
\put(595,545){\line( 0,-1){ 45}}
\put(595,545){\makebox(0.4444,0.6667){\tenrm .}}
\put(635,300){\line( 0,-1){ 30}}
\put(670,335){\line( 0,-1){ 50}}
\put(700,360){\line( 0,-1){ 50}}
\put(575,475){\line( 0,-1){ 90}}
\put(545,470){\line( 0,-1){ 30}}
\put(605,475){\line( 0,-1){ 35}}
\put(670,740){\line( 0,-1){105}}
\put(640,705){\line( 0,-1){ 65}}
\put(705,705){\line( 0,-1){ 65}}
\put(705,705){\makebox(0.4444,0.6667){\tenrm .}}
\put(725,420){\line( 0,-1){ 30}}
\put(755,475){\line( 0,-1){ 90}}
\put(725,540){\line( 0,-1){ 30}}
\put(755,600){\line( 0,-1){110}}
\put(780,425){\line( 0,-1){ 35}}
\put(780,590){\line( 0,-1){ 35}}
\put(370,500){\makebox(0,0)[lb]{\raisebox{0pt}[0pt][0pt]{\frtnrm $-$}}}
\put(370,640){\makebox(0,0)[lb]{\raisebox{0pt}[0pt][0pt]{\frtnrm $-$}}}
\put(345,260){\makebox(0,0)[lb]{\raisebox{0pt}[0pt][0pt]{\frtnrm $-$}}}
\put(360,320){\makebox(0,0)[lb]{\raisebox{0pt}[0pt][0pt]{\frtnrm $-$}}}
\put(400,350){\makebox(0,0)[lb]{\raisebox{0pt}[0pt][0pt]{\frtnrm $-$}}}
\put(370,590){\makebox(0,0)[lb]{\raisebox{0pt}[0pt][0pt]{\frtnrm $-$}}}
\put(400,310){\makebox(0,0)[lb]{\raisebox{0pt}[0pt][0pt]{\frtnrm $+$}}}
\put(395,540){\makebox(0,0)[lb]{\raisebox{0pt}[0pt][0pt]{\frtnrm $+$}}}
\put(350,535){\makebox(0,0)[lb]{\raisebox{0pt}[0pt][0pt]{\frtnrm $+$}}}
\put(410,685){\makebox(0,0)[lb]{\raisebox{0pt}[0pt][0pt]{\frtnrm $+$}}}
\put(370,705){\makebox(0,0)[lb]{\raisebox{0pt}[0pt][0pt]{\frtnrm $+$}}}
\put(320,690){\makebox(0,0)[lb]{\raisebox{0pt}[0pt][0pt]{\frtnrm $+$}}}
\end{picture}
\caption{ The crossed box. (a) Feynman diagram, (b) skeleton graphs , and (c)
l.c.t.-ordered diagrams.}
\label{fig03}
\end{center}
\end{figure}

The difference between the flat box and the crossed box is clearly visible
in the sign patterns, for the imaginary parts of the denominators,
one encounters when going around the loop.
In the flat box one
encounters the sign patterns $+---$, $++--$ and $+++-$. In the crossed box,
however, the sign patterns are $+---$, $+-+-$ and $+++-$. When there are two
poles on either side of the real $k^-$axis, two sign changes occur, which can
be seen as "bends" in the internal line, from backward to forward and vice
versa. The skeleton graphs with signatures $+---$
and $+++-$ are treated in the same way as the corresponding flat ones.
The case $+-+-$ leads to four l.c.t.-orderings, which we explain now.
There are two possible orderings of the two vertices with incoming external
lines. Having chosen one ordering, we follow one of the internal lines until we
reach a vertex with an outgoing external line. Either of the two vertices
with outgoing lines can come first.
This gives a total of four l.c.t.-ordered diagrams.
The corresponding diagrams are depicted in fig.~\ref{fig03}.
We choose again $p^+>l^+$.
The algebra for the $+-+-$ case, given by the interval
 $q^+-l^+< k^+< l^+$, gives the residue

\begin{equation}
 \frac{ W_{2,2}(H_4,H_2 | H_1, H_3)} {\Delta(H_4, H_2, H_1, H_3)} =
 \frac{1}{(H_4-H_1)}
 \left( \frac{W_{1,2}(H_2 | H_1, H_3)}{\Delta(H_2, H_1, H_3)} +
        \frac{W_{2,1}(H_4, H_2 | H_3)}{\Delta(H_4, H_2, H_3)} \right)
 \label{eq2.23}
\end{equation}
that can be reduced to
\begin{equation}
{1 \over (H_4-H_1)(H_2-H_3)(H_4-H_3)} + { 1 \over (H_4-H_1)(H_2-H_3)(H_2-H_1)}
{}.
 \label{eq2.24}
\end{equation}

In order to expose the four l.c.t.-orderings, we write the two energy
denominators next to the vertices with incoming lines as sums of two terms ,
{\it e.g.},
\begin{eqnarray}
{1 \over (H_4-H_1)(H_2-H_3)} & = & {1 \over (H_2-H_3)(H_4+H_2-H_3-H_1)}
 \nonumber \\
                             & + & {1 \over (H_4-H_1)(H_4+H_2-H_3-H_1)}
 \label{eq2.25}
\end{eqnarray}

Inserting these two denominators in the above expression gives four
l.c.t.-ordered diagrams:
\begin{eqnarray}
D^{2}_{\times} & = & \frac{i}{\phi}
 \frac{1}{(H_2-H_3)(H_4+H_2-H_3-H_1)(H_4-H_3)} \nonumber \\
D^{3}_{\times} & = & \frac{i}{\phi}
 \frac{1}{(H_4-H_1)(H_4+H_2-H_3-H_1)(H_4-H_3)} \nonumber \\
D^{4}_{\times} & = & \frac{i}{\phi}
 \frac{1}{(H_2-H_3)(H_4+H_2-H_3-H_1)(H_2-H_1)} \nonumber \\
D^{5}_{\times} & = & \frac{i}{\phi}
 \frac{1}{(H_4-H_1)(H_4+H_2-H_3-H_1)(H_1-H_2)}.
 \label{eq2.26}
\end{eqnarray}

The l.c.t-ordered diagrams $D^{1}$ and $D^{6}$ are obtained in a similar way
as in the case of the box diagram.
Now we can easily answer the question why the number of l.c.t.-ordered diagrams
is smaller than the number of ordinary time-ordered diagrams. In the latter
case, a loop with 4 vertices leads to $4!=24$ diagrams. For the
l.c.t.-ordered diagrams the number is reduced, because of the smaller number of
poles. We can also interpret it as the {\it spectrum condition}.
The spectrum condition
restricts the plus-component of any momentum on any line, internal as well as
external, to non-negative values. Internal lines with negative plus-momentum
correspond to poles with positive imaginary part and are interpreted as
anti-particles. By reversing the direction of the momenta on these internal
lines while
maintaining four-momentum conservation, these lines can be again associated
with particles. So, conservation of plus-momentum then
provides the limitation of possible diagrams. The number of diagrams however,
does depend on the external momenta; it is four in the case of the box
diagram and six for the crossed box diagram.

\subsection{General case}
\label{GeneralCase}
In a typical Feynman diagram on encounters several single-particle propagators.
Following the same line of reasoning as used in the two examples given above,
one will find that the
corresponding poles in the loop variable $k^-$ are located at different sides
of the real $k^-$-axis, depending on the value of $k^+$. To illustrate this
fully general property, we consider a one-loop diagram with $N$ vertices, $N$
internal and $N$ external lines. The number of incoming lines is $N_1$; there
are
$N_2 = N - N_1$ outgoing lines.
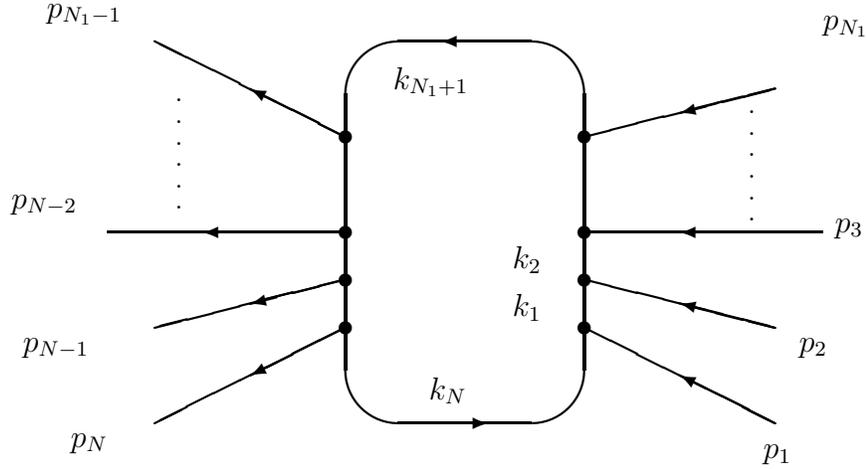
\begin{figure}
\begin{center}
\setlength{\unitlength}{0.012500in}%
\begin{picture}(355,198)(110,545)
\thicklines
\put(360,640){\circle*{6}}
\put(360,680){\circle*{6}}
\put(260,640){\circle*{6}}
\put(260,620){\circle*{6}}
\put(260,600){\circle*{6}}
\put(360,600){\circle*{6}}
\put(360,620){\circle*{6}}
\put(260,680){\circle*{6}}
\put(310,640){\oval(100,160){}}
\put(360,640){\line( 1, 0){100}}
\put(260,640){\line(-1, 0){100}}
\multiput(190,695)(0.00000,-9.00000){6}{\makebox(0.4444,0.6667){\tenrm .}}
\put(460,640){\vector(-1, 0){ 60}}
\put(260,640){\vector(-1, 0){ 60}}
\put(300,560){\vector( 1, 0){ 20}}
\put(320,720){\vector(-1, 0){ 20}}
\put(360,600){\line( 2,-1){ 80}}
\put(360,620){\line( 4,-1){ 80}}
\put(360,680){\line( 4, 1){ 80}}
\multiput(430,690)(0.00000,-9.00000){6}{\makebox(0.4444,0.6667){\tenrm .}}
\put(260,680){\line(-2, 1){ 80}}
\put(260,620){\line(-4,-1){ 80}}
\put(260,600){\line(-2,-1){ 80}}
\put(260,600){\vector(-2,-1){ 40}}
\put(260,620){\vector(-4,-1){ 40}}
\put(260,680){\vector(-2, 1){ 40}}
\put(440,700){\vector(-4,-1){ 40}}
\put(440,600){\vector(-4, 1){ 40}}
\put(440,560){\vector(-2, 1){ 40}}
\put(465,640){\makebox(0,0)[lb]{\raisebox{0pt}[0pt][0pt]{\twlrm $p_3$}}}
\put(460,725){\makebox(0,0)[lb]{\raisebox{0pt}[0pt][0pt]{\twlrm $p_{N_1}$}}}
\put(135,730){\makebox(0,0)[lb]{\raisebox{0pt}[0pt][0pt]{\twlrm $p_{N_1-1}$}}}
\put(295,570){\makebox(0,0)[lb]{\raisebox{0pt}[0pt][0pt]{\twlrm $k_N$}}}
\put(280,700){\makebox(0,0)[lb]{\raisebox{0pt}[0pt][0pt]{\twlrm $k_{N_1+1}$}}}
\put(330,605){\makebox(0,0)[lb]{\raisebox{0pt}[0pt][0pt]{\twlrm $k_1$}}}
\put(330,625){\makebox(0,0)[lb]{\raisebox{0pt}[0pt][0pt]{\twlrm $k_2$}}}
\put(145,550){\makebox(0,0)[lb]{\raisebox{0pt}[0pt][0pt]{\twlrm $p_N$}}}
\put(125,590){\makebox(0,0)[lb]{\raisebox{0pt}[0pt][0pt]{\twlrm $p_{N-1}$}}}
\put(120,650){\makebox(0,0)[lb]{\raisebox{0pt}[0pt][0pt]{\twlrm $p_{N-2}$}}}
\put(450,590){\makebox(0,0)[lb]{\raisebox{0pt}[0pt][0pt]{\twlrm $p_2$}}}
\put(435,545){\makebox(0,0)[lb]{\raisebox{0pt}[0pt][0pt]{\twlrm $p_1$}}}
\end{picture}
\caption{One-loop Diagram \label{loop}}
\end{center}
\end{figure}
Suppose we call our integration variable $k$ and identify it with $k_N$
(see fig. \ref{loop}). Four-momentum conservation takes the form
\begin{equation}
 k^\mu_i = k^\mu + K^\mu_i (p^\mu_1, ..., p^\mu_N),
 \label{eq10}
\end{equation}
where the functions $K^\mu_i (p^\mu_1, ..., p^\mu_N)$ are linear. It is obvious
that for arbitrary, but fixed external momenta all $k^+_i < 0$ ($k^+_i > 0)$
for $k^+ \rightarrow \infty$ ($k^+ \rightarrow -\infty$). So, we can divide the
real $k^+$-axis into different regions, a semi-infinite region where all
$k^+_i < 0$, another where all $k^+_i > 0$ and $N-1$ finite regions where some
of the $k^+_i$ are positive, the others being negative.

We will use here again the concept of a skeleton graph, as we did in the
cases of the box and crossed box diagrams. For
any Feynman diagram with given values of the external momenta, and for every
interval in the loop variable $k^+$, one draws a graph that is topologically
equivalent to the original Feynman diagram and has its internal lines graded
either + or $-$, corresponding to the signs of the imaginary part of the
poles $H_i$. In
the one-loop case we thus find for $k^+ \rightarrow -\infty$ the skeleton graph
with all lines graded $+$; there are $N-1$ skeleton graphs with lines graded
$+$ as well as lines graded $-$ and, finally, for $k^+ \rightarrow \infty$,
a skeleton graph with all lines graded $-$.
{}From our discussion of the causal single-particle propagator it
becomes immediately clear that lines graded $-$ ($+$) correspond to particles
moving forward (backward) in $x^+$-evolution. This justifies the terminology we
adopt: if two vertices in a skeleton graph are connected by a line with
internal
momentum say $k^+_i > 0$, then the vertex from which the momentum $k^\mu_i$ is
flowing is said to be {\it earlier} than the vertex into which $k^\mu_i$ is
flowing.

\begin{figure}
\begin{center}
\setlength{\unitlength}{0.012500in}%
\begin{picture}(410,92)(80,680)
\thicklines
\put(140,720){\circle*{10}}
\put(115,730){early}
\put( 80,720){\line( 1, 0){ 60}}
\put(140,720){\line(3,2){ 60}}
\put(140,720){\vector(3,2){ 30}}
\put(200,680){\vector(-3,2){ 30}}
\put(430,720){\circle*{10}}
\put(430,730){late}
\put(490,720){\line(-1, 0){ 60}}
\put(430,720){\line(-3,-2){ 60}}
\put(430,720){\vector(-3,-2){ 30}}
\put(370,760){\vector( 3,-2){ 30}}
\put(140,720){\line( 3, -2){ 60}}
\put(370,760){\line( 3,-2){ 60}}
\put(205,680){\makebox(0,0)[lb]{\raisebox{0pt}[0pt][0pt]{\twlrm $k_{n-1}$}}}
\put(205,760){\makebox(0,0)[lb]{\raisebox{0pt}[0pt][0pt]{\twlrm $k_n$}}}
\put( 80,740){\makebox(0,0)[lb]{\raisebox{0pt}[0pt][0pt]{\twlrm $p_n$ }}}
\put(165,750){\makebox(0,0)[lb]{\raisebox{0pt}[0pt][0pt]{\twlrm $-$}}}
\put(160,680){\makebox(0,0)[lb]{\raisebox{0pt}[0pt][0pt]{\twlrm $+$}}}
\put(465,730){\makebox(0,0)[lb]{\raisebox{0pt}[0pt][0pt]{\twlrm $p_m$}}}
\put(385,760){\makebox(0,0)[lb]{\raisebox{0pt}[0pt][0pt]{\twlrm $k_{m-1}$}}}
\put(395,680){\makebox(0,0)[lb]{\raisebox{0pt}[0pt][0pt]{\twlrm $k_m$}}}
\put(405,740){\makebox(0,0)[lb]{\raisebox{0pt}[0pt][0pt]{\twlrm $-$}}}
\put(410,690){\makebox(0,0)[lb]{\raisebox{0pt}[0pt][0pt]{\twlrm $+$}}}
\put(480,680){\vector(1,0){60}}
\put(480,685){time}
\end{picture}
\caption{The sign change in $\Im H_i$ correspond to early or late vertices.
Both lines go in the same time direction. The sign of $\Im H_i$ is opposite to
the sign of $k_i^+$.\label{earlylate}}
\end{center}
\end{figure}
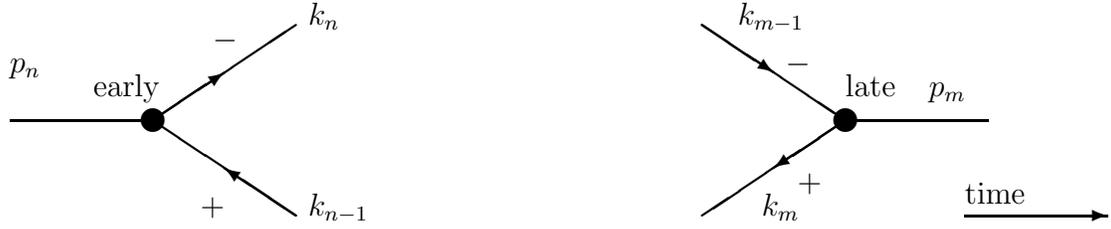

Apparently, each skeleton graph corresponds to a (partial) l.c.t.-ordering of
the vertices in a Feynman diagram. The graphs with all lines graded + or all
graded $-$ correspond to a cyclic l.c.t.-ordering of the vertices that
contradicts logic.
Fortunately, these graphs are associated with the situation that all poles in
$k^-$ are lying at one side of the real $k^-$-axis, in which case the
amplitude vanishes. In all other graphs there is at least one vertex with an
outgoing internal line with positive plus-momentum and an incoming internal
line with a negative
plus-momentum, and at least one vertex where the situation is reversed. The
former vertices are called {\it early}, the latter {\it late} vertices. A sign
change in the skeleton graph corresponds with an early or a late vertex,
the other vertices are called {\it trivial}. (see fig.~\ref{earlylate})
 If only one early
and one late vertex are present in a given skeleton graph, the partial ordering
is complete after the trivial vertices are ordered. This was the case for the
flat box diagram. Otherwise, the different early vertices must be ordered with
respect to each other and with respect to the late vertices.
This additional ordering produces several l.c.t.-ordered diagrams
associated with a single skeleton graph. In this way, a single Feynman diagram
gives rise to a number of consistently l.c.t.-ordered diagrams. At this stage
one reverses the directions of the four-momenta $k^\mu_i$ on all those lines
$i$ where $k^+_i < 0$. One sees immediately that as a result early vertices
have only outgoing internal momenta, whereas late ones have only incoming
internal momenta.

We use the late vertices in a different way than the
early vertices. Starting from an early vertex, all vertices on the lines
going out from this vertex are ordered relatively to this vertex. However,
vertices lying on different lines are not yet ordered relative to each other.
For two lines starting at different early vertices and connected at a late
vertex, we can fix the relative ordering of the intermediate vertices, since
all vertices on both lines  must occur before the late vertex.
When all late vertices have been encountered, their relative ordering
determines
the complete ordering of the full diagram.

The last point to be clarified is the form of the amplitudes
corresponding to l.c.t.-ordered diagrams. We consider again the one-loop
diagram and write the covariant amplitude as follows:
\begin{equation}
FD= \int \frac{{\rm d}^4 k}{(2 \pi)^{4}}
 \frac{1}{[k^2_1 - m^2_1+i \epsilon]\cdots [k^2_N - m^2_N+i \epsilon]}
 \label{eq11}
\end{equation}
where for the sake of simplicity we have put all vertex functions equal to
unity. Using eq.~(\ref{eq10}) we can write for a typical factor in the
denominator:
\begin{eqnarray}
 k^2_i - m^2_i + i \epsilon & = & 2 k^+_i \left( k^-_i -
 \frac{k^2_{i \perp} + m^2_i - i \epsilon }{ 2 k^+_i} \right) \nonumber \\
 & = & 2 k^+_i \left( k^- + K^-_i -
 \frac{(k + K_i)^2_{\perp} + m^2_i - i \epsilon}{ 2 k^+_i} \right) \nonumber \\
 & \equiv & 2 k^+_i ( k^- - H_i)
 \label{eq12}
\end{eqnarray}
The poles in $k^-$, $H_j$, are functions
of the kinematical components of $k^\mu$ and the external momenta $p^\mu_j$.
The
imaginary part of $H_i$, $\Im H_i$, is determined by the sign of $k^+_i$. Now
suppose that for given external momenta $k^+$ is such that $m$ pole
positions are located in the upper half plane ($\Im H_i >0$) and $n=N-m$ in the
lower half plane ($\Im H_j <0$). In order to simplify the discussion, we
renumber the lines such that $\Im H_i >0$ for $1 \leq i \leq m$, $\Im H_j >0$
for $m+1 \leq j \leq N=m+n$. Consider the $k^-$-integral by itself:
\begin{equation}
 D_{m,n} = \int \frac{{\rm d} k^{-}}{2  \pi 2^N  k^+_1\cdots k^+_N}
 \frac{1}{[k^- - H_1]\cdots [k^--H_N]}
 \label{eq13}
\end{equation}
After performing the integral by closing the contour in either $\Im k^- > 0$ or
$\Im k^- < 0$ one obtains a rational function of the $H_i$'s:
\begin{equation}
 D_{m,n} = \frac{i}{2^N k^+_1 \cdots k^+_N}
 \frac{W_{m,n}(H_1, \cdots, H_m|H_{m+1}\cdots H_N)} {\Delta (H_1, \cdots, H_N)}
 \label{eq14}
\end{equation}
Details on the functions $W_{m,n}$ and $\Delta$ are given in
sect.~\ref{Chapter6}.
As we argued before, the $k^+$ integration
splits into $N+1$ intervals, $(-\infty, k^+(0)), ( k^+(0), k^+(1))$, $\cdots$,
$(k^+(N-1), +\infty)$, where the boundaries $k^+(i)$ are defined such that in
the interval $(k^+(r-1), k^+(r))$ there are, for finite $r$, precisely $r$
$H_i$'s with $\Im H_i > 0$. In
each of the finite intervals one skeleton graph is present corresponding to
one $k^-$-integral $D_{m,n}$. For either $n$ or $m$ different from 1, $D_{m,n}$
does not have the desired form eq.~(\ref{eq8}).
The full Feynman diagram is recovered by summing $D_{m,n}$ over $m$ from 1 to
$N-1$, integrating over $k^+$ over the appropiate finite interval, and over
$k_\perp$. A further reduction,
corresponding to the transition from the partially ordered skeleton graphs to
the completely ordered diagrams must be performed. The heuristics that help us
to do so is provided by the space-time concepts. Take any early vertex and
identify it with an {\it event} at some l.c.t. $x^+ = \tau_0$.
Suppose $H_1$ and $H_{m+1}$ are the poles corresponding to the outgoing and
incoming internal lines resp. at this vertex. The intermediate state with
momenta $k_1$ and $k_{m+1}$ corresponds to the l.c.-energy denominator
$H_1 - H_{m+1}$. We can convince ourselves that this is correct if the
reversal of four momenta on lines with $k^+_j < 0$ is effected. We find
\begin{equation}
H_1 - H_{m+1}=
 P^- - \frac{k^2_{1\perp} + m^2_1}{2 k^+_1}
            - \frac{k^2_{m+1\perp} + m^2_{m+1}}{2 k^+_{m+1}} .
 \label{eq15}
\end{equation}
The identification of $P^-$ becomes clear when one uses the surface $\tau$ to
cut the internal and external lines. We can close the surface by cutting all
the
external lines attached to the piece of the Feynman diagram out of which the
$p^+$ momentum flows.
The momentum flow through this first cut plane $\tau_0$ equals the flow through
the second cut plane $\tau$ since energy-momentum is conserved locally in
a Feynman diagram.
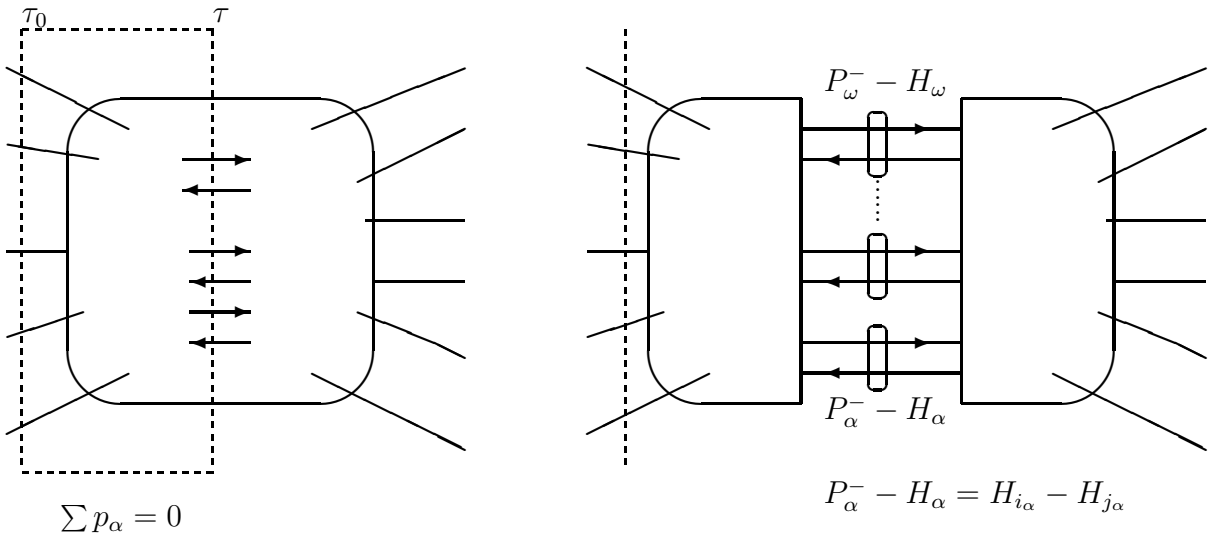
\begin{figure}
\begin{center}
\label{fig02a}
\setlength{\unitlength}{0.0080in}%
\begin{picture}(785,419)(90,376)
\put(625,755){$ P^-_\omega-H_\omega$}
\thicklines
\put(805,610){\line( 5,-2){ 50}}
\put(815,630){\line( 1, 0){ 40}}
\put(805,695){\line( 2, 1){ 70}}
\put(775,730){\line( 5, 2){100}}
\put(855,630){\line( 1, 0){ 20}}
\put(855,590){\line( 2,-1){ 20}}
\put(855,530){\line( 2,-1){ 20}}
\put(775,570){\line( 2,-1){ 80}}
\put(855,670){\line( 1, 0){ 20}}
\put(810,670){\line( 1, 0){ 45}}
\put(625,540){\makebox(0,0)[lb]{\raisebox{0pt}[0pt][0pt]{\frtnrm
$P^-_\alpha-H_\alpha$}}}
\put(715,650){\oval(200,200)[br]}
\put(715,650){\oval(200,200)[tr]}
\put(660,720){\oval(12,42)}
\put(660,640){\oval(12,42)}
\put(660,580){\oval(12,42)}
\put(660,695){\makebox(0.4444,0.6667){\tenrm .}}
\put(550,730){\line(-2, 1){ 80}}
\put(530,710){\line(-6, 1){ 59.189}}
\put(520,610){\line(-3,-1){ 49.500}}
\put(550,570){\line(-2,-1){ 80}}
\multiput(495,795)(0.00000,-8.02817){36}{\line( 0,-1){  4.014}}
\put(610,750){\line( 0,-1){200}}
\put(715,750){\line( 0,-1){200}}
\put(610,730){\vector( 1, 0){ 85}}
\put(715,710){\vector(-1, 0){ 90}}
\put(715,630){\vector(-1, 0){ 90}}
\put(610,590){\vector( 1, 0){ 85}}
\put(715,570){\vector(-1, 0){ 90}}
\put(510,650){\line(-1, 0){ 40}}
\put(100,800){$\tau_0$}
\put(225,800){$\tau$}
\multiput(100,795)(0.00000,-8.02817){36}{\line( 0,-1){  4.014}}
\multiput(225,795)(0.00000,-8.02817){36}{\line( 0,-1){  4.014}}
\multiput(225,795)(-8.02817,0.0){16}{\line(-1,0){  4.014}}
\multiput(225,505)(-8.02817,0.0){16}{\line(-1,0){  4.014}}
\put(170,730){\line(-2, 1){ 80}}
\put(150,710){\line(-6, 1){ 59.189}}
\put(130,650){\line(-1, 0){ 40}}
\put(140,610){\line(-3,-1){ 49.500}}
\put(170,570){\line(-2,-1){ 80}}
\put(125,470){$\sum p_\alpha =0$}
\put(320,610){\line( 5,-2){ 50}}
\put(330,630){\line( 1, 0){ 40}}
\put(320,695){\line( 2, 1){ 70}}
\put(290,730){\line( 5, 2){100}}
\put(370,630){\line( 1, 0){ 20}}
\put(370,590){\line( 2,-1){ 20}}
\put(370,530){\line( 2,-1){ 20}}
\put(290,570){\line( 2,-1){ 80}}
\put(370,670){\line( 1, 0){ 20}}
\put(325,670){\line( 1, 0){ 45}}
\put(625,485){$P^-_\alpha-H_\alpha = H_{i_\alpha}- H_{j_\alpha}$}
\put(610,650){\oval(200,200)[tl]}
\put(610,650){\oval(200,200)[bl]}
\put(230,650){\oval(200,200)}
\put(610,650){\vector( 1, 0){ 85}}
\put(695,730){\line( 1, 0){ 20}}
\put(625,710){\line(-1, 0){ 15}}
\put(695,650){\line( 1, 0){ 20}}
\put(625,630){\line(-1, 0){ 15}}
\put(695,590){\line( 1, 0){ 20}}
\put(625,570){\line(-1, 0){ 15}}
\put(660,690){\makebox(0.4444,0.6667){\tenrm .}}
\put(660,685){\makebox(0.4444,0.6667){\tenrm .}}
\put(660,680){\makebox(0.4444,0.6667){\tenrm .}}
\put(660,675){\makebox(0.4444,0.6667){\tenrm .}}
\put(660,670){\makebox(0.4444,0.6667){\tenrm .}}
\put(205,710){\vector( 1, 0){ 45}}
\put(250,690){\vector(-1, 0){ 45}}
\put(210,650){\vector( 1, 0){ 40}}
\put(250,630){\vector(-1, 0){ 40}}
\put(210,610){\vector( 1, 0){ 40}}
\put(250,590){\vector(-1, 0){ 40}}
\end{picture}
\caption{Illustration of the cut-procedure in a general n-leg diagram.
 For each cut area there is four-momentum conservation in a Feynman diagram. We
can use this to make general statements for an intermediate state. The
minus-momentum transferred across the intermediate state is equal to the
incoming minus-momentum. The sum of energy denominators contains the  sum of
all  external minus-momenta minus the sum of all on-shell  values of the
particle lines. Loop momenta drop out because these go in and out of the cut
area.}
 \end{center}
\end{figure}
The flow of $P^-$ through the internal lines equals the initial flow into the
diagram minus the flow through the external lines cut at $\tau$. The loop
momenta do not contribute because they go into the $\tau$ cut plane as well as
out of the cut plane, so there is no net contribution of these momenta.
(We stress here that this interpretation is correct only after reversal of the
four momenta on the lines with $\Im H_j > 0$.) For brevity we call denominators
of the form $H_i - H_j = P^- - H_0(i,j)$ {\it energy denominators}. These cut
planes can be interpreted as equal-time surfaces.

Now the strategy is clear. For every skeleton graph one uses the surface
$x^+ = \tau$ to cut lines that give rise to energy denominators. That this is
possible is the content of our proof of equivalence. Indeed, as we demonstrate
in sect.~\ref{Chapter6}, we have
\begin{eqnarray}
 \frac{W_{m,n}(H_1, \cdots, H_m|H_{m+1}, \cdots, H_N)}
 {\Delta (H_1, \cdots, H_N)} &=&
 \frac{1}{H_1 -H_{m+1}}   \times
 \label{eq16} \\
 \left ( \frac{W_{m-1,n}(H_2, \cdots, H_m|H_{m+1}, \cdots, H_N) }
 {\Delta (H_2, \cdots, H_m,H_{m+1}, \cdots, H_N)}   \right.
& + &  \left.
   \frac{W_{m,n-1}(H_1, \cdots, H_m|H_{m+2}, \cdots, H_N) }
 {\Delta (H_1, \cdots, H_m,H_{m+2}, \cdots, H_N)} \right)  \nonumber
\end{eqnarray}
That this identity leads to a recurrence follows from the fact that the two
terms in the r.h.s. of eq.(\ref{eq16}) have the same form as the original one.
The reduction stops if either $n$ or $m$ is reduced to 1.

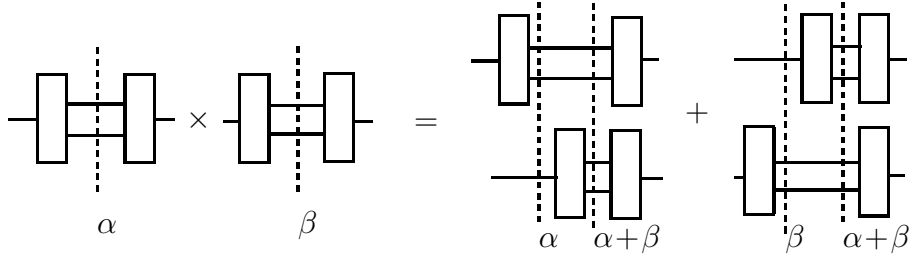
\begin{figure}
\begin{center}
\setlength{\unitlength}{0.007500in}%
\begin{picture}(626,165)(99,646)
\thicklines
\put(653,741){\framebox(19,60){}}
\put(693,741){\framebox(19,60){}}
\put(693,662){\framebox(19,60){}}
\put(613,663){\framebox(19,60){}}
\put(633,699){\line( 1, 0){ 60}}
\put(632,680){\line( 1, 0){ 61}}
\put(673,780){\line( 1, 0){ 21}}
\put(673,758){\line( 1, 0){ 20}}
\multiput(681,811)(0.00000,-7.85366){20}{\line( 0,-1){  3.927}}
\put(681,640){$\alpha\!+\!\beta$}
\multiput(641,809)(0.00000,-8.15385){20}{\line( 0,-1){  4.077}}
\put(641,640){$\beta$}
\put(653,771){\line(-1, 0){ 48}}
\put(613,689){\line(-1, 0){  8}}
\put(713,690){\line( 1, 0){ 11}}
\put(713,771){\line( 1, 0){ 12}}
\put(120,699){\framebox(19,60){}}
\put(181,699){\framebox(19,60){}}
\put(261,699){\framebox(19,60){}}
\put(320,700){\framebox(19,60){}}
\put(442,740){\framebox(19,60){}}
\put(521,739){\framebox(19,60){}}
\put(520,660){\framebox(19,60){}}
\put(481,661){\framebox(19,60){}}
\put(139,740){\line( 1, 0){ 42}}
\put(140,718){\line( 1, 0){ 41}}
\put(280,739){\line( 1, 0){ 40}}
\put(281,719){\line( 1, 0){ 39}}
\put(462,779){\line( 1, 0){ 58}}
\put(462,758){\line( 1, 0){ 59}}
\put(501,699){\line( 1, 0){ 19}}
\put(501,679){\line( 1, 0){ 19}}
\multiput(161,779)(0.00000,-8.00000){13}{\line( 0,-1){  4.000}}
\put(161,650){$\alpha$}
\multiput(301,780)(0.00000,-8.08000){13}{\line( 0,-1){  4.040}}
\put(301,650){$\beta$}
\multiput(507,810)(0.00000,-8.00000){20}{\line( 0,-1){  4.000}}
\put(507,640){$\alpha\!+\!\beta$}
\multiput(469,811)(0.00000,-7.85366){20}{\line( 0,-1){  3.927}}
\put(469,640){$\alpha$}
\put(120,729){\line(-1, 0){ 21}}
\put(200,729){\line(1,0){15}}
\put(261,728){\line(-1, 0){ 12}}
\put(339,728){\line( 1, 0){ 14}}
\put(482,689){\line(-1, 0){ 46}}
\put(441,770){\line(-1, 0){ 19}}
\put(541,771){\line( 1, 0){ 11}}
\put(540,688){\line( 1, 0){ 15}}
\put(223,723){\makebox(0,0)[lb]{\raisebox{0pt}[0pt][0pt]{\twlrm $\times$}}}
\put(382,722){\makebox(0,0)[lb]{\raisebox{0pt}[0pt][0pt]{\twlrm =}}}
\put(571,728){\makebox(0,0)[lb]{\raisebox{0pt}[0pt][0pt]{\twlrm +}}}
\end{picture}
\caption{The product of two time orderings is the sum of the relative
orderings.}\label{simultaneous}
\end{center}
\end{figure}

There remains one loose end that we tie up now. If several early vertices
occur,
we have to consider all l.c.t.-orderings of them. (This happens if $n \geq 2$.)
Moreover, we may have to consider l.c.t.-orderings such that some late vertices
occur before some, but not all, early vertices. In all those cases the number
of internal lines cut by an $x^+ = \tau$ surface is greater than two, but
always even: for every line going into this surface there is a corresponding
outgoing line. We have seen that a pair of lines, one incoming, the other
outgoing, that connect in an early (or late, for that matter) vertex, gives
rise to an energy denominator, say $H_i - H_j$. When two pairs of such lines
occur, there will be two energy denominators, which we call {\it simultaneous
parts}. A simple example illustrates this. Let the two early vertices be
$\alpha$ and $\beta$, and $p^\mu_\alpha$ and $p^\mu_\beta$ the momenta on the
two corresponding incoming external lines. The reduction algorithm gives two
factors, one corresponding to the vertex $\alpha$, of the form
$1/(H_{i_\alpha}-H_{j_\alpha})$, the other being $1/(H_{i_\beta}-H_{j_\beta})$.
As before, we can rewrite such factors, upon reversing the backward flowing
momenta, in the form $1/(P^-(\beta) - H_0(\beta))$ and
$1/(P^-(\alpha) - H_0(\alpha))$, where $P^-(\gamma), \gamma=\alpha, \beta$ is
the total net external minus momentum flowing into vertex $\gamma$. A simple
algebraic identity
\begin{eqnarray}
{1 \over P^-(\alpha) - H_0(\alpha)} \times {1 \over P^-(\beta) - H_0(\beta)}
 &=&
 \label{eq17} \\
{1 \over P^-(\alpha) + P^-(\beta) - H_0(\alpha) - H_0(\beta)} &\times  &
\left( {1 \over P^-(\alpha) - H_0(\alpha)} +{ 1 \over P^-(\beta) - H_0(\beta)}
\right) \nonumber
\end{eqnarray}
combines the two factors in the correct way.
The first factor can be rewritten as $1/(P^-(\alpha \cup \beta) - H_0
(\alpha \cup \beta) )$, which we recognize as an energy denominator for the
intermediate state with the four internal lines $i_\alpha$, $j_\alpha$,
$i_\beta$ and $j_\beta$. The combination $1/(P^-(\alpha) - H_0(\alpha))
(P^-(\alpha \cup \beta) - H_0(\alpha \cup \beta) )$ corresponds to the
l.c.t.-ordered diagram where the vertex $\alpha$ comes before vertex $\beta$,
the
other part of the r.h.s. of eq.(\ref{eq17}) corresponding of course to vertex
$\beta$ preceding vertex $\alpha$ (see fig.~\ref{simultaneous}). Clearly, the
splitting formula
works also recursively, so it applies to any number of pairs of internal lines.

With this observation we end our general discussion of the reduction algorithm.
We stress that the l.c.t.-ordered language used here has heuristic value only,
but does not replace a strict proof. The algebraic details are provided in
section~\ref{Chapter6}.

\section{Spin-1/2 particles}
\label{Chapter3}
In the previous sections we dealt with scalar particles only, thus avoiding
complications due to summations over spin degrees of freedom in intermediate
states. Here, we discuss these complications for spin-1/2 particles. The
reduction algorithm in this case is partly identical to the algorithm for
scalar particles. However, we now have to include in our treatment not only
the energy denominators, but also the numerators.
Consider the Feynman propagator for a single particle. The spin sum
$p\!\!\!/ + m$ depends on $p^-$, so we have to account for that when we define
the skeleton graphs corresponding to a Feynman diagram.
It has been argued before \cite{KS70} that one can split the Feynman propagator
into two pieces, one that is independent of $p^-$, the {\it instantaneous
part},
and another piece, the {\it propagating part}, where $p^-$ occurs in the
denominator only:
\begin{equation}
{p\!\!\!/+m \over p^2-m^2+i \epsilon}= {p\!\!\!/_{on-shell}+m \over p^2-m^2+ i
\epsilon} + {\gamma^+ \over 2 p^+} = \sum_\alpha { u^{(\alpha)}(p) \otimes
\bar u^{(\alpha)}(p) \over p^2-m^2 + i \epsilon} + {\gamma^+ \over 2 p^+} ,
 \label{eq3.1}
\end{equation}
with the obvious definition $p^\mu_{on-shell} = (p^-_{on-shell}, p^+,
p_\perp$),
where $p^-_{on-shell} = (p^2_\perp + m^2)/(2p^+)$. The spin sum
$ \sum_\alpha  u^{(\alpha)}(p) \otimes \bar u^{(\alpha)}(p)$ runs over a
complete basis in spin space, {\it viz}, particles and anti-particles.

In order to illustrate the differences between the purely scalar case and the
situation where spin-1/2 particles occur, we discuss in the next subsection
the flat box diagram with two bosons and two fermions.

\subsection{Example: fermion box diagram}
\begin{figure}
\begin{center}
\setlength{\unitlength}{0.0090in}%
\begin{picture}(660,865)(45,-70)
\thicklines
\put( 60,780){\line( 1,-1){ 20}}
\put( 80,760){\line( 1, 0){ 60}}
\put(140,760){\line( 1, 1){ 20}}
\put( 60,680){\line( 1, 1){ 20}}
\put( 80,700){\line( 1, 0){ 60}}
\put(140,700){\line( 1,-1){ 20}}
\multiput( 80,760)(0.00000,-8.57143){8}{\makebox(0.4444,0.6667){\tenrm .}}
\multiput(140,760)(0.00000,-8.57143){8}{\makebox(0.4444,0.6667){\tenrm .}}
\put(200,780){\line( 1,-1){ 20}}
\put(220,760){\line( 1, 0){ 60}}
\put(280,760){\line( 1, 1){ 20}}
\put(200,680){\line( 1, 1){ 20}}
\put(220,700){\line( 1, 0){ 60}}
\put(280,700){\line( 1,-1){ 20}}
\multiput(220,760)(0.00000,-8.57143){8}{\makebox(0.4444,0.6667){\tenrm .}}
\multiput(280,760)(0.00000,-8.57143){8}{\makebox(0.4444,0.6667){\tenrm .}}
\multiput(340,700)(4.28571,8.57143){8}{\makebox(0.4444,0.6667){\tenrm .}}
\multiput(400,700)(-4.28571,8.57143){8}{\makebox(0.4444,0.6667){\tenrm .}}
\put(320,680){\line( 1, 1){ 20}}
\put(340,700){\line( 1, 0){ 60}}
\put(320,780){\line( 5,-2){ 50}}
\put(370,760){\line( 5, 2){ 50}}
\put(400,700){\line( 1,-1){ 20}}
\multiput(460,760)(4.28571,-8.57143){8}{\makebox(0.4444,0.6667){\tenrm .}}
\multiput(490,700)(4.28571,8.57143){8}{\makebox(0.4444,0.6667){\tenrm .}}
\put(440,680){\line( 5, 2){ 50}}
\put(490,700){\line( 5,-2){ 50}}
\put(440,780){\line( 1,-1){ 20}}
\put(460,760){\line( 1, 0){ 60}}
\put(520,760){\line( 1, 1){ 20}}
\multiput(580,760)(-5.00000,-10.00000){3}{\makebox(0.4444,0.6667){\tenrm .}}
\multiput(570,740)(0.00000,-10.00000){3}{\makebox(0.4444,0.6667){\tenrm .}}
\multiput(570,720)(5.00000,-10.00000){3}{\makebox(0.4444,0.6667){\tenrm .}}
\multiput(580,760)(5.00000,-10.00000){3}{\makebox(0.4444,0.6667){\tenrm .}}
\multiput(590,740)(0.00000,-10.00000){3}{\makebox(0.4444,0.6667){\tenrm .}}
\multiput(590,720)(-5.00000,-10.00000){3}{\makebox(0.4444,0.6667){\tenrm .}}
\put(560,780){\line( 1,-1){ 20}}
\put(580,760){\line( 1, 1){ 20}}
\put(560,680){\line( 1, 1){ 20}}
\put(580,700){\line( 1,-1){ 20}}
\put(180,640){\line( 1,-1){ 20}}
\put(200,620){\line( 1, 0){ 60}}
\put(260,620){\line( 1, 1){ 20}}
\put(180,540){\line( 2, 1){ 40}}
\put(220,560){\line( 1, 0){ 20}}
\put(240,560){\line( 2,-1){ 40}}
\multiput(200,620)(1.81818,-5.45455){12}{\makebox(0.4444,0.6667){\tenrm .}}
\multiput(260,620)(-1.81818,-5.45455){12}{\makebox(0.4444,0.6667){\tenrm .}}
\put(300,640){\line( 2,-1){ 40}}
\put(340,620){\line( 1, 0){ 40}}
\put(380,620){\line( 1, 1){ 20}}
\put(300,540){\line( 1, 1){ 20}}
\put(320,560){\line( 1, 0){ 40}}
\put(360,560){\line( 2,-1){ 40}}
\multiput(340,620)(-1.81818,-5.45455){12}{\makebox(0.4444,0.6667){\tenrm .}}
\multiput(380,620)(-1.81818,-5.45455){12}{\makebox(0.4444,0.6667){\tenrm .}}
\multiput(440,560)(4.28571,8.57143){8}{\makebox(0.4444,0.6667){\tenrm .}}
\multiput(500,560)(-4.28571,8.57143){8}{\makebox(0.4444,0.6667){\tenrm .}}
\put(420,540){\line( 1, 1){ 20}}
\put(440,560){\line( 1, 0){ 60}}
\put(420,640){\line( 5,-2){ 50}}
\put(470,620){\line( 5, 2){ 50}}
\put(500,560){\line( 1,-1){ 20}}
\multiput( 80,500)(4.28571,-8.57143){8}{\makebox(0.4444,0.6667){\tenrm .}}
\multiput(110,440)(4.28571,8.57143){8}{\makebox(0.4444,0.6667){\tenrm .}}
\put( 60,420){\line( 5, 2){ 50}}
\put(110,440){\line( 5,-2){ 50}}
\put( 60,520){\line( 1,-1){ 20}}
\put( 80,500){\line( 1, 0){ 60}}
\put(140,500){\line( 1, 1){ 20}}
\put(180,420){\line( 1, 1){ 20}}
\put(200,440){\line( 1, 0){ 40}}
\put(240,440){\line( 2,-1){ 40}}
\put(180,520){\line( 4,-1){ 80}}
\put(260,500){\line( 1, 1){ 20}}
\multiput(260,500)(-4.28571,-4.28571){15}{\makebox(0.4444,0.6667){\tenrm .}}
\multiput(260,500)(-1.81818,-5.45455){12}{\makebox(0.4444,0.6667){\tenrm .}}
\put(300,420){\line( 1, 1){ 20}}
\put(320,440){\line( 4,-1){ 80}}
\put(300,520){\line( 2,-1){ 40}}
\put(340,500){\line( 1, 0){ 40}}
\put(380,500){\line( 1, 1){ 20}}
\multiput(320,440)(1.81818,5.45455){12}{\makebox(0.4444,0.6667){\tenrm .}}
\multiput(380,500)(-4.28571,-4.28571){15}{\makebox(0.4444,0.6667){\tenrm .}}
\multiput(490,500)(-4.00000,-4.00000){6}{\makebox(0.4444,0.6667){\tenrm .}}
\multiput(470,480)(-2.50000,-5.00000){5}{\makebox(0.4444,0.6667){\tenrm .}}
\multiput(460,460)(0.00000,-6.66667){4}{\makebox(0.4444,0.6667){\tenrm .}}
\multiput(460,440)(4.00000,4.00000){6}{\makebox(0.4444,0.6667){\tenrm .}}
\multiput(480,460)(2.50000,5.00000){5}{\makebox(0.4444,0.6667){\tenrm .}}
\multiput(490,480)(0.00000,6.66667){4}{\makebox(0.4444,0.6667){\tenrm .}}
\put(420,420){\line( 2, 1){ 40}}
\put(460,440){\line( 3,-1){ 60}}
\put(420,520){\line( 4,-1){ 70.588}}
\put(490,500){\line( 3, 2){ 30}}
\put( 60,640){\line( 2,-1){ 40}}
\put(100,620){\line( 1, 0){ 20}}
\put(120,620){\line( 2, 1){ 40}}
\put( 60,540){\line( 1, 1){ 20}}
\put( 80,560){\line( 1, 0){ 60}}
\put(140,560){\line( 1,-1){ 20}}
\multiput(100,620)(-1.81818,-5.45455){12}{\makebox(0.4444,0.6667){\tenrm .}}
\multiput(120,620)(1.81818,-5.45455){12}{\makebox(0.4444,0.6667){\tenrm .}}
\put( 60,400){\line( 2,-1){ 40}}
\put(100,380){\line( 1, 0){ 20}}
\put(120,380){\line( 2, 1){ 40}}
\put( 60,300){\line( 1, 1){ 20}}
\put( 80,320){\line( 1, 0){ 60}}
\put(140,320){\line( 1,-1){ 20}}
\multiput(100,380)(-1.81818,-5.45455){12}{\makebox(0.4444,0.6667){\tenrm .}}
\multiput(120,380)(1.81818,-5.45455){12}{\makebox(0.4444,0.6667){\tenrm .}}
\multiput(205,320)(4.28571,8.57143){8}{\makebox(0.4444,0.6667){\tenrm .}}
\multiput(265,320)(-4.28571,8.57143){8}{\makebox(0.4444,0.6667){\tenrm .}}
\put(185,300){\line( 1, 1){ 20}}
\put(205,320){\line( 1, 0){ 60}}
\put(185,400){\line( 5,-2){ 50}}
\put(235,380){\line( 5, 2){ 50}}
\put(265,320){\line( 1,-1){ 20}}
\put( 60,280){\line( 1,-1){ 20}}
\put( 80,260){\line( 1, 0){ 60}}
\put(140,260){\line( 1, 1){ 20}}
\put( 60,180){\line( 2, 1){ 40}}
\put(100,200){\line( 1, 0){ 20}}
\put(120,200){\line( 2,-1){ 40}}
\multiput( 80,260)(1.81818,-5.45455){12}{\makebox(0.4444,0.6667){\tenrm .}}
\multiput(140,260)(-1.81818,-5.45455){12}{\makebox(0.4444,0.6667){\tenrm .}}
\multiput(200,260)(4.28571,-8.57143){8}{\makebox(0.4444,0.6667){\tenrm .}}
\multiput(230,200)(4.28571,8.57143){8}{\makebox(0.4444,0.6667){\tenrm .}}
\put(180,180){\line( 5, 2){ 50}}
\put(230,200){\line( 5,-2){ 50}}
\put(180,280){\line( 1,-1){ 20}}
\put(200,260){\line( 1, 0){ 60}}
\put(260,260){\line( 1, 1){ 20}}
\put(180, 60){\line( 1, 1){ 20}}
\put(200, 80){\line( 1, 0){ 40}}
\put(240, 80){\line( 2,-1){ 40}}
\put(180,160){\line( 4,-1){ 80}}
\put(260,140){\line( 1, 1){ 20}}
\multiput(260,140)(-4.28571,-4.28571){15}{\makebox(0.4444,0.6667){\tenrm .}}
\multiput(260,140)(-1.81818,-5.45455){12}{\makebox(0.4444,0.6667){\tenrm .}}
\put(300, 60){\line( 1, 1){ 20}}
\put(320, 80){\line( 4,-1){ 80}}
\put(300,160){\line( 2,-1){ 40}}
\put(340,140){\line( 1, 0){ 40}}
\put(380,140){\line( 1, 1){ 20}}
\multiput(320, 80)(1.81818,5.45455){12}{\makebox(0.4444,0.6667){\tenrm .}}
\multiput(380,140)(-4.28571,-4.28571){15}{\makebox(0.4444,0.6667){\tenrm .}}
\multiput(490,140)(-4.00000,-4.00000){6}{\makebox(0.4444,0.6667){\tenrm .}}
\multiput(470,120)(-2.50000,-5.00000){5}{\makebox(0.4444,0.6667){\tenrm .}}
\multiput(460,100)(0.00000,-6.66667){4}{\makebox(0.4444,0.6667){\tenrm .}}
\multiput(460, 80)(4.00000,4.00000){6}{\makebox(0.4444,0.6667){\tenrm .}}
\multiput(480,100)(2.50000,5.00000){5}{\makebox(0.4444,0.6667){\tenrm .}}
\multiput(490,120)(0.00000,6.66667){4}{\makebox(0.4444,0.6667){\tenrm .}}
\put(420, 60){\line( 2, 1){ 40}}
\put(460, 80){\line( 3,-1){ 60}}
\put(420,160){\line( 4,-1){ 70.588}}
\put(490,140){\line( 3, 2){ 30}}
\put( 60, 40){\line( 2,-1){ 40}}
\put(100, 20){\line( 1, 0){ 40}}
\put(140, 20){\line( 1, 1){ 20}}
\put( 60,-60){\line( 1, 1){ 20}}
\put( 80,-40){\line( 1, 0){ 40}}
\put(120,-40){\line( 2,-1){ 40}}
\multiput(100, 20)(-1.81818,-5.45455){12}{\makebox(0.4444,0.6667){\tenrm .}}
\multiput(140, 20)(-1.81818,-5.45455){12}{\makebox(0.4444,0.6667){\tenrm .}}
\put(280, 40){\line( 2,-1){ 40}}
\put(320, 20){\line( 1, 0){ 40}}
\put(360, 20){\line( 1, 1){ 20}}
\put(280,-60){\line( 1, 1){ 20}}
\put(300,-40){\line( 1, 0){ 40}}
\put(340,-40){\line( 2,-1){ 40}}
\multiput(320, 20)(-1.81818,-5.45455){12}{\makebox(0.4444,0.6667){\tenrm .}}
\multiput(360, 20)(-1.81818,-5.45455){12}{\makebox(0.4444,0.6667){\tenrm .}}
\put(320,355){\line( 1, 0){ 20}}
\put(320,345){\line( 1, 0){ 20}}
\put(320,235){\line( 1, 0){ 20}}
\put(320,225){\line( 1, 0){ 20}}
\put(200, -5){\line( 1, 0){ 20}}
\put(200,-15){\line( 1, 0){ 20}}
\put(540,115){\line( 1, 0){ 20}}
\put(540,105){\line( 1, 0){ 20}}
\put(380,400){\line( 2,-1){ 40}}
\put(420,380){\line( 1, 0){ 20}}
\put(440,380){\line( 2, 1){ 40}}
\put(380,300){\line( 1, 1){ 20}}
\put(400,320){\line( 1, 0){ 60}}
\put(460,320){\line( 1,-1){ 20}}
\multiput(420,380)(-1.81818,-5.45455){12}{\makebox(0.4444,0.6667){\tenrm .}}
\multiput(440,380)(1.81818,-5.45455){12}{\makebox(0.4444,0.6667){\tenrm .}}
\put(380,280){\line( 1,-1){ 20}}
\put(400,260){\line( 1, 0){ 60}}
\put(460,260){\line( 1, 1){ 20}}
\put(380,180){\line( 2, 1){ 40}}
\put(420,200){\line( 1, 0){ 20}}
\put(440,200){\line( 2,-1){ 40}}
\multiput(400,260)(1.81818,-5.45455){12}{\makebox(0.4444,0.6667){\tenrm .}}
\multiput(460,260)(-1.81818,-5.45455){12}{\makebox(0.4444,0.6667){\tenrm .}}
\put( 60,160){\line( 3,-1){ 60}}
\put(120,140){\line( 1, 0){ 20}}
\put(140,140){\line( 1, 1){ 20}}
\put(160, 60){\line(-3, 1){ 60}}
\put(100, 80){\line(-1, 0){ 20}}
\put( 80, 80){\line(-1,-1){ 20}}
\multiput(140,140)(-3.33333,-5.00000){13}{\makebox(0.4444,0.6667){\tenrm .}}
\multiput(120,140)(-3.33333,-5.00000){13}{\makebox(0.4444,0.6667){\tenrm .}}
\put(580,160){\line( 3,-1){ 60}}
\put(640,140){\line( 1, 0){ 20}}
\put(660,140){\line( 1, 1){ 20}}
\put(680, 60){\line(-3, 1){ 60}}
\put(620, 80){\line(-1, 0){ 20}}
\put(600, 80){\line(-1,-1){ 20}}
\multiput(660,140)(-3.33333,-5.00000){13}{\makebox(0.4444,0.6667){\tenrm .}}
\multiput(640,140)(-3.33333,-5.00000){13}{\makebox(0.4444,0.6667){\tenrm .}}
\put(170,360){\line( 0,-1){ 20}}
\put(160,350){\line( 1, 0){ 20}}
\put(170,120){\line( 0,-1){ 20}}
\put(160,110){\line( 1, 0){ 20}}
\put(290,120){\line( 0,-1){ 20}}
\put(280,110){\line( 1, 0){ 20}}
\put(410,120){\line( 0,-1){ 20}}
\put(400,110){\line( 1, 0){ 20}}
\put(175,240){\line( 0,-1){ 20}}
\put(165,230){\line( 1, 0){ 20}}
\put(640,140){\circle*{6}}
\put(660,140){\circle*{6}}
\put(660,141){\line(-1,0){20}}
\put(602, 81){\circle*{6}}
\put(622,82){\line(-1,0){20}}
\put(441,200){\circle*{6}}
\put(441,201){\line(-1,0){21}}
\put(420,200){\circle*{6}}
\put(419,380){\circle*{6}}
\put(441,381){\line(-1,0){22}}
\put(441,380){\circle*{6}}
\put(622, 81){\circle*{6}}
\put(320,-40){\vector(-1, 0){ 10}}
\put(340, 20){\vector(-1, 0){ 10}}
\put(120, 20){\vector(-1, 0){ 10}}
\put(105,-40){\vector(-1, 0){ 10}}
\put( 95, 80){\vector(-1, 0){ 10}}
\put(130,140){\vector(-1, 0){ 10}}
\put(225, 80){\vector(-1, 0){ 10}}
\put(360,140){\vector(-1, 0){ 10}}
\put(655,140){\vector(-1, 0){ 10}}
\put(615, 80){\vector(-1, 0){ 10}}
\put(435,200){\vector(-1, 0){ 10}}
\put(435,320){\vector(-1, 0){ 10}}
\put(240,320){\vector(-1, 0){ 10}}
\put(110,260){\vector(-1, 0){ 10}}
\put(230,260){\vector(-1, 0){ 10}}
\put(110,320){\vector(-1, 0){ 10}}
\put(225,440){\vector(-1, 0){ 10}}
\put(430,260){\vector(-1, 0){ 10}}
\put(435,380){\vector(-1, 0){ 10}}
\put(115,380){\vector(-1, 0){ 10}}
\put(115,200){\vector(-1, 0){ 10}}
\put(115,500){\vector(-1, 0){ 10}}
\put(110,560){\vector(-1, 0){ 10}}
\put(115,620){\vector(-1, 0){ 10}}
\put(230,620){\vector(-1, 0){ 10}}
\put(235,560){\vector(-1, 0){ 10}}
\put(360,620){\vector(-1, 0){ 10}}
\put(345,560){\vector(-1, 0){ 10}}
\put(470,560){\vector(-1, 0){ 10}}
\put(360,500){\vector(-1, 0){ 10}}
\put(500,760){\vector(-1, 0){ 10}}
\put(375,700){\vector(-1, 0){ 10}}
\put(250,760){\vector(-1, 0){ 10}}
\put(110,760){\vector(-1, 0){ 10}}
\put(115,700){\vector(-1, 0){ 10}}
\put(255,700){\vector(-1, 0){ 10}}
\put(190,665){\framebox(420,130){}}
\put( 45,410){\framebox(495,240){}}
\put( 45,-70){\framebox(650,480){}}
\put(100,675){\makebox(0,0)[lb]{\raisebox{0pt}[0pt][0pt]{\twlrm (a) }}}
\put(615,670){\makebox(0,0)[lb]{\raisebox{0pt}[0pt][0pt]{\twlrm (b)}}}
\put(545,420){\makebox(0,0)[lb]{\raisebox{0pt}[0pt][0pt]{\twlrm (c)}}}
\put(705,-70){\makebox(0,0)[lb]{\raisebox{0pt}[0pt][0pt]{\twlrm (d)}}}
\put(360,775){\makebox(0,0)[lb]{\raisebox{0pt}[0pt][0pt]{\twlrm $\gamma^+$}}}
\put(480,675){\makebox(0,0)[lb]{\raisebox{0pt}[0pt][0pt]{\twlrm $\gamma^+$}}}
\put(570,675){\makebox(0,0)[lb]{\raisebox{0pt}[0pt][0pt]{\twlrm $\gamma^+$}}}
\put(570,775){\makebox(0,0)[lb]{\raisebox{0pt}[0pt][0pt]{\twlrm $\gamma^+$}}}
\end{picture}
\caption
{(a) The Feynman diagram, (b) the corresponding diagrams with on-shell
spinor projections or instantaneous parts, (c) the skeleton graphs,
(only the tilted box leads to two l.c.t.-ordered diagrams) (d) the summed
l.c.t-ordered diagrams, which yields the adjusted blink propagators.
 \label{fig3.1}
}
\end{center}
\end{figure}
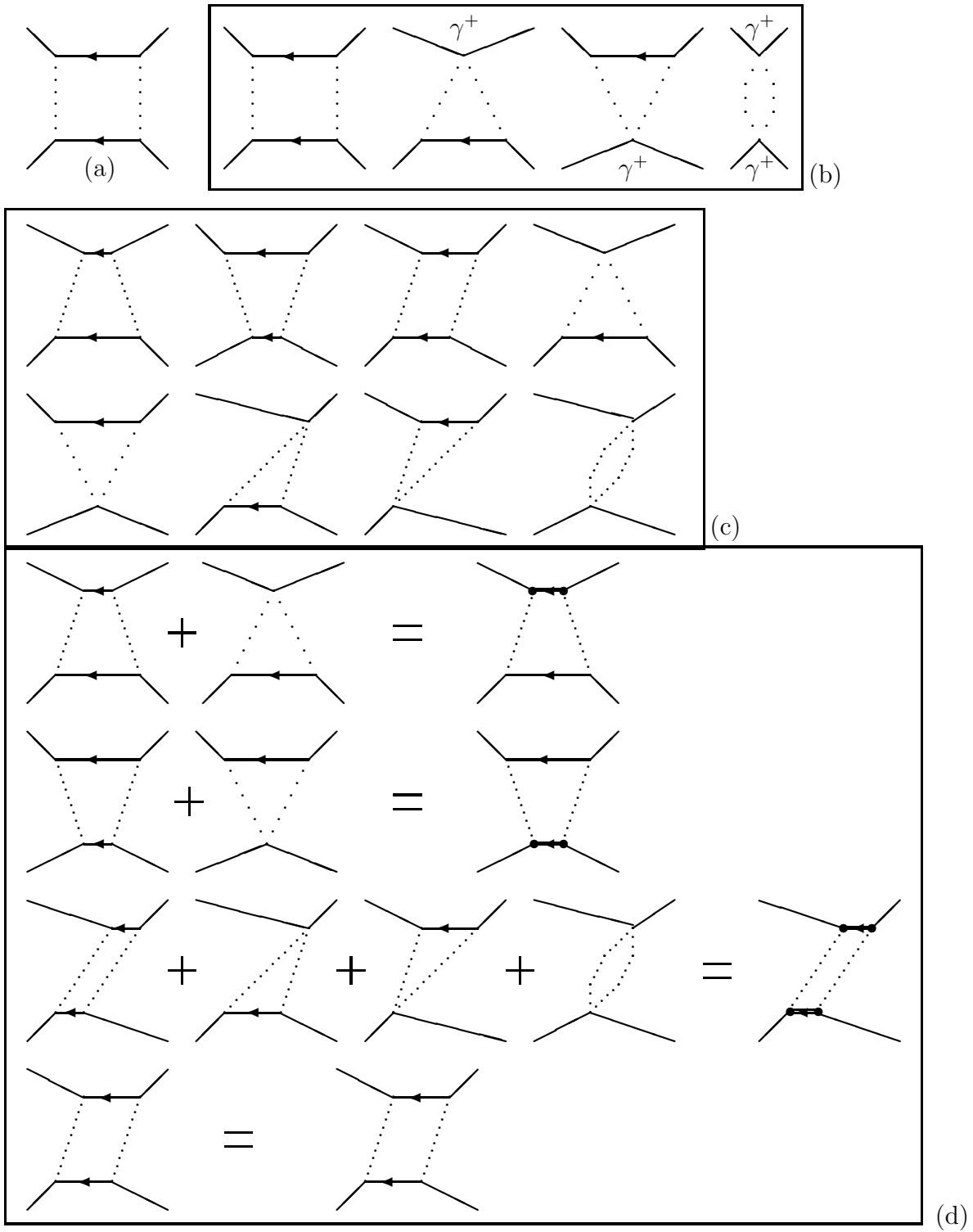
Consider the case of two spin-1/2 fermions exchanging spinless bosons.
A Feynman diagram for the fourth order in perturbation theory is shown in
fig.~\ref{fig3.1}, (a). The momenta are defined similar to the scalar case,
 ~fig.~\ref{box}. Before the associated skeleton graph can be
drawn, one must split the fermion propagators for the intermediate states
into the two parts: instantaneous and propagating.
This results in four different diagrams shown in fig.~\ref{fig3.1} (b).
In these four, the $p^-$-dependence of the propagators is present in the
denominators only. Therefore, one can apply the methods described in the
previous
section immediately, since the numerator does not depend on the integration
variable. Doing so, one obtains the eight skeleton graphs drawn in
panel (c). These graphs form the basis of the splitting into l.c.t-ordered
diagrams. After this has been achieved one can combine certain diagrams
into a single diagram by adding the propagating part to the instantaneous
part of the fermion propagator. This regrouping of diagrams is represented
graphically in the last panel of this figure. The formulae associated with the
four final diagrams are

\begin{eqnarray}
 D^{(1)} & = & \int {{\rm d} k^+ {\rm d}^2 k_\perp \over (2 \pi)^3\phi}
 \frac{\gamma^a \Lambda_3 \gamma^b \otimes \gamma^c \Lambda_1 \gamma^d}
 {(P^- - H_0(1,4)) (P^- - H_0(1,3)) (P^- - H_0(1,2))}
 \nonumber \\
 & + & \int {{\rm d} k^+ {\rm d}^2 k_\perp \over (2\pi)^3\phi}
 \frac{\gamma^a \gamma^+ \gamma^b \otimes \gamma^c \Lambda_1 \gamma^d}
 {(P^- - H_0(1,4)) (P^- - H_0(1,2))}
 \nonumber \\
 & = & \int {{\rm d} k^+ {\rm d}^2 k_\perp \over (2\pi)^3\phi}
 \frac{\gamma^a \Omega_3 \gamma^b \otimes \gamma^c \Lambda_1 \gamma^d}
 {(P^- - H_0(1,4)) (P^- - H_0(1,3)) (P^- - H_0(1,2))} ,
 \nonumber \\
 D^{(2)} & = & \int {{\rm d} k^+ {\rm d}^2 k_\perp \over (2\pi)^3\phi}
 \frac{\gamma^a \Lambda_3 \gamma^b \otimes \gamma^c \Omega_1 \gamma^d}
 {(P^- - H_0(1,4)) (P^- - H_0(1,3)) (P^- - H_0(2,3))}  ,
 \nonumber \\
 D^{(3)} & = & \int {{\rm d} k^+ {\rm d}^2 k_\perp \over (2\pi)^3\phi}
 \frac{\gamma^a \Omega_3 \gamma^b \otimes \gamma^c \Omega_1 \gamma^d}
 {(P^- - H_0(1,4)) (P^- - H_0(2,4)) (P^- - H_0(2,3))} ,
 \nonumber \\
 D^{(4)} & = & \int {{\rm d} k^+ {\rm d}^2 k_\perp \over (2\pi)^3\phi}
 \frac{\gamma^a \Lambda_3 \gamma^b \otimes \gamma^c \Lambda_1 \gamma^d}
 {(P^- - H_0(1,4)) (P^- - H_0(1,3)) (P^- - H_0(2,3))} .
 \label{eq3.6}
\end{eqnarray}

The objects $\Lambda$ and $\Omega$ are defined as follows:
\begin{equation}
 \Lambda_i = k\!\!\!/_{i, on} + m_i ,
 \label{eq3.7}
\end{equation}
and
\begin{equation}
 \Omega_i =  k\!\!\!/_i + m_i = \Lambda_i + \gamma^+ (k^-_i - k^-_{on}).
 \label{eq3.8}
\end{equation}
The on-shell values of the minus components have been defined before, see
eq.~(\ref{eq5}). The energy denominator $P^- - H_0(1,2)$ is of course
\begin{eqnarray}
 P^- - H_0(1,2) & = & p^- + q^- - k^-_{1,on} - k^-_{2,on} - q^-_{on} \nonumber
\\
               & = & p^- - \frac{k^2_{1\perp} + m^2_1}{2 k^+_1}
 - \frac{k^2_{2\perp} + m^2_2}{2 k^+_2} ,
 \label{eq3.9}
\end{eqnarray}
and the other ones are defined similarly. The phase-space element $\phi$
has also been given before,  below eq.~(\ref{eq2.4}).

The flat fermion box shows clearly the peculiar complications caused by spin.
Because the numerators and the denominators of the fermion propagators depend
both linearly on the integration variable $k^-$, one has to perform a
Laurent expansion in order to identify the pole terms. This leads to a number
of "intermediate amplitudes", equal to $2^F$, where $F$ is the number of
internal fermion lines. These amplitudes give rise to skeleton graphs that can
be reduced to l.c.t.-ordered diagrams in the, by now, familiar way. The
 l.c.t.-ordered diagrams in which an element $\gamma^+ / 2 k^+_i$ occurs, are
special, because associated to every one of them there occurs a diagram where
the
element $\gamma^+ / 2 k^+_i$ is replaced by $\Lambda_i/(2k_i^+(P^--H_0))$. This
happens only in
those cases where the on-shell value $H_i = (k^2_i + m_i) /2 k^+_i$ occurs in
a single energy denominator. These are the states that begin and end with the
creation and the annihilation of the same particle. We call an internal line
with this property a
{\it blink}. In the case considered above: $D^{(1)}$ contains $H_3$ in only
one factor in the denominator, the same holds for $D^{(2)}$ and $H_1$, whereas
$D^{(3)}$ contains the blinks $H_1$ and $H_3$. If a blink occurs, one can
recombine, after the l.c.t.-ordering has been performed, the propagating part
and the instantaneous part into a complete propagator. This is done in the
diagrams $D^{(1)}$, $D^{(2)}$ and $D^{(3)}$ eq.~(\ref{eq3.6}), and illustrated
in
fig.~\ref{fig3.1}(d), where the thick lines beginning and terminating in  dots
symbolize complete propagators.

\subsection{Including the instantaneous terms}

By now, it is relatively easy to formulate the general reduction algorithm. It
has four steps.
\begin{itemize}
 \item[(i)] For a given Feynman diagram, perform the Laurent expansion of
the fermion propagator, {\it i.e.}, split the propagator into a propagating
part and an instantaneous part;

\item[(ii)] Determine the skeleton graphs for all diagrams obtained in step
(i);

\item[(iii)] Perform the reduction of all skeleton graphs in exactly the same
way as it was done in the scalar case;

\item[(iv)] Identify the blinks and sum the diagrams corresponding to the same
blink in order to obtain amplitudes with complete spin sums.

\end{itemize}

\begin{figure}
\begin{center}
\setlength{\unitlength}{0.007500in}%
\begin{picture}(740,160)(80,600)
\thicklines
\multiput(170,620)(0,8){20}{\line(0,1){4}}
\put( 80,680){\line( 1, 0){180}}
\put( 80,670){\line( 1, 0){180}}
\put( 80,660){\line( 1, 0){180}}
\put( 80,640){\line( 1, 0){180}}
\put(360,680){\line( 1, 0){180}}
\put(360,670){\line( 1, 0){180}}
\put(360,660){\line( 1, 0){180}}
\put(360,640){\line( 1, 0){180}}
\put(640,680){\line( 1, 0){180}}
\put(640,670){\line( 1, 0){180}}
\put(640,660){\line( 1, 0){180}}
\put(640,640){\line( 1, 0){180}}
\put(200,720){\line( 3,-1){ 60}}
\put(140,720){\line(-3,-1){ 60}}
\put( 80,760){\line( 3,-2){ 60}}
\put(140,720){\line( 1, 0){ 60}}
\put(200,720){\line( 3, 2){ 60}}
\multiput(730,620)(0,8){20}{\line(0,1){4}}
\put(740,760){$G_i+g_i$}
\put(180,760){$G_i$}
\put(760,722){\circle*{8}}
\put(701,723){\circle*{8}}
\put(320,710){\line( 0,-1){ 20}}
\put(310,700){\line( 1, 0){ 20}}
\put(440,750){\line( -6,1){ 79.460}}
\put(440,750){\line( 0,-1){ 40}}
\put(440,710){\line( 5,-1){100}}
\put(440,710){\line( 2, 1){100}}
\put(440,750){\line(-5,-3){ 80.882}}
\put(435,730){\line( 1, 0){ 10}}
\put(445,760){$g_i$}
\put(580,705){\line( 1, 0){ 20}}
\put(580,695){\line( 1, 0){ 20}}
\put(760,720){\line( 3,-1){ 60}}
\put(640,700){\line( 3, 1){ 60}}
\put(700,725){\line(-5, 3){ 59.559}}
\put(760,725){\line( 5, 3){ 59.559}}
\put(700,720){\line( 1, 0){ 60}}
\put(700,725){\line( 1, 0){ 60}}
\end{picture}
\caption{The state begins and ends with the creation and annihilation of the
same particle. The singularity in this diagram cancels against the same
singularity in the instantaneous diagram.} \label{onestate}
\end{center}
\end{figure}
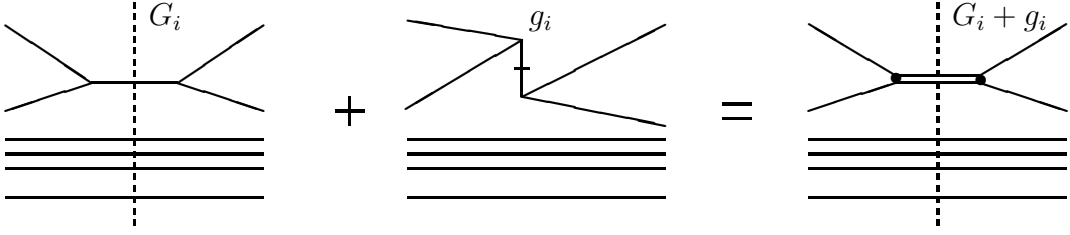

In order to understand why we recommend step (iv) we consider the general case.
Let $k^\mu_i$ be the four momentum of a blink. The two corresponding diagrams,
partly shown in fig.~\ref{onestate}, contain the factors\footnote{We define
$\gamma_\perp = (0, \gamma_1,\gamma_2,0)$, $k_\perp= (0,k_1,k_2,0)$ and
$\gamma_\perp \cdot k_\perp = -(\gamma_1 k_1+\gamma_2 k_2)$.}
\begin{eqnarray}
 G_i & = & \frac{1}{2 k^+_i} \frac{\Lambda_i}{P^- - ( \cdots H_i \cdots)} ,
 \nonumber \\
 \Lambda_i & = & \gamma^+ H_i + \gamma^- k^+_i + \gamma_\perp \cdot k_\perp
 + m_i ,
 \nonumber \\
 H_i & = & ( k^2_{i \perp} + m_i ) / 2 k^+_i ,
 \nonumber \\
 g_i & = & \gamma^+ / 2 k^+_i.
 \label{eq3.10}
\end{eqnarray}
We see that both $G_i$ and $g_i$ become singular at $k^+_i = 0$. These two
singularities appear to cancel. We can see this most clearly if we realize that
the denominator $P^-  - ( \cdots H_i \cdots)$ can be rewritten as $k^-_i -
H_i$.
If $k^+_i \rightarrow 0$, then $H_i \rightarrow \infty$, so we see that in this
limit $G_i$ behaves as follows:
\begin{equation}
 G_i \stackrel{k^+_i \rightarrow 0}{\sim} \frac{1} {2 k^+_i} \frac{\gamma^+
H_i}
 {k^-_i - H_i} \sim - \frac{\gamma^+}{2 k^+_i } = -g_i .
 \label{eq3.11}
\end{equation}
Therefore, the sum of the two contributions is finite for $k^+_i \rightarrow
0$.
\begin{eqnarray}
  G_i + g_i & = &
  \frac{1}{2 k^+_i}
 \frac{\gamma^+ H_i + \gamma^- k^+_i + \gamma_\perp \cdot k_\perp + m_i}
 {k^-_i - H_i} + \frac{\gamma^+}{2 k^+_i}
 \nonumber \\
&= &
 \frac{\gamma^+ k^-_i + \gamma^- k^+_i + \gamma_\perp \cdot k_\perp + m_i}
 {2 k^+_i (k^-_i - H_i)} \nonumber \\
 & \matrix{ \ _{k^+_i \rightarrow 0}\cr \sim \cr \cr }&
  - \frac{\gamma^+ k^-_i + \gamma_\perp \cdot k_\perp + m_i}{k^2_{i\perp} +
m^2_i} .
 \label{eq3.12}
\end{eqnarray}
This expression appears {\it after} the integration of the energy variable.
Then
$k_i^-$ is a function of the external variables only and represents the
$p^-$-flow through the intermediate state under consideration.
\subsection{General case}
We have seen that singularities in l.c.t-ordered diagrams cancel
similar singularities in instantaneous terms. The instantaneous terms
might contain other divergences that are cancelled by lower-order terms with
additional instantaneous terms. The question remains whether this procedure
ends,
or whether we are left with terms which contain only instantaneous
singularities
that do not cancel each other. In the section on divergent contour
integration we show that the proper treatment of the shift of poles to
infinity removes all singularities from each residue.  So after the
recombination of terms we won't have a residual term in the form of
instantaneous
parts.\\
Although we are not concerned with gauge theories explicitly, we note that most
of these terms drop in a gauge theory with the (naive) light-cone gauge, and in
theories with scalar and
pseudo-scalar coupling, due to $\gamma^+\gamma^+=\gamma^+ \gamma^5\gamma^+=0=
\gamma^+ \gamma^i \gamma^+$.

\section{Multi-loop diagrams}
\label{Chapter4}
The extension of the reduction algorithm from Feynman diagrams with one loop to
Feynman diagrams with several loops is not difficult, but there are some points
that need to be clarified. The loop integrations can be done one after another
since the
structure of a l.c.t.-ordered diagram is not essentially different from a
Feynman diagram. We will illustrate the procedure with a simple example in the
section below.

\subsection{Two-loop diagram}

\begin{figure}
\begin{center}
\setlength{\unitlength}{0.012500in}%
\begin{picture}(280,186)(80,545)
\thicklines
\put( 80,640){\line( 1, 0){ 60}}
\put(140,640){\line( 1, 1){ 80}}
\put(220,720){\line( 1,-1){ 80}}
\put(300,640){\line( 1, 0){ 60}}
\put(300,640){\line(-1,-1){ 80}}
\put(220,560){\line(-1, 1){ 80}}
\put(220,560){\vector( 0, 1){ 80}}
\put(200,700){\vector(-1,-1){ 20}}
\put(160,620){\vector( 1,-1){ 20}}
\put(280,620){\vector(-1,-1){ 20}}
\put(240,700){\vector( 1,-1){ 20}}
\put(220,720){\line( 0,-1){ 80}}
\put(190,630){\makebox(0.4444,0.6667){\tenrm .}}
\put(320,640){\vector( 1, 0){ 20}}
\put(100,640){\vector( 1, 0){ 20}}
\put( 80,645){\makebox(0,0)[lb]{\raisebox{0pt}[0pt][0pt]{\twlrm $p$}}}
\put(145,680){\makebox(0,0)[lb]{\raisebox{0pt}[0pt][0pt]{\twlrm $k-p$}}}
\put(155,585){\makebox(0,0)[lb]{\raisebox{0pt}[0pt][0pt]{\twlrm $k$}}}
\put(270,585){\makebox(0,0)[lb]{\raisebox{0pt}[0pt][0pt]{\twlrm $q$}}}
\put(260,685){\makebox(0,0)[lb]{\raisebox{0pt}[0pt][0pt]{\twlrm $p+q$}}}
\put(180,605){\makebox(0,0)[lb]{\raisebox{0pt}[0pt][0pt]{\twlrm 1}}}
\put(185,665){\makebox(0,0)[lb]{\raisebox{0pt}[0pt][0pt]{\twlrm 2}}}
\put(210,635){\makebox(0,0)[lb]{\raisebox{0pt}[0pt][0pt]{\twlrm 3}}}
\put(255,660){\makebox(0,0)[lb]{\raisebox{0pt}[0pt][0pt]{\twlrm 4}}}
\put(250,600){\makebox(0,0)[lb]{\raisebox{0pt}[0pt][0pt]{\twlrm 5}}}
\put(350,645){\makebox(0,0)[lb]{\raisebox{0pt}[0pt][0pt]{\twlrm $p$}}}
\put(225,620){\makebox(0,0)[lb]{\raisebox{0pt}[0pt][0pt]{\twlrm $k+q$}}}
\put(130,645){\makebox(0,0)[lb]{\raisebox{0pt}[0pt][0pt]{\twlrm a}}}
\put(220,545){\makebox(0,0)[lb]{\raisebox{0pt}[0pt][0pt]{\twlrm b}}}
\put(215,725){\makebox(0,0)[lb]{\raisebox{0pt}[0pt][0pt]{\twlrm c}}}
\put(300,645){\makebox(0,0)[lb]{\raisebox{0pt}[0pt][0pt]{\twlrm d}}}
\end{picture}
\caption{The two-loop diagram.}\label{fig.twoloop}
\end{center}
\end{figure}
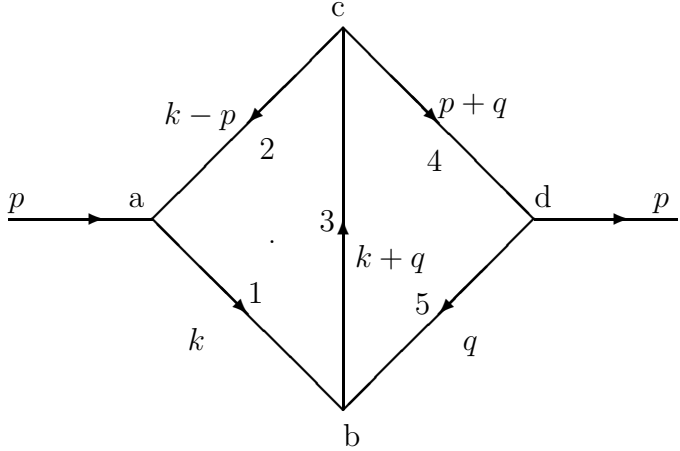
We consider the scalar diagram with two loops, depicted in
fig.~\ref{fig.twoloop}.
The corresponding integral is :
\begin{equation}
 D = \int {{\rm d}q^- {\rm d}k^- \over (2\pi)^2 \phi} {1 \over (k^-
-H_1)(k^--H_2)
 (k^-+q^--H_3)(q^--H_4)(q^--H_5)}
 \label{eq4.1}
\end{equation}
Where the phase factor $\phi = 2^5k^+(k^+-p^+)(k^++q^+)q^+(q^++p^+)$ and the
 poles are given by:
 \begin{equation}
\begin{array}{rccl}
 H_1 & = & {k_\perp^2 +m^2 \over 2 k^+} &-
 {i \epsilon \over k^+}  \\
 H_2 & = &  p^- + {(p_\perp -k_\perp)^2 +m^2 \over 2 (k^+-p^+)}& -
 {i \epsilon \over k^+-p^+}  \\
 H_3 & = &  {(k_\perp +q_\perp)^2 +m^2 \over 2 (k^++q^+)}& -
 {i \epsilon \over k^++q^+}  \\
 H_4 & = & -p^- +{ (q_\perp+p_\perp)^2 +m^2 \over 2 (q^++p^+)} &  -
 {i \epsilon \over q^++p^+} \\
 H_5 & = & {q_\perp^2 +m^2 \over 2 q^+} & -
 {i \epsilon \over q^+}  \ .
 \label{eq4.2}
\end{array}
\end{equation}
\begin{figure}
\begin{center}
\setlength{\unitlength}{0.012500in}%
\begin{picture}(295,268)(70,490)
\thicklines
\put(160,660){\circle*{6}}
\put(160,520){\vector( 0, 1){220}}
\put( 80,660){\vector( 1, 0){280}}
\put( 80,560){\line( 1, 0){280}}
\put(260,740){\line( 0,-1){220}}
\put( 80,740){\line( 1,-1){220}}
\put(260,520){\line( 0,-1){ 20}}
\put(160,520){\line( 0,-1){ 20}}
\put(300,520){\line( 1,-1){ 20}}
\put(280,680){\makebox(0,0)[lb]{\raisebox{0pt}[0pt][0pt]{\twlrm $\{-----\}$}}}
\put(280,600){\makebox(0,0)[lb]{\raisebox{0pt}[0pt][0pt]{\twlrm $\{----+\}$}}}
\put(285,540){\makebox(0,0)[lb]{\raisebox{0pt}[0pt][0pt]{\twlrm $\{---++\}$}}}
\put(265,490){\makebox(0,0)[lb]{\raisebox{0pt}[0pt][0pt]{\twlrm $\{--+++\}$}}}
\put(175,535){\makebox(0,0)[lb]{\raisebox{0pt}[0pt][0pt]{\twlrm $\{-++++\}$}}}
\put( 80,535){\makebox(0,0)[lb]{\raisebox{0pt}[0pt][0pt]{\twlrm $\{+++++\}$}}}
\put( 75,605){\makebox(0,0)[lb]{\raisebox{0pt}[0pt][0pt]{\twlrm $\{+++-+\}$}}}
\put( 65,670){\makebox(0,0)[lb]{\raisebox{0pt}[0pt][0pt]{\twlrm $\{+++--\}$}}}
\put( 75,750){\makebox(0,0)[lb]{\raisebox{0pt}[0pt][0pt]{\twlrm $\{++---\}$}}}
\put(175,680){\makebox(0,0)[lb]{\raisebox{0pt}[0pt][0pt]{\twlrm $\{-+---\}$}}}
\put(180,645){\makebox(0,0)[lb]{\raisebox{0pt}[0pt][0pt]{\twlrm $\{-+--+\}$}}}
\put(162,570){\makebox(0,0)[lb]{\raisebox{0pt}[0pt][0pt]{\twlrm $\{-++-+\}$}}}
\put(220,610){\makebox(0,0)[lb]{\raisebox{0pt}[0pt][0pt]{\bf I}}}
\put(185,585){\makebox(0,0)[lb]{\raisebox{0pt}[0pt][0pt]{\bf I$\!$I}}}
\put(170,740){\makebox(0,0)[lb]{\raisebox{0pt}[0pt][0pt]{\twlrm $q^+$}}}
\put(365,660){\makebox(0,0)[lb]{\raisebox{0pt}[0pt][0pt]{\twlrm $k^+$}}}
\put(250,745){\makebox(0,0)[lb]{\raisebox{0pt}[0pt][0pt]{\twlrm $k^+=p^+$}}}
\put(365,560){\makebox(0,0)[lb]{\raisebox{0pt}[0pt][0pt]{\twlrm $q^+=-p^+$}}}
\put(325,505){\makebox(0,0)[lb]{\raisebox{0pt}[0pt][0pt]{\twlrm $k^++q^+=0$}}}
\end{picture}
\caption{The imaginary signs of $\{H_1,H_2,H_3,H_4,H_5\}$ for the different
sectors in $k^+\otimes q^+$ space. Only the inner regions {\bf I} and {\bf
I$\!$I} correspond to integrals and skeleton graphs.}
 \label{fig.sector}
 \end{center}
\end{figure}
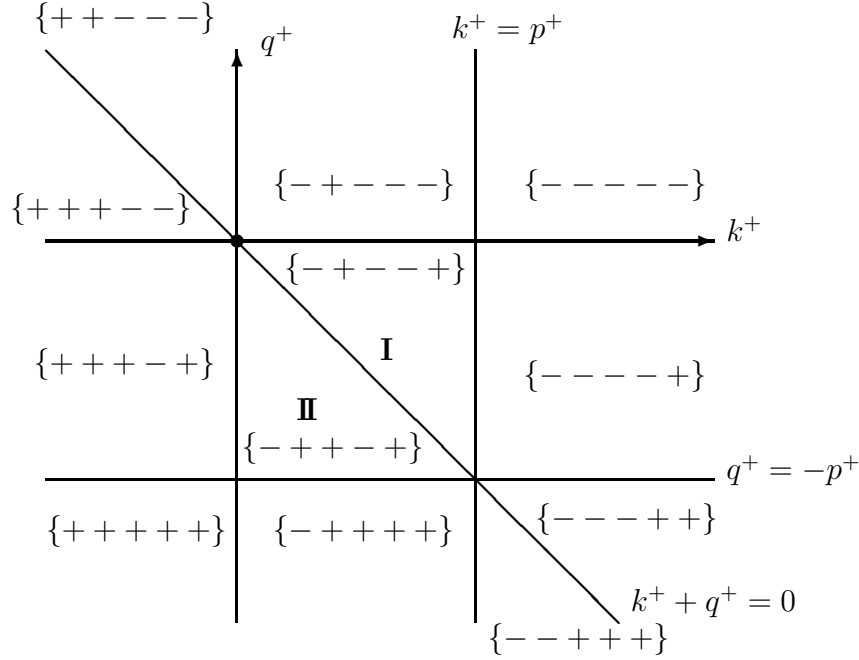
There are twelve sectors in $k^+ \otimes q^+$-space corresponding to twelve
skeleton graphs. These sectors are depicted in fig.~\ref{fig.sector},
where also the signatures of the skeleton graphs are shown. The amplitude $D$
vanishes if either the integral over $k^-$ or the one over $q^-$ vanishes. The
former happens if $\Im H_1 , \Im H_2$ and $\Im H_3$ have the same sign, the
latter if this happens for $\Im H_3 , \Im H_4$ and $\Im H_5$.
We read off from fig.~\ref{fig.sector} that there are two sectors remaining,
denoted as {\bf I} and {\bf I$\!$I}. In sector {\bf I} we have
$H_1, H_3, H_4<0$ and $H_2, H_5>0$.
The reduction algorithm gives the l.c.t.-ordered diagram {\bf I}
\begin{figure}
\begin{center}
\setlength{\unitlength}{0.012500in}%
\begin{picture}(460,95)(60,605)
\thicklines
\put( 60,660){\line( 1, 0){ 40}}
\put(100,660){\line( 2, 1){ 80}}
\put(180,700){\line( 1,-1){ 40}}
\put(220,660){\line(-2,-1){ 80}}
\put(140,620){\line(-1, 1){ 40}}
\put(220,660){\line( 1, 0){ 40}}
\put(180,700){\line(-1,-2){ 40}}
\put(320,660){\line( 1, 0){ 40}}
\put(360,660){\line( 1, 1){ 40}}
\put(400,700){\line( 2,-1){ 80}}
\put(480,660){\line(-1,-1){ 40}}
\put(440,620){\line(-2, 1){ 80}}
\put(400,700){\line( 1,-2){ 40}}
\put(480,660){\line( 1, 0){ 40}}
\put(210,605){\makebox(0,0)[lb]{\raisebox{0pt}[0pt][0pt]{\bf I}}}
\put(470,610){\makebox(0,0)[lb]{\raisebox{0pt}[0pt][0pt]{\bf I$\!$I}}}
\put( 60,670){\makebox(0,0)[lb]{\raisebox{0pt}[0pt][0pt]{\it p}}}
\put(115,680){\makebox(0,0)[lb]{\raisebox{0pt}[0pt][0pt]{\it p-k}}}
\put(105,630){\makebox(0,0)[lb]{\raisebox{0pt}[0pt][0pt]{\it k}}}
\put(200,690){\makebox(0,0)[lb]{\raisebox{0pt}[0pt][0pt]{\it p-q}}}
\put(185,625){\makebox(0,0)[lb]{\raisebox{0pt}[0pt][0pt]{\it q}}}
\put(165,655){\makebox(0,0)[lb]{\raisebox{0pt}[0pt][0pt]{\it k-q}}}
\put(320,665){\makebox(0,0)[lb]{\raisebox{0pt}[0pt][0pt]{\it p}}}
\put(360,685){\makebox(0,0)[lb]{\raisebox{0pt}[0pt][0pt]{\it p-k}}}
\put(445,685){\makebox(0,0)[lb]{\raisebox{0pt}[0pt][0pt]{\it p-q}}}
\put(385,625){\makebox(0,0)[lb]{\raisebox{0pt}[0pt][0pt]{\it k}}}
\put(465,630){\makebox(0,0)[lb]{\raisebox{0pt}[0pt][0pt]{\it q}}}
\put(390,660){\makebox(0,0)[lb]{\raisebox{0pt}[0pt][0pt]{\it q-k}}}
\end{picture}
\caption{The two l.c.t-ordered diagrams that follow from the two-loop
         diagram.}
 \label{fig.lcttwoloop}
 \end{center}
\end{figure}
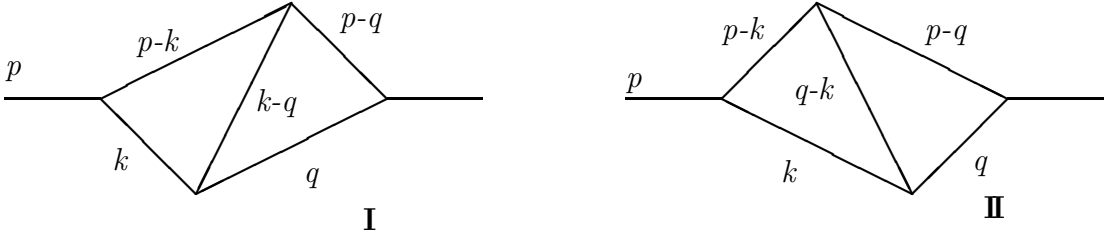
of fig.~\ref{fig.lcttwoloop}, with l.c.t. ordering $a<b<c<d$. In sector
{\bf I$\!$I} we
have $H_1, H_4 < 0$ and $H_2, H_3, H_5 > 0$. The corresponding
l.c.t. ordering is $a<c<b<d$. The only difference between the two diagrams is
the sign of $\Im H_3$, that is linked to the two different l.c.t. orderings of
the vertices $b$ and $c$. For the sake of completeness we give the algebraic
expressions for the two l.c.t. ordered diagrams. Upon integration over $k^-$
we obtain:
\begin{eqnarray}
 D_{\bf I} & = &   {i \over 2 \pi}\int {{\rm d}q^- \over \phi} {1 \over
 (H_2 -H_1)(q^- +H_2 - H_3 )(q^- - H_4)( q^- -H_5)} \nonumber \\
 D_{\bf I\!I} &=& {i \over 2 \pi } \int {{\rm d}q^- \over \phi} {1 \over
 (H_2 -H_1)(q^- +H_1 - H_3 )(q^- - H_4)( q^- -H_5)}.
 \label{eq4.3}
\end{eqnarray}
The $q^-$ integration is straightforward and gives the result:
\begin{eqnarray}
 D_{\bf I} & = & {1 \over \phi} {1 \over (H_2-H_1)(H_2 + H_4 -H_3)
 (H_4-H_5)} \nonumber \\
 D_{\bf I\!I} &=& {1 \over \phi} {1 \over (H_2-H_1)(-H_5 -H_1 +H_3)
 (H_4-H_5)}
 \label{eq4.4}
\end{eqnarray}
After reversing the directions of the lines corresponding to negative $k^+_i$,
{\it viz}, $k_2 = k-p$ and $k_5 = q$ in $D_{\bf I}$ and $k_2$, $k_3 = k+q$ and
$k_4 = p+q$ in $D_{\bf I\!I}$ we obtain the l.c.t.-ordered diagrams depicted in
fig.~\ref{fig.lcttwoloop}. As before, every factor in the denominators of
eq.~(\ref{eq4.4}) can again be written in the form $P^- - H_0(i,j)$, where
$H_0(i,j)$ is the sum of the energies on the lines $i$ and $j$.

We see again that the integrations over $k^+$ and $q^+$ are limited to finite
regions. After reversing the lines with negative $\Im H_i$, one sees that the
diagrams obtained have the spectrum property.

\subsection{General multi-loop diagrams}

In an arbitrary Feynman diagram with $L$ loops, one must first identify the
independent integration variables, say $q^-_1, \dots, q^-_L$. Then one can
characterize the different types of pole positions in an $L$-dimensional space
with coordinates $(q^+_1, \dots, q^+_L)$. The different signatures
$(\Im H_1, \dots , \Im H_N)$ divide this space into a number of sectors, each
sector
being associated with its particular skeleton graph. The sectors where the pole
positions in all variables $q^-_i$ are distributed over both half planes,
$\Im q^-_i>0$ and $\Im q^-_i<0$ resp., are necessarily finite. This is so,
because for any loop $i$, all poles $H_k$ occurring in this loop will
have $\Im H_k < 0$ ($\Im H_k > 0$) if the integration variable $q^+_i$ goes
to infinity ($-$infinity). Therefore, the sectors in $(q^+_1, \dots,
q^+_L)$-space
which are semi-infinite in either of the $q^+_i$ do not contribute to
at least one of the integrals over the $q^-_i$ variables.

So, in general we will have a finite number of skeleton graphs that each gives
rise to a finite number of l.c.t.-ordered diagrams. Each and every one of them
has the spectrum property. In the case of spin-1/2 particles, one can duplicate
this algorithm, provided the full Feynman diagram, containing $F$ fermion
lines,
is first split into $2^F$ intermediate diagrams according to the division of
the spin-1/2 propagator into instantaneous and propagating parts. Then the
reduction algorithm is applied to each of the intermediate diagrams, giving
rise
to the appropriate skeleton graphs and finally to the l.c.t.-ordered diagrams,
as was demonstrated in the one-loop case in the previous subsection. Of course,
in the multi-loop case blinks may occur as well as in the one-loop case. They
are treated in exactly the same way as before. Thus we see that the
multi-loop Feyman diagrams, although algebraically more involved than the
one-loop cases, can be reduced to l.c.t.-ordered diagrams using precisely the
same algorithm as was used for one-loop Feyman diagrams.

A final remark concerning the $i\epsilon$-prescription is in order here.
It is used to define the deformed integration contours in all variables $q^-_i$
simultaneously. After the residue theorem is applied to perform the contour
integrals, the real parts of the poles are substituted in formulae like
eqs. (\ref{eq4.3}, \ref{eq4.4}) ($\epsilon=0$). If one would substitute
complex poles in eq. \ref{eq4.3}, ambiguities might arise in the values of
the imaginary parts of the poles in $q^-$.
\section{Technical difficulties}
\label{Chapter5}
In the previous sections we dealt with the equivalence between Feynman diagrams
and l.c.t-ordered diagrams when the integration over $k^-$ is well-defined.
There are two types of special cases were the $k^-$ integration is not
well-defined. We can best illustrate these with simple examples.
\begin{figure}
\begin{center}
\setlength{\unitlength}{0.007000in}%
\begin{picture}(660,180)(40,600)
\thicklines
\put(160,720){\circle{80}}
\put( 80,720){\line( 1, 0){ 40}}
\put(200,720){\line( 1, 0){ 40}}
\put(540,780){\line( 1, 0){ 80}}
\put(620,780){\line( 0,-1){100}}
\put(620,680){\line( 1, 0){ 80}}
\multiput(620,780)(4.00000,-8.00000){6}{\makebox(0.4444,0.6667){\tenrm .}}
\multiput(640,740)(0.00000,-10.00000){3}{\makebox(0.4444,0.6667){\tenrm .}}
\multiput(640,720)(-4.00000,-8.00000){6}{\makebox(0.4444,0.6667){\tenrm .}}
\put(610,730){\line( 1, 0){ 20}}
\put(560,780){\vector( 1, 0){ 20}}
\put(640,680){\vector( 1, 0){ 20}}
\put( 40,730){\makebox(0,0)[lb]{\raisebox{0pt}[0pt][0pt]{\twlrm $p^+\!=\!0$}}}
\put(145,765){\makebox(0,0)[lb]{\raisebox{0pt}[0pt][0pt]{\twlrm $p-k$}}}
\put(155,660){\makebox(0,0)[lb]{\raisebox{0pt}[0pt][0pt]{\twlrm $k$}}}
\put(140,600){\makebox(0,0)[lb]{\raisebox{0pt}[0pt][0pt]{\twlrm (a)}}}
\put(610,600){\makebox(0,0)[lb]{\raisebox{0pt}[0pt][0pt]{\twlrm (b)}}}
\end{picture}
\caption{Two cases where the energy integral is ambiguous: (a) If the
$k^+$ momentum along the loop is constant. (b) If there is one pole left in the
the $k^-$-integration; the pole of the boson
accompanied by the instantaneous part of the fermion propagator.}
\label{divergent}
\end{center}
\end{figure}
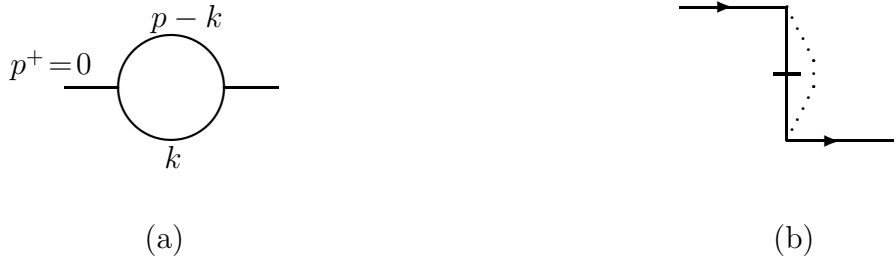
Consider a scalar loop like in $\phi^3$ theory. If the external line has
positive
$p^+$ then the integration domain in $k^+$ is the interval $0<  k^+ < p^+$
(see fig.~\ref{divergent} (a)). One may wonder what will happen if $p^+=0$,
because in that case the measure of the integration interval is zero.
The poles in the two propagators cross the real axis in the $k^-$-plane for the
same value of $k^+$: $k^+ = 0$. If this diagram is finite, there must occur a
delta-like contribution in $k^+$. In cases where $p^2 \leq 0$, one can approach
$p^+=0$ by performing a Lorentz-transformation (that, however, does not belong
to
the stability group of the null plane) and take the limit.
Such a Lorentz-transformation is always possible for a space-like external
momentum, and there are situations where the momenta on three space-like
external
lines can be transformed to have $p^+=0$ simultaneously.
In other cases, like in (generalized) tadpoles, the measure of the integration
domain is rigorously zero.
We will consider a general approach which holds in all cases and give the
same answer as a covariant calculation in the limit $p^+=0$. Tadpoles
have a close relation with the ordering of operators in the Hamiltonian,
therefore we see that the $\delta (k^+)$ contributions have a relation with the
ordering.

The other case where the $k^-$ integration is ill-defined, occurs if at most
one pole is present in the $k^-$ integration. This happens for all
Feynman diagrams with at most one boson propagator and at least one fermion
propagator in the loop. Then the intermediate diagrams with an instantaneous
part
of the fermion propagator combined with a boson propagator needs
regularization.
Other examples are diagrams with at least two instantaneous terms, which also
lead to divergent integrals.
The first order correction to the fermion self energy in a theory of scalar
bosons
and fermions with Yukawa coupling, is a simple example
(see fig. \ref{divergent} (b)). The fermion propagator has an instantaneous
part
such that the only $k^-$-dependence of the integrand resides in the boson
propagator. This integral is not defined,
so we need a way to deal with this type of integrals in a consistent way.
We are primarily interested in a treatment which does not interfere with the
algebraic rules, so the regularization must be a linear operation. In addition
we require it to be homogeneous in the integration variables.
In both cases, one where tadpoles are present and the other where instantaneous
parts give rise to infinities, we are lead by covariance in our choice of
regularization.
Other arguments do not restrict the regularization to a unique method,
while covariance does.
\subsection{"Zero modes" from energy integration.}
\label{zeromodes}
One of the integrals which show the presence of zero modes in a time-ordered
formulation has been discussed already by Yan  \cite{Y73}:
\begin{eqnarray}
\lim_{k^+\to 0}
\int{\rm d}p^- {1 \over (2(p^+-k^+)p^- -m^2+i \epsilon)(2 p^+p^--m^2+
i\epsilon)} &=& \nonumber \\
\int {\rm d}p^- {1 \over (2 p^+ p^- -m^2 + i \epsilon)^2}
&\stackrel{\mbox{\tiny covariant}}{=} & {i \pi \delta (p^+) \over m^2}  \ .
\label{yaneq}
\end{eqnarray}
For $p^+\not = 0$ there is one double pole either above or below the real axis
so it was concluded that, since $p^+=0$ is an unphysical value (no free state
can acquire $p^+=0$), the integral should vanish.
A careful analysis shows that eq.~(\ref{yaneq}) is an ambiguous expression,
so one can get any value (including the covariant answer) and
one needs to choose a regularization to get a well-defined integral.
(The proper covariant value was obtained by Yan by taking the
limits $p^+\downarrow 0$ and $ p^+ \uparrow 0$ in a special way.)

For $p^+=0$ the $p^-$ integral diverges, so $(p^+ \to 0, p^- \to \infty)$
is the ambiguous point. If $p^+$ moves along the real axis and crosses $p^+=0$,
the poles move through infinity and end up on the other side of the real axis.
To deal with all singularities of this type at the same
time, we introduce the variable $u = 1/p^-$ and study a general case:
\[
 D_n = \int {\rm d} p^- {1 \over (2 p^+_1 p^- - H_1^\perp +i \epsilon) (
2 p^+_2 p^- - H_2^\perp  + i \epsilon) \cdots (2 p^+_n p^- -  H^\perp_n + i
\epsilon) } =
\]
\begin{equation}
 \int {\rm d}u { u^{n-2} \over (2 p^+_1 - (H^\perp_1 - i \epsilon)u) \cdots
 (2 p^+ _n - (H^\perp_n - i\epsilon) u)}  \ .
 \label{eq5.2}
\end{equation}
The integrand goes to zero like $u^{-2}$ for $u \to \pm \infty$, therefore the
integral is well-defined, unless the integrand has a singularity at $u=0$.
So the only divergence can occur if $u \to 0$ which gives a finite contribution
only if all $p^+$-momenta vanish at the same time.
The poles in the variable $p^-$ that moved to infinity, now correspond to poles
in the variable $u$ that cross the real axis at $u=0$ when either of the
variables
$p^+_i$ is zero. If all $p^+_i$ happen to be equal, the integrand is singular
at
$u=0$.
(The first example, eq.~(\ref{yaneq}), is the special case with
$n=2;H_1^\perp =H_2^\perp = m^2$.)
This gives a finite contribution to the integral of $D$ over $p^+$,
on support $p^+=0$, thus $D$ contains a delta function in $p^+$.

The $u$ coordinate regularization replaces all other arguments we
might have to deal with this "zero-mode" problem. The choice
of regularization
determines the integral uniquely.
Instead of treating the general expression, eq.~(\ref{eq5.2}), we regularize
the case of a single pole in the integrand and use an algebraic relation to
obtain
the general expression. Consider the integral
\begin{equation}
 D_1 = \int {\rm d} u {1 \over u (2 p^+ - (H^\perp-i \epsilon) u)}.
 \label{eq5.3}
\end{equation}
This expression is ambiguous for two reasons: it has a pole at $u=0$ and a
double pole occurring for $p^+=0\wedge u=0$. The first ambiguity we remove by
adding a small imaginary part to one factor $u$ coming from the Jacobian.
In order to obtain a covariant expression we must do this symmetrically:
\begin{equation}
{1 \over u } \to { 1 \over 2} \left( {1 \over u+i\delta } + {1 \over u-
i\delta} \right).
 \label{eq5.5}
\end{equation}
We split the integral into two pieces; one just above the real axis and the
other just below it. We do not give the singularities some strict nature
(like principal value), which would lead to the square of the principal value
for $u^{-2}$.
Generally we treat the energy as a complex variable (since each pole
corresponds
to a particle), and the kinematical variables are treated geometrically.
The choice eq.~(\ref{eq5.5}) separates $p^+>0$ from $p^+<0$ for all positive
values of $\epsilon$ and $\delta $. Thus we find for the regularized integral:
\begin{eqnarray}
 D^{\rm Reg}_1 & = & \int{\rm d} u{1 \over 2} \left({1 \over u+i \delta } +
{1 \over u- i\delta} \right){1 \over (2 p^+ -(H^\perp-i \epsilon) u)}
\nonumber \\
 & = &  - { \pi i \theta (p^+)  \over 2 p^+ +i (H^\perp- i \epsilon) \delta} +
  { \pi i  \theta (-p^+)  \over 2 p^+ -i (H^\perp- i \epsilon) \delta}
\nonumber \\
 & =  & - { 2 \pi i \theta(p^+) \over 2 p^+ +i (H^\perp- i \epsilon) \delta}.
 \label{eq5.6}
\end{eqnarray}
For $p^+<0$ we reversed its sign to obtain the last line.
Integrating $D^{\rm Reg}_1$ over $p^+$ from 0 to a cutoff $\Lambda$ and taking
the limit $\epsilon \to 0$, gives
$\pi (\ln (H^\perp) + \ln \delta + \pi i /2 - \ln 2 \Lambda )$.
We shall see that the constant part $\ln \delta + i \pi /2-\ln 2\Lambda$ drops
if we have two or more energy denominators. Using the algebraic relation
\begin{equation}
{1 \over \prod_{j=1}^n(2 p^+ p^- - H_i^\perp +i  \epsilon)}=
\sum_{k=1}^n {1 \over (2 p^+ p^- - H_k^\perp + i \epsilon)\prod_{j\not =k}
(H^\perp_k - H^\perp_j)}
 \label{eq5.7}
\end{equation}
the regularized integral becomes:
\begin{equation}
 {\rm R}\!\!\!\!\!\int {\rm d} p^- {1 \over (2 p^+ p^- - H_1^\perp +i \epsilon)
 \cdots (2 p^+ p^- -  H^\perp_n + i \epsilon) } =
\sum_{k=1}^n {i \pi \delta (p^+) \ln (H^\perp_k) \over \prod_{j\not =k}
(H^\perp_k - H^\perp_j)}\ .
 \label{eq5.9}
\end{equation}
The function $\delta(p^+)$ appears here, because the integral is strictly zero
for $p^+\not =0$, although integration over $p^+$ gives a finite result.
The result eq.~(\ref{eq5.9}) can also be obtained as a limiting case of
eq.~(\ref{eq5.2}) in the simultaneous limits $p^+_i \to p^+ , \forall i$.
One can check that eq.~(\ref{eq5.9}) is the same as the covariant result,
using a Wick rotation such that $2 p^+ p^- -p^2_\perp \to -|p|^2$.
The limit for $H_i\to  H_j$ is well-defined.
The constant term has dropped since
\begin{equation}
 \sum_{k=1}^n { 1 \over \prod_{j\not =k} (H^\perp_k - H^\perp_j)}=0 \ ,
 \label{eq5.10}
\end{equation}
which follows from the fundamental theorem of algebra.

We emphasize here that these zero modes appear in loops where the $p^+$
momentum
is constant along the loop.
Zero-modes can be interpreted as an infinite number of states (around $p^+=0$)
which are infinitely off-shell ($p^-_{on}= \infty $), and thus have zero
probability for propagation over a finite distance. The combination of both
gives a finite contribution. This is reminiscent of the ultra-violet
divergences,
where the large number of high-energy states give an infinite contribution.

Although zero-modes are needed to obtain the covariant answer, they remain
slightly artificial, which can be seen from the configurations were they
occur. It seems that nature is telling us that the high density of states
for high energy causes trouble: divergent integrals appear which have
to be regularized, and zero-modes. Both result from singularities on
the light-cone.

\subsection{Divergences in the fermion loop}

In the section on diagrams containing fermions we stated that they could be
reduced to l.c.t.-ordered diagrams, provided no additional singularities
would occur.
In this section we deal with these singularities. We state that the
regularization proposed here removes all of them. There remains one point to
clarify, {\it i.e.}, whether the method of regularization does indeed produce
a covariant result. The latter point, however, will not be discussed in this
paper.

We now have the tools to deal with the singularities in the fermion loop.
Earlier (sect. \ref{Chapter3}) we saw that a blink combines two singular terms
in such a way that we
get a non-singular expression. However, in  general the low-order terms, with
several instantaneous contributions, are singular by themselves.  We will show
that the contribution from the contour at infinity leads to these
singularities.
After subtraction of the latter the singularities are gone and
a proper  recombination of terms will remove apparent singularities.

For Feynman diagrams with at most one boson propagator in a loop there are
singular parts in the contour integration. Even if the diagram would be
convergent in the ordinary sense, {\it i.e.}, in the
covariant or instant-form, it will still be divergent in $k^-$-integration.
The singular behavior of the fermion propagator on the light front is to blame.

Also in the case of bosons an ambiguity occurred (see sect. \ref{zeromodes}).
The result did depend on whether the contour was closed in
the lower half plane or in the upper half plane.
This problem could be resolved by choosing a particular combination of
contours.
We considered two contours, one consisting of the real axis and a semicircle in
the upper half plane, the other one has the semicircle in the lower half plane.
The integral over the real $k^-$-axis was replaced by the average of the
integrals over these two contours.
This regularization turned out to have a number of desirable properties.

In the case of fermion loops the problem is more complex. A straightforward
application of the residue theorem gives results that depends strongly
on the choice of the contour.
The integral is also more divergent than one would
expect from naive power counting, and has non-covariant singularities.
At the origin of these problems lies the contribution of the (semi-) circle at
infinity to the contour integral. We have to subtract this contribution.
In a sense we propose a regularization of the contour integral.

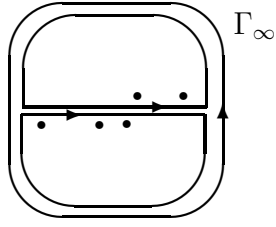
\begin{figure}
\begin{center}
\setlength{\unitlength}{0.00800in}%
\begin{picture}(140,140)(70,610)
\thicklines
\put(138,677){\oval(120,120)[bl]}
\put(138,677){\oval(120,120)[br]}
\put( 78,677){\line( 1, 0){120}}
\put( 98,677){\vector( 1, 0){ 20}}
\put(139,682){\oval(120,120)[tr]}
\put(139,682){\oval(120,120)[tl]}
\put( 79,682){\line( 1, 0){120}}
\put(159,682){\vector( 1, 0){ 15}}
\put(140,680){\oval(140,140)}
\put(184,689){\circle*{6}}
\put(154,689){\circle*{6}}
\put( 91,670){\circle*{6}}
\put(147,671){\circle*{6}}
\put(129,670){\circle*{6}}
\put(210,675){\vector( 0, 1){ 10}}
\put(210,730){\makebox(0,0)[lb]{\raisebox{0pt}[0pt][0pt]{ $\Gamma_\infty$}}}
\end{picture}
\end{center}
\caption{The difference between the upper-plane contour and the lower-plane
contour is the contour at infinity $\Gamma_\infty$.\label{contour}}
\end{figure}

For an integral that converges on the real axis, it does not matter whether we
close the contour in the upper or
the lower half-plane. This means that the sum of all residues is zero:
\begin{equation}
\sum _i {\rm Res}_i =0.
\end{equation}
If the integration is divergent we have contributions at infinity, which add
differently to the two contours (see fig. \ref{contour}).
Therefore the two contours give different
results. We denote this difference by $\Gamma_\infty$. Then we have
\begin{equation}
\sum _i {\rm Res}_i = \Gamma_\infty.
\end{equation}
To solve this problem we "regularize" the residues by subtracting fractions
of $\Gamma_\infty$ from them:
\begin{equation}
\sum _i {\rm Res}^{\rm reg}_i = \sum _i ({\rm Res}_i- \alpha_i
 \Gamma_\infty )=0 , \;\ \ \  \sum_i \alpha_i=1 .
 \label{eq71}
\end{equation}
We then find the desired result that closing the contour in the upper half
plane
gives the same result as closing it in the lower half plane.
We will see that the $\alpha_i$'s can be chosen such that the singularities
cancel.

The $k^-$-integral corresponding to a general Feynman diagram with a fermion
loop can be written as:
\begin{equation}
 D_n = \int {{\rm d}k^- \over 2^n k^+_1\cdots k_n^+ } {(\gamma^+ k^- + \beta_1)
\cdots  (\gamma^+ k^- + \beta_n) \over (k^- -H_1)\cdots (k^- -H_n)}.
\end{equation}
We wrote only the structure that depends on $k^-$ and $k^+$ explicitly.
We can consider one residue at a time, since the singularities of one residue
are not canceled by another residue.
The residue of the pole $(k^--H_j)^{-1}$, and the contribution of the
contour at infinity are:
\begin{eqnarray}
{\rm Res}_j&=&{1 \over 2^n k^+_1\cdots k_n^+ }{(\gamma^+ H_j + \beta_1) \cdots
(\gamma^+ H_j + \beta_n) \over (H_j -H_1)\cdots (H_j -H_n)}   \\
\Gamma_\infty &=& { 1 \over 2^n k^+_1\cdots k_n^+ } \left( (\sum_{i=1}^n  H_i)
({\gamma^+  \cdots  \gamma^+})+ (\beta_1  \cdots  \gamma^+) +
 \cdots + (\gamma^+  \cdots  \beta_n) \right)
\end{eqnarray}
The latter can best be calculated by
changing the integration variable to $u=(k^-)^{-1}$ and integrating
over a circle around  the origin with a small radius, $|u|=\epsilon$.
The residue Res$_j$ has only singularities if $k^+_j \to 0$, since then
$H_j \to \infty$. In the limit $k^+_j \to 0$ only a few terms survive.
The surviving structure is precisely the contribution from the contour at
infinity, which is independent of $j$
\begin{equation}
\lim_{k^+_j \to 0} {\rm Res}_j=
  \Gamma_\infty \ .
\end{equation}
If we decompose the contribution from the contour at infinity now in such a way
that each term contains only one singular point ($k^+_j \to 0$), we can
subtract these terms from the residues with the same
singularities, and the resulting regularized residues are finite. If we define
the quantities $q^+_j$ and $H_j^\perp$ as follows: $k^+_j= k^+-q^+_j$ and
$H_j^\perp= 2 k^+_j H_j$, then we find that the following regularized residues
are finite for $k^+_j \to 0$
\[
{\rm Res}^{\rm reg}_j = {\rm Res}_j - \left( {1 \over 2^n(k^+- q^+_j)
\prod(q^+_j-q^+_i)}H_j + \right.
\]
\[
\left. \sum_l{H_l^\perp \over 2^{n+1} (k^+ - q^+_j)(q^+_j-q^+_l)
 \prod(q^+_j-q^+_i)} + {H_j \over 2^n (q^+_j-q^+_l)\prod  (q^+_l-q^+_i)}
\right) (\gamma^+  \cdots  \gamma^+ )+
\]
\begin{equation}
{1 \over 2^n(k^+- q^+_j) \prod(q^+_j-q^+_i)} (( \beta_1 \cdots \gamma^+) +
\cdots + ( \gamma^+  \cdots  \beta_n) )
\label{eq75}
\end{equation}
Since each residue is now regular we know that there is a combination of
l.c.t.-ordered diagrams where each term is convergent. The blink procedure
tells
us how to do this step-by-step. The argument above tells us that there are
no singularities left.

We killed two birds with one stone: we resolved the ambiguity in the contour
integration and removed the singularities. These two problems are intimately
related since the singularity comes from the pinching of the contour at
infinity between the "pole at infinity" and an ordinary pole.

Another advantage of this procedure is that only the physical sectors
are non-zero, a property that would be destroyed by ordinary contour
integration. It allows us to keep a simpler view of causality and
unitarity, where each line is associated with a particle moving forward
in l.c.-time.
In non-physical sectors all the poles are on the same side of the axis.
Then formula (\ref{eq71}) tells us that the result is zero.
We do not know whether this regularization  leads to the same amplitude as
the covariant calculations.

\subsubsection{Example; vacuum polarization}
We will illustrate the procedure from the previous section with an example. The
simplest diagram is the vacuum polarization diagram; a closed fermion
loop with two fermion propagators.
\begin{equation}
D^{\mu\nu}= \int {{\rm d}k^- \over 2 \pi i}
{Tr[\gamma^\mu(\gamma^+ k^- + \beta_1)\gamma^\nu (\gamma^+k^- + \beta_2)]
\over  4 k^+_1k^+_2 ( k^--H_1)(k^--H_2)}.
\end{equation}
We deal with the physical sector so we can assume, without loss of generality
$\Im H_1>0$ and $\Im H_2<0$. The result depends on the way the contour is
closed. We will close the contour in the upper half plane. In terms of
l.c.t.-ordered diagrams we have the ordinary diagram, with two propagating
fermions, and the diagram with the instantaneous
part associated with the pole $H_2$ in the lower half plane.
Together they give the residue of the pole $H_1$, as expected.
\begin{equation}
D_1^{\mu\nu}= {Tr[\gamma^\mu(\gamma^+ H_1 + \beta_1)\gamma^\nu
(\gamma^+H_1 + \beta_2)] \over  4 k^+_1k^+_2 ( H_1-H_2)}.
\end{equation}
If we would have chosen to close the contour in the lower half plane the
result would be different. The difference is the result of a finite
contribution of the semicircles.
We still have to subtract the fraction of the contour at infinity which is
given by
\ref{eq75}:
\begin{equation}
\alpha_1\Gamma_\infty=
{Tr[\gamma^\mu \gamma^+\gamma^\nu \gamma^+] \over 4 k^+_1(k^+_2-k^+_1)} \left(
H_1 + {k^+_2 H_2 \over k^+_2-k^+_1} + {k^+_1 H_1 \over k^+_1-k^+_2}\right)+
{Tr[\gamma^\mu \beta_1\gamma^\nu \gamma^+] \over 4 k^+_1(k^+_2-k^+_1)}+
{Tr[\gamma^\mu \gamma^+ \gamma^\nu \beta_2] \over 4 k^+_1(k^+_2-k^+_1)}.
\end{equation}
The l.c.t.-ordered diagram obtained has no singularities, and is independent
of the direction in which the contour is closed, since it is symmetric in
$H_1$ and $H_2$:
\begin{equation}
D^{\mu\nu}_1-\alpha_1\Gamma_\infty =D_{\rm reg}^{\mu\nu}=
{Tr[\gamma^\mu(\left( {k^+_2H_2 -k^+_1H_1 \over k^+_2 -k^+_1} \right)
\gamma^++\beta_1)\gamma^\nu( \left( {k^+_2H_2 -k^+_1H_1 \over k^+_2 -k^+_1}
\right) \gamma^++\beta_2) ] \over 4 k^+_1 k^+_2 (H_1 - H_2)}.
\end{equation}
This result could not be obtained if we would have taken a combination
of an upper-plane semicircle and a lower-plane semicircle, this would
give a singular, and thus an ambiguous, result. The contribution of the contour
at infinity should be decomposed in a unique way to obtain a regular
expression.\\
We will not calculate the amplitude here since the diagram is divergent and
the comparison is with other results is spoiled by renormalization. After
this regularization the $k^+$-integration is automatically finite, since
the domain is finite and the integrand is regular.

\subsection{Dynamical spin}
In a Hamiltonian theory we separate kinematical variables from dynamical ones.
Starting from one equal-l.c.t surface, we evolve in the time direction to
the next equal-l.c.t. surface. In a covariant formulation, different
interaction points could be on the same equal-l.c.t surface. Constraints
should,
in a Hamiltonian formulation, account for these parts. A straightforward
interpretation does not exist, the constraints "evolve" with the order in
perturbation theory, and so does the physical Hilbert-space (constraints are
working
on the Hilbert space). We will illustrate this point with a Hilbert-space
interpretation of the instantaneous interaction $\gamma^+/2p^+$.

The $(p^+)^{-1}$ singularity is ambiguous as it stands. We need a way to
look at this part such that we avoid additional divergences which seem to
appear.  Note that the spinor-matrix elements of $\gamma^+ $  are the same
as those of $p^+$. Therefore we might suspect that the occurrence of
$\gamma^+$ suppresses the singularity. To make this apparent we
use the completeness of the physical Hilbert space. The physical
Hilbert space is spanned by the free states. Therefore we can write
the identity operator as the sum over all states:
\begin{equation}
\openone =  \sum_{\alpha} \int {\rm d}^3 p\  |u^{(\alpha)}(p) \rangle \langle
u^{(\alpha)}(p) |.
 \label{eq3.2}
\end{equation}
We use this abstract notation because we don't want to bother with
conventions which are not relevant. The normalization follows from
the idempotency  of the identity operator ($\openone \cdot \openone
=\openone$).
\begin{equation}
\langle u^{(\alpha)}(p) |u^{(\beta )}(p') \rangle = \bar  u^{(\alpha)}(p)
u^{(\beta )}(p') \langle p |p' \rangle = \delta^{\alpha \beta} \delta^3 (p-p')
{}.
 \label{eq3.3}
\end{equation}
We can project $\gamma^+/(2p^+)$  onto the physical states by applying the
identity operator on both sides:
\begin{eqnarray}
\openone \cdot  {\gamma^+ \over 2 p^+} \cdot \openone & =& \sum_{\alpha \beta}
\int {
\rm d}^3 p' {\rm d}^3 p
 |u^{(\alpha)}(p')\rangle \langle u^{(\alpha)}(p') |
{\gamma^+ \over 2 p^+} |u^{(\beta)}(p)\rangle \langle u^{(\beta)}(p) |
\nonumber \\
&=& \sum_{\alpha \beta}\int {\rm d}^3 p' {\rm d}^3 p
|u^{(\alpha)}(p')\rangle \delta^{\alpha \beta} {\delta^3 (p-p')  \over 2 m}
\langle u^{(\beta)}(p) |=  {\openone \over 2 m}
 \label{eq3.4}
\end{eqnarray}
There is no mixing between upper and lower components because they are
spectrally separated.
\begin{equation}
1\!\!1 = \left( {p\!\!\!/+ m \over 2m }  + {- p\!\!\!/+m \over 2 m} \right)
\theta (p^+)
 \label{eq3.5}
\end{equation}
Wherever $\gamma^+ /(2 p^+)$ appears we can replace it by $1 /(2m)$, since
eq.~(\ref{eq3.4}) is an operator identity on our space. The instantaneous
interaction can be interpreted as
nothing but a point interaction, $1/(2p^+)$ being the phase space that goes
with it and $\gamma^+$ the vertex.

However, these arguments do not hold in a Feynman diagram. The spin plays
a dynamical role. In contrast to the instant-form dynamics, where
all components of the angular momentum are kinematical, in front-form
dynamics only the $z$-component of the spin is a kinematical operator.
The other components are involved in the interactions.
It turns out that we can combine the $\gamma^+/(2p^+)$ singularity with the
$1/p^+$ singularity that appears in a corresponding l.c.t.-ordered diagram with
a propagating fermion line, such that the singularities of the two cancel.
Thus we find that in a Hamiltonian approach to light-front field theory,
there is an intimate interplay between the singularities of the propagators and
a singular piece of the interaction. Interestingly enough, this piece occurs
even for free particles.
In old-fashioned ordinary time-ordered perturbation theory, {\it i.e.},
instant form dynamics, the amplitude of propagation depends on the off-shell
energy only, not on the polarization.
\subsection{Analyticity and covariance}

A Feynman amplitude is an analytical function of scalar objects like
$p^2_i$, $p_i \cdot p_j$. Often, the real values of the scalars are the
boundary
values of the complex domain on which this function is defined. We use these
arguments in order to be able to apply the residue theorem to integration over
the loop variables  and to perform Fourier transformations.
All contour integrals are finite (if we don't pinch the contour)
and coincide for integrals convergent on the real axis with the integral
along this axis. (A coordinate transformation $y=x^{-1}$, as used in the
section on zero modes, doesn't alter the results, since it maps the real axis
on the real axis.) We cannot use the exponential ${\rm e}^{i\,kx}$ to
improve the convergent along the semicircle of the contour (Jordan's lemma).

If we integrate over one coordinate separately, manifest covariance is lost.
In the case of integration over ordinary energy this was not much of a problem
since we can consider $\vec p^{\,2}+m^2$ as real and then analyticity in $p^0$
is directly related to analyticity in $p_\mu p^\mu$. In light-front coordinates
the situation is more complicated since $p^2-m^2=2 p^- p^+ - (p_\perp^2+m^2)$.
For real values of the scalar object the complex values of $p^-$ and $p^+$ are
related ($(p_\perp^2+m^2)$ is real). The coordinates are each others complex
conjugate. After Wick-rotation ($p^0 \to -i p^0$) this remains almost exactly
true:
$p^+ \to p, \; p^- \to  - \bar{p}$. For a strip along the real axis we have:
 $p^- \Im p^+ = -p^+ \Im p^-$.
 Singularities that occur in a complex function can be regularized, but
 the relation between the conjugate variables restricts the possibilities of
 regularization. Singularities of an integer order (like $1/x$) cannot be
 integrated by parts. But one can approach these singularities in parametric
 space $x^\alpha ;\ \alpha \to -1$.
 The advantage of this dimensional regularization is that it does not
 interfere with algebraic
 rules; the regularized distributions satisfy the same relations as the
 singular ones, which is of great important for complex, analytical functions.
For regularization of complex distributions
one subtracts these poles as function of the order but with a fixed difference
 between the order  of the singularity of $p$ with respect to $\bar p$:
$p^\alpha \bar p^{\alpha+k}$ with a fixed $k$ \cite{Generf}.

We will follow a simpler approach with the same result.
To avoid complications we define distributions of covariant objects only.
Analyticity of the covariant object tells us the relation
between $p^+$ and $p^-$ at regularization. A homogeneous distribution is
given by the partial integration of a singular, but integrable, function:
\begin{equation}
 \left({1 \over p^+},\phi \right)= \int {\rm d} p^- {1 \over p^+} \phi =
 \int {\rm d} p^- \left[ {\partial \over \partial p^+} \ln (p^+p^-) \right]
 \phi .
\end{equation}
We need $p^+p^-$ for positive imaginary values. So we take the cut
of the logarithm along the negative imaginary axis, therefore the logarithm
has an imaginary part of the form $i \pi \theta (-p^+p^-)$.
The homogeneous distribution is
\begin{equation}
{1 \over [p^+]} \equiv
{\partial \over \partial p^+}( \ln |p^+p^-|+ i \pi \theta ( - p^+p^-))=
{1 \over p^+ + {i \epsilon } \sigma(p^-)}=
 {\rm PV} {1 \over p^+} - {i \pi } \sigma (p^-) \delta (p^+) .
\end{equation}
The result is  that
$\partial_{p^+}\partial_{p^-} \ln(p^+p^-)=\partial_{p^-} (p^+)^{-1} =
 2 \pi i \delta(p^-) \delta(p^+)$,
which is nothing but the Mandelstam-Leibrandt regularization. (The $i$
can be accounted for through Wick rotating the $z$-variable.) After Fourier
transformation the distribution $\ln (p^+p^- +i \epsilon)$ becomes a singular
function which contains the
intersection of the light cone with the null-planes $x^+=0 \vee x^-=0$:
 ${\cal F} [\ln (p^+p^- +i \epsilon)] =
 { i \delta^2(x_\perp) \over x^+ x^- - i \epsilon} =
 - \delta^2(x) \delta^1 (x^+ x^-) + i \delta^2(x_\perp) {\rm PV }
 {1 \over x^+x^-}$.
This fact is part of the reason why there  exists  confusion about the
instantaneous term in the fermion propagator. In a Feynman diagram the
integrands are treated as meromorphic functions in the complex plane.
The real part is automatically complemented with a imaginary part.
If we use a Hamiltonian approach we have constraints which relate real parts
to real parts and we express one in terms of the other, therefore we might
loose
some information concerning the behavior of the imaginary parts. To put it in
other words:
the off-shell behavior comes naturally in a Feynman diagram and this does not
always happen in a l.c.t.-ordered diagram.
This is another reason why we chose to define light-front field theory in close
connection with a formulation in terms of Feynman diagrams.

\section{Proof of equivalence}
\label{Chapter6}
In this section we have collected several technical aspects of the proof of
equivalence. In sect.~\ref{Chapter2} we gave some examples, in order to
illustrate the general procedure. Therefore, we do not give any examples here.

\subsection{ Energy integration}
We present here the proof of the basic theorem on the integration over $p^-$.
First we discuss the case of a single loop. The proof for several loops
comes next.
\subsubsection{One loop}
The integration of a Feynman diagram over one energy loop variable $p^-$ gives
the following expression the $p^+$-interval corresponding to
$\{ (\Im H_i> 0 \wedge i \leq m) \vee  ( \Im H_i< 0 \wedge i > m) \}$:
\begin{eqnarray}
FD(\vec H )=  \int {{\rm d}p^-\over 2 \pi}{1 \over 2^N p^+_1\cdots p^+_N [p^--
H_1]
 \cdots [p^--H_N]}= \rule{20mm}{0mm}
\nonumber \\
\frac{i  (-1)^{N+1}  }{ 2^N(p^+_1 \cdots p^+_N)\Delta(H_1,\cdots,H_N) }\left|
\matrix{ H_1^{N-2}& \cdots & H_1^2 & H_1 &0 & 1 \cr
 H_2^{N-2}& \cdots & H_2^2 & H_2 &0 & 1 \cr
 . &  &  & & .& . \cr
 H_m^{N-2}& \cdots & H_m^2 & H_m &0 & 1 \cr
 H_{m+1}^{N-2}& \cdots & H_{m+1}^2 & H_{m+1} &1 & 0 \cr
 . &   &  & & .& . \cr
 H_{N}^{N-2}& \cdots & H_{N}^2 & H_{N} &1 & 0 \cr } \right|
 \label{eq83}
\end{eqnarray}
The last factor is a complicated mixed symmetric polynomial in the $H_i$'s that
we denote by $(-1)^{n}W_{m,n}(H_1,\cdots,H_m|H_{m+1},\cdots,H_{n+m})$.
(The phase factor $ (-1)^{N+1}$ is introduced to simplify the final
expressions.)
$\Delta$ is the Vandermonde determinant
($n\geq 2$):
\begin{eqnarray}
\Delta(x_1\cdots,x_n)& =& {\rm det}[\Delta_{ij}] \\
 \Delta_{ij} & = & x^{n-j}_i
\end{eqnarray}
A well known result is
\begin{equation}
\Delta(x_1\cdots,x_n) = \prod_{i<j}^{n-1,n}(x_i-x_j)
\end{equation}
See for instance MacDonald or Fulton \& Harris \cite{symmpol} for properties of
the Vandermonde determinant.\\
\noindent {\it Proof}

Depending on the (fixed) values of the (kinematical) $p^+_i$'s some poles are
above and some are below the real axis.
The integral is computed as $2 \pi i$ times the sum of the residues.
The residue of pole $p^- = H_i$ can be
written as
\begin{equation}
  (-1)^{i+1}{ \Delta(H_1,\cdots,H_{i-1},H_{i+1},\cdots,H_N) \over
2^N(p^+_1\cdots p^+_N)\Delta (H_1,\cdots, H_N)}
\end{equation}
We can add a line and a column to the determinant in the numerator:
\begin{equation}
(-1)^{i+1}
\left| \matrix{ H_1^{N-2}& \cdots & H_1^2 & H_1 &1  \cr
 H_2^{N-2}& \cdots & H_2^2 & H_2 & 1 \cr
 . &  &  & & . \cr
 H_{i-1}^{N-2}& \cdots & H_{i-1}^2 & H_{i-1} &1  \cr
 H_{i+1}^{N-2}& \cdots & H_{i+1}^2 & H_{i+1} &1  \cr
 . &   &  & & . \cr
 H_{N}^{N-2}& \cdots & H_{N}^2 & H_{N} &1  \cr } \right|=(-1)^{N+1}
\left| \matrix{ H_1^{N-2}& \cdots & H_1^2 & H_1 &1 & 0 \cr
 H_2^{N-2}& \cdots & H_2^2 & H_2 &1 & 0 \cr
 . &  &  & & .& . \cr
 H_{i-1}^{N-2}& \cdots & H_{i-1}^2 & H_{i-1} &1 & 0 \cr
 H_i^{N-2}& \cdots & H_i^2 & H_i &1 & 1 \cr
 H_{i+1}^{N-2}& \cdots & H_{i+1}^2 & H_{i+1} &1 & 0 \cr
 . &   &  & & .& . \cr
 H_{N}^{N-2}& \cdots & H_{N}^2 & H_{N} &1 & 0 \cr } \right|
 \label{eq85}
 \end{equation}
The final formula is obtained by adding determinants of type eq.~(\ref{eq85})
which amounts to just adding their last columns. This gives the result stated
in the theorem.
\subsubsection{Several loops}
In the case of several loops we integrate loop by loop. We must use the residue
theorem such that the order of integration does not change the result.
In general, the momenta of the particles on the internal lines will be linear
combinations of the integration variables, say $k_i = \sum_k \alpha^k_i p_k$,
where $k^\mu_i$ is the four momentum on line $i$ and $p^\mu_k$ is the
integration variable. One has the freedom to choose the latter such that
$\alpha^k_i$ is either +1, $-$1 or 0.

\noindent {\bf Theorem} Multi-dimensional energy integration \\
An unambiguous expression is:
\begin{equation}
\int {\rm d} p^-_1 {\rm d} p^-_2 \cdots {\rm d} p^-_m \prod_{i=1}^n
 (\sum \alpha_i^k p^-_k + H_i)^{-1}= \rule{90mm}{0mm}
\end{equation}
 \[
 (2 \pi i)^m\sum_{ \matrix{ \{ j_1,j_2,\cdots,j_m \} \cr [\alpha^1\cdots
\alpha^k]_{j_1\cdots j_k}
\not =0 \cr }} {1 \over [ \alpha^1 \alpha^2 \cdots \alpha^m]_{j_1\cdots j_m}}
\prod_{i \not = j_r}^n \left( {[ \alpha^1 \alpha^2 \cdots \alpha^m H
]_{j_1\cdots j_m i}
\over [ \alpha^1 \alpha^2 \cdots \alpha^m]_{j_1\cdots j_m} } \right) ^{-1}
\]
The antisymmetrized product $[ \alpha^1 \cdots \alpha^m]_{j_1\cdots j_m}$ is
the
determinant of the matrix
\begin{equation}
 (\alpha)_{j_1\cdots j_m} \equiv \alpha = \left( \matrix{
 \alpha^1_{j_1} & \cdots & \alpha^1_{j_m} \cr
  \vdots        &        &    \vdots    \cr
 \alpha^m_{j_1} & \cdots & \alpha^m_{j_m} \cr } \right)
 \label{eq86a}
\end{equation}
The inverse of the matrix $\alpha$ will be denoted by $\bar{\alpha}$.
The poles $\{j_1,\cdots,j_m\}$ that are included in the sum have for all
values of $p^-_i$ the correct imaginary sign of $H_j/\alpha^i_j$, because this
sign is
determined by the plus-components of the integration variables.
Before we start to integrate we first determine inside which contours the
different poles lie, then drop the $i \epsilon$-description.
This is clearly an invariant formulation, so the order of integration can be
altered.\\
\noindent {\it Proof}

An unambiguous way to define the integration is to shift the
integration contour slightly into the complex plane and leave the poles on the
real axis.
The poles are determined by the following set of linear equations
\begin{equation}
 \sum_i \alpha^i_k p^-_i + H_{j_k} = 0,\ \  \; i,k = 1, \cdots m,
 \label{eq100}
\end{equation}
which have the solution
\begin{equation}
 \bar{p}^-_i = - \sum_k \bar{\alpha}^{k}_i H_{k} .
 \label{101}
\end{equation}
Next the multidimensional integral is written in terms of these new variables.
The Jacobian of the transformation is $\det \bar{\alpha} = 1/ \det \alpha =
1/ [\alpha^1 \cdots \alpha^m]_{j_1 \cdots j_m}$. Subsequently, we apply the
residue theorem to every $\bar{p}^-$-integration.
We substitute the new variables, and find that the pole part of
the integral factorizes:
\begin{equation}
 \int \prod_{j=1}^n {\rm d}z_j {f(z) \over (\alpha_1^j z_j- \gamma_1) \cdots
 (\alpha_m^j z_j - \gamma_m) }= {\rm det}[\bar \alpha]
 \int {\rm d}y_1{1 \over y_1\!-\!\bar \alpha_1^j \gamma_j} \cdots \int {\rm d}
 y_m { 1 \over y_m \!-\! \bar \alpha_m^j \gamma_j} \ f(\bar \alpha y).
 \label{eq102}
\end{equation}
We used the summation convention.
Note that the integral is independent of the choice of integration variables.

This type of multi-dimensional complex integration is not related to topology,
so the deformation of the contour might change the result. The torus
obtained by closing the different contours depends on the choice of
coordinates. To avoid ambiguities we take an algebraic view instead.

\subsection{ Recursion formula }

The recursion formula is the basis of the proof of equivalence. It tells us how
to take out
of any Feynman diagram the building block of a l.c.t-ordered diagram: an energy
denominator. This happens without changing the structure of the algebraic form
of the reduced Feynman diagram, so we can apply this formula, a number of times
(the recursion). The final result, obtained upon the last application of the
recursion
formula can immediately be evaluated. The final object is just a piece of a
l.c.t.-ordered diagram ($TOD$): a product of energy denominators.

The recursion formula allows us to consecutively pull energy denominators out
of
$FD(\vec H)$ in order to obtain a sum of $TOD$'s.

\noindent {\bf Theorem}
The following identity is true for any $m$ and $n$ ($N=m+n$):
\begin{equation}
{W_{m,n}(H_1,\cdots,H_m|H_{m+1},\cdots,H_N) \over \Delta(H_1,\cdots,H_N)}=
{1 \over H_1-H_{m+1}} \times
 \label{eq86}
\end{equation}
\[
\left(
{W_{(m-1),n}(H_2,\cdots,H_m|H_{m+1},\cdots,H_N) \over \Delta(H_2,\cdots,H_N)}+
{W_{m,(n-1)}(H_1,\cdots,H_m|H_{m+2},\cdots,H_N) \over \Delta(H_1,\cdots
,H_{m},H_{m+2},\cdots,H_N)} \right)
\]
\noindent {\it Remark}

The reduction step removes two poles, $H_1$ and $H_{m+1}$, and combines them
into a single energy denominator, $H_1 - H_{m+1}$. The second factor of the
r.h.s. of eq.~(\ref{eq86}) consists of two terms, both of which contain one
pole less than the original form.
The factor $(H_1-H_{m+1})^{-1}$ is incorporated in the $TOD$.
At the first stage, $W_{m,n} / \Delta$ is a structure that is directly related
to a Feynman diagram. After taking steps in the reduction algorithm objects
with
the same structure are obtained, but these objects are not in the same way
associated with (possibly different) Feynman diagrams.
The last step in the algorithm is given by:
\begin{equation}
W_{1,n}(y|x_1,\cdots,x_n)=W_{n,1}(x_1,\cdots , x_n |y)= \Delta(x_1,\cdots,x_n)
\end{equation}
First we prove the formula (\ref{eq86}) and then we show how to carry
out the reduction.\\
\noindent {\it Proof}

First express $W_{m,n}$ in terms of a determinant as in eq.~(\ref{eq83}).
Then perform the usual manipulations with determinants: take linear
combinations of rows or columns.
If we subtract the first row from the rows 2 to $m$, the
$m+1^{st}$ row from the other rows $\{ m+2,\cdots,m+n \}$, and expand the
determinant with respect to the last two columns, we obtain:
\begin{equation}
W_{m,n}(x_1,\cdots,x_m|y_1,\cdots,y_{n})=
(-1)^{N}\left| \matrix{  x_2^{K}-x_1^{K}& \cdots & x_2-x_1 \cr
                  \vdots & &  \vdots \cr
 x_m^{K}-x_1^{K}& \cdots & x_m-x_1 \cr
y_2^{K}-y_1^{K}& \cdots & y_2-y_1 \cr
 \vdots & &  \vdots \cr
y_n^{K}-y_1^{K}& \cdots &  y_n-y_1 \cr } \right|
 \label{eq89}
\end{equation}
where $K=n+m-2=N-2$. We can add a row and a column to the determinant to
obtain:
\begin{equation}
{(-1)^{m+1} \over x_1-y_1}
\left| \matrix{  x_2^{K}-x_1^{K}& \cdots &x_2-x_1 &0 \cr
                  \vdots & &  \vdots  &\vdots \cr
 x_m^{K}-x_1^{K}& \cdots  & x_m-x_1 &0 \cr
 x_1^{K}-y_1^{K} & \cdots & x_1-y_1 & x_1-y_1 \cr
y_2^{K}-y_1^{K}& \cdots & y_2-y_1  &0\cr
 \vdots & &  \vdots  &\vdots \cr
y_n^{K}-y_1^{K}& \cdots & y_n-y_1  &0\cr } \right|
 \label{eq90}
\end{equation}
We split the determinant into two parts by adding and subtracting in the last
column the column $(x_2-x_1,\cdots,x_m-x_1,0,\cdots,0)$. Then we subtract the
last column from the next to last column in both determinants to obtain
\begin{equation}
{(-1)^{m+1} \over x_1-y_1}  \left(
\left| \matrix{  x_2^{K}-x_1^{K}& \cdots &0  &x_2-x_1  \cr
                  \vdots & &  \vdots   & \vdots \cr
 x_m^{K}-x_1^{K}& \cdots  &0&  x_m-x_1  \cr
 x_1^{K}-y_1^{K} & \cdots & 0 & x_1-y_1 \cr
y_2^{K}-y_1^{K}& \cdots & y_2-y_1  &0\cr
 \vdots & &  \vdots  &\vdots \cr
y_n^{K}-y_1^{K}& \cdots & y_n-y_1  &0\cr } \right|-
\left| \matrix{  x_2^{K}-x_1^{K}& \cdots &0  &x_2-x_1  \cr
                  \vdots & &  \vdots   & \vdots \cr
 x_m^{K}-x_1^{K}& \cdots  &0&  x_m-x_1  \cr
 x_1^{K}-y_1^{K} & \cdots &  x_1-y_1&0 \cr
y_2^{K}-y_1^{K}& \cdots & y_2-y_1  &0\cr
 \vdots & &  \vdots  &\vdots \cr
y_n^{K}-y_1^{K}& \cdots & y_n-y_1  &0\cr } \right| \right)
 \label{eq91}
\end{equation}
We add the $m$-th row to the rows above it in the first determinant, and
subtract it from the rows below it in the second determinant. The result is
\begin{equation}
{(-1)^{m+1} \over x_1-y_1}
\left| \matrix{  x_2^{K}-y_1^{K}& \cdot\!\!\!\cdot &0  &x_2-y_1  \cr
                  \vdots & &  \vdots   & \vdots \cr
 x_m^{K}-y_1^{K}& \cdot\!\!\!\cdot  &0&  x_m-y_1  \cr
 x_1^{K}-y_1^{K} & \cdot\!\!\!\cdot & 0 & x_1-y_1 \cr
y_2^{K}-y_1^{K}& \cdot\!\!\!\cdot & y_2-y_1  &0 \cr
 \vdots & &  \vdots  &\vdots \cr
y_n^{K}-y_1^{K}& \cdot\!\!\!\cdot & y_n-y_1  &0\cr } \right|+
{(-1)^m \over x_1-y_1}
\left| \matrix{  x_2^{K}-x_1^{K}& \cdot\!\!\!\cdot &0  &x_2-x_1  \cr
                  \vdots & &  \vdots   & \vdots \cr
 x_m^{K}-x_1^{K}& \cdot\!\!\!\cdot  &0&  x_m-x_1  \cr
 x_1^{K}-y_1^{K} & \cdot\!\!\!\cdot &  x_1-y_1&0 \cr
y_2^{K}-x_1^{K}& \cdot\!\!\!\cdot & y_2-x_1  &0\cr
 \vdots & &  \vdots  &\vdots \cr
y_n^{K}-x_1^{K}& \cdot\!\!\!\cdot & y_n-x_1  &0\cr } \right|
\label{eq92}
\end{equation}
Let $M=n+m-3$ and define for any $k$ the symmetric function $\phi$ by the
relation $x^k-y^k = (x-y)\phi^{k-1}(x,y)$. The rows contain one of the factors
$x_i-x_1$, $y_j-x_1$, $y_j-y_1$ or $x_i-x_1$, that can be divided out.
The product of these factors are written as the ratio of two
Vandermonde determinants, to obtain from eq.~(\ref{eq92}):
{\small
\begin{equation}
{ (\!-\!1)^{m\!+\!1}\over x_1\!-\!y_1}\left({\Delta(x_1\cdot\!\cdot  x_m,y_2
\cdot\!\cdot y_n,y_1) \over \Delta( x_1\cdot\!\cdot x_m,y_2 \cdot\!\cdot y_n) }
\left| \matrix{  \phi^{M}(x_2,y_1)& \cdot\!\!\!\cdot &0  &1  \cr
                  \vdots & &  \vdots   & \vdots \cr
 \phi^{M}(x_m,y_1)& \cdot\!\!\!\cdot  &0&  1  \cr
 \phi^{M}(x_1,y_1) & \cdot\!\!\!\cdot & 0 & 1 \cr
\phi^{M}(y_2,y_1)& \cdot\!\!\!\cdot & 1  &0\cr
 \vdots & &  \vdots  &\vdots \cr
\phi^{M}(y_n,y_1)& \cdot\!\!\!\cdot & 1  &0\cr } \right|\!+\!
{\Delta(x_2 \cdot\!\cdot y_n,x_1) \over \Delta(x_2 \cdot\!\cdot y_n)}
\left| \matrix{  \phi^{M}(x_2,x_1)& \cdot\!\!\!\cdot &0  &1  \cr
                  \vdots & &  \vdots   & \vdots \cr
 \phi^{M}(x_m,x_1)& \cdot\!\!\!\cdot  &0&  1  \cr
 \phi^{M}(y_1,x_1) & \cdot\!\!\!\cdot & 1 & 0 \cr
\phi^{M}(y_2,x_1)& \cdot\!\!\!\cdot & 1  &0\cr
 \vdots & &  \vdots  &\vdots \cr
\phi^{M}(y_n,x_1)& \cdot\!\!\!\cdot & 1  &0\cr } \right| \right)
 \label{eq93}
\end{equation}}
The dependence of the first determinant on $y_1$ is only apparent.
 The same is true for the dependence on $x_1$ of the second determinant.
We can easily see this through matrix multiplication:
{\small
\begin{equation}
\left( \matrix{ x_1^{n-2} & \cdot\!\!\!\cdot & x_1 & 1 &1 \cr
                \vdots & & \vdots & \vdots & \vdots \cr
               x_m^{n-2} & \cdot\!\!\!\cdot & x_m & 1 &1 \cr
                x_{m+1}^{n-2} & \cdot\!\!\!\cdot & x_{m+1} & 1 &0 \cr
                \vdots & & \vdots & \vdots & \vdots \cr
               x_n^{n-2} & \cdot\!\!\!\cdot & x_n & 1 &0 \cr 	} \right)
\left( \matrix{ 1 & 0 & \cdot\!\!\!\cdot & 0  & 0\cr
               y & 1 & \ddots & \vdots & \vdots \cr
               \vdots & \ddots & \ddots & 0 &\vdots  \cr
               y^{n-2} & \cdot\!\!\!\cdot & \ \ y &1 &0 \cr
                0 & \cdot\!\!\!\cdot & \ \ 0 & 0 & 1\cr} \right)=
\left( \matrix{ \phi^{n-2}(x_1,y) & \cdot\!\!\!\cdot & x_1\!+\!y & 1 &1 \cr
                \vdots & & \vdots & \vdots & \vdots \cr
               \phi^{n-2}(x_m,y) & \cdot\!\!\!\cdot & x_m\!+\!y & 1 &1 \cr
                \phi^{n-2}(x_{m\!+\!1},y) & \cdot\!\!\!\cdot &
x_{m\!+\!1}\!+\!y & 1 &0 \cr
                \vdots & & \vdots & \vdots & \vdots \cr
               \phi^{n-2}(x_n,y) & \cdot\!\!\!\cdot & x_n\!+\!y & 1 &0 \cr
} \right)
 \label{eq94}
\end{equation}}
The determinant of the second matrix at the l.h.s. is 1, so the determinant of
the first matrix at the l.h.s. is the same as the determinant of the matrix at
the r.h.s..
Removing the $y_1$ dependence in the first determinant and the $x_1$ dependence
in the second we get two familiar objects:
{\small
\begin{equation}
{(-1)^n \over x_1-y_1} \left(
{- \Delta(x_1\cdots  \cdots y_n) \over \Delta( x_1\cdots x_m,y_2 \cdots y_n) }
\left| \matrix{  x_1^{M}& \cdots& x_1 &0  &1  \cr
                  \vdots & &  \vdots  & \vdots & \vdots \cr
 x_m^{M}& \cdots  & x_m&0&  1  \cr
y_2^{M}& \cdots & y_2&1  &0\cr
 \vdots & &  \vdots &\vdots &\vdots \cr
y_n^{M}& \cdots & y_n&1  &0\cr } \right|+
{\Delta(x_1\cdots y_n) \over \Delta(x_2 \cdots y_n)}
\left| \matrix{  x_2^{M}& \cdots &x_2&0  &1  \cr
                  \vdots & &   \vdots  &\vdots& \vdots \cr
 x_m^{M}& \cdots  & x_m&0&  1  \cr
 y_1^{M} & \cdots &y_1& 1 & 0 \cr
 \vdots & &  \vdots  & \vdots& \vdots \cr
y_n^{M}& \cdots & y_n &1  &0\cr } \right| \right)
\label{eq95}
\end{equation}}
These determinants are  nothing but $W$-functions, but now with
less arguments than we started with. So we can write eq.~(\ref{eq95}) as
follows
\begin{equation}
{\Delta(x_1 \cdots y_n) \over x_1-y_1}\left( {W_{m,n-1}(x_1\cdots x_m|y_2\cdots
y_n) \over \Delta( x_1\cdots x_m,y_2 \cdots y_n) }+ {W_{m-1,n}(x_2\cdots
x_m|y_1\cdots y_n) \over \Delta( x_2\cdots  y_n) } \right)
 \label{eq96}
\end{equation}
If we divide the whole expression by $\Delta(x_1\cdots y_n)$ we get the
reduction formula.
\subsection{Reduction algorithm}
We will now show that the application of the formula derived above gives us
parts of l.c.t-ordered diagrams. First we need to specify the structure of a
l.c.t.-ordered diagram.\\
\noindent {\bf Definition} Loop connection tuple

The loop connection tuple $\vec H=(H_1,\cdots, H_n)$ is an ordered set
of objects related to the  propagator denominators $p^2_j-m^2_j+i \epsilon=
2 p_j^+(p^- - H_j)$ or $H_j=p^- - p^-_j +{p^2_{j\perp}+m_j^2 \over 2 p_j^+}$.
The ordering of the tuple corresponds to consecutive ordering of the internal
lines in the corresponding loop in a Feynman diagram.

We will use the terminology of "lines" when we mean the corresponding momentum
or energy, or state. There is some arbitrariness in the definition of the
momenta in the loop, but
in the objects of interest $H_i-H_j$ this arbitrariness is gone because they
are invariant under a shift of the loop momentum ($p^\mu \to p^\mu + a^\mu$).
The expression $H_i-H_j$ is the total incoming
$\sum P^-_{ext}$ momentum minus the on-shell values of the minus-momenta of the
internal lines $p^-_{i,on-shell}+p^-_{j,on-shell}$, calculated with the help of
$p^+_{i,j}$ and $p^\perp_{i,j}$. (See also sect.~\ref{GeneralCase}.)\\
\noindent {\bf Definition} Backward and forward

A line $i$ of the loop connection tuple is going backward if the object
$H_i$ has a positive imaginary part and is going forward if it has negative
imaginary part.
Thus in the Feynman diagram above, eq.~(\ref{eq83}), $H_1,\cdots,H_m$ are going
forward and $H_{m+1},\cdots,H_N$ are going backward.
The sign of the imaginary part is opposite to the sign of the on-shell
energy of the particle, therefore this definition of backward and forward
coincides with the causality condition: positive-energy particles go forward in
l.c. time and negative-energy particles go backward in l.c. time.\\
\noindent {\bf Definition} Early, late and trivial events

An early event is a vertex between a backward and a forward going line, a late
event is a vertex between a forward and a backward going line, if one goes
around the loop in the order corresponding to the connection tuple. All other
vertices are trivial events. There are  equal numbers of early and late
events.\\
\noindent {\bf Definition} Flat loop diagrams

A flat loop diagram has one early (and thus one late) event.\\
\noindent {\bf Definition} Crossed loop diagrams

A crossed loop has more than one early event.\\
\noindent {\bf Definition} Skeleton graph

A skeleton graph is the tuple of signs of the objects $H_i$ of a connection
tuple. It is given by the mapping
$\{ H_1,H_2,\cdots,H_n \} \to \{ \sigma ({\Im} H_1), \sigma ({\Im } H_2),
\cdots , \sigma ({\Im} H_n) \}$. The function $\sigma$ is the sign function.

For different external momenta in the Feynman graph we have a different set of
skeleton graphs, and for each Feynman diagram there are a number of skeleton
graphs.
For each sector (associated with a specific number of poles above and below the
real axis) of the loop momentum $p^+$ there is a skeleton graph, thus
for a loop with $n$ lines there are $n-1$ skeleton graphs.

The skeleton graph already tells us the general features of the l.c.t.-ordered
diagrams which are contained in a Feynman diagram, because it tells us the
direction of the internal lines. This is used as our guide how to take
"time-slices" of the Feynman diagram. The direction of a line tells us in which
order we can encounter events (vertices). Early and late events correspond to
sign changes in the skeleton graph.\\
\noindent {\bf Definition} Causally connected events

Two events are causally connected if they lie on the same loop and there are
neither early nor late events lying in between.

So, two causally connected
events are connected by parts of a loop that consist of lines that are either
all forward or all backward.\\
\noindent {\it Remark}

Clearly, it makes sense to say that causally connected events are ordered in
l.c. time. If we follow a loop in the direction given by the orientation
defined
by its connection tuple, then we will encounter forward and backward going
lines. If two vertices are connected by a forward line, they are said to be
ordered in l.c. time in the same way as they are ordered in the loop. Otherwise
their order in l.c. time is the inverse of their order in the loop. This
partial ordering, which  is given by the skeleton graph, is obviously not
complete. Only causally connected events are mutually ordered this way, but
not with respect to other events.

Note that we don't make any statements here about reducible Feynman diagrams,
which is a completely different story. Our causally unconnected parts connect
up
a later time, so they are parts of the same irreducible Feynman diagram.\\
\noindent {\bf Definition} Simultaneous

Two parts of a skeleton graph are said to be {\it simultaneous} if they do not
share events that are causally connected.\\
\noindent {\it Remark}

The flat box that we discussed in sect.~\ref{Chapter2} consisted of a late and
an early vertex, connected by two distinct parts of the loop, one consisting of
lines graded $+$, the other of lines graded $-$. The relative l.c.t.-ordering
of
the events on these two parts is not necessarily determined by the skeleton
graph, but application of the reduction formula, eq.~(\ref{eq86}), produced
immediately the two possible l.c.t. orderings. The diagrams found showed the
expected energy denominators.

In situations where two simultaneous parts occur, the reduction formula does
not
provide immediately the l.c.t.-ordered diagrams. An example was given in
sect.~\ref{CrossedBox}. From the point of view of l.c. time ordering, one
expects diagrams to occur corresponding to all relative l.c.t. orderings of
simultaneous parts. In momentum-energy language this means that diagrams with
certain energy denominators should occur. Indeed, this is the content of the
next theorem.\\
\noindent {\bf Theorem} Simultaneous parts come in all combinations.

\noindent {\it Remark}

The proof of this theorem relies again on a recursion. First we suppose that we
have two simultaneous parts, that are already ordered by themselves, but not
mutually. Both parts correspond to sets of energy denominators, say
$\{ \alpha \}$ and $\{ \beta \}$. So we have the two distinct $TOD$'s
$\Pi \alpha^{-1}_i$ and $\Pi \beta^{-1}_j$. In this language the content of the
theorem can be written as follows,
 $\{  \{ i_k,j_k\}_k | [\{ i_k,j_k\}\not = \{i_{k+1},j_{k+1}\}] \wedge [k<l
\Rightarrow (i_k \leq i_l) \wedge (j_k \leq j_l)] \}$
\begin{equation}
\prod_{i=1}^m\alpha_i^{-1} \prod_{j=1}^n \beta_j^{-1} = \sum_{ {{\rm
all}\ i_k,j_k   }}\prod_{k=1}^{m+n}(\alpha_{i_k} +\beta_{j_k})^{-1} .
 \label{eq110}
\end{equation}
\noindent {\it Proof}

We apply the formula:
\begin{equation}
\prod_{i=1}^m\alpha_i^{-1} \prod_{j=1}^n \beta_j^{-1}
= {1 \over \alpha_m+\beta_n} \times
\left[ \prod_{i=1}^{m-1} \alpha_i^{-1} \prod_{j=1}^n \beta_j^{-1}+
\prod_{i=1}^m\alpha_i^{-1} \prod_{j=1}^{n-1} \beta_j^{-1} \right]
 \end{equation}
We can apply this algorithm recursively
to obtain $(^{n+m}_{\ m})$ \ $TOD$'s with
energy denominators of the form $\sum \alpha_i + \sum \beta_j$.
If the product consists of more than two $TOD$'s we can apply this algorithm
recursively, to two $TOD$'s at a time, since it is associative.

The expressions obtained are composite energy denominators:
\begin{equation}
(P^-(\alpha) - H_0(\alpha)) +(P^-(\beta) -H_0(\beta ))=  P^-(\alpha \cup \beta)
- H_0(\alpha \cup \beta ).
\end{equation}
\subsection{Reduction of Feynman diagrams}

The previous sections were mainly concerned with algebraic identities. Now we
turn to the general strategy of the reduction.
The reduction algorithm for a flat loop is straightforward as we noted when we
discussed simultaneous parts. The trivial vertices come in all orderings of
vertices on the forward line ($\Im H<0$)  with respect to those on the backward
line ($\Im H>0$).
The vertices on one line are already ordered with respect to each other by
the skeleton graph.
Starting with the early event one can reduce the lines next to the early event.
 This algorithm ends and gives $2^{N-2}\ TOD$'s if the loop is flat for all
skeleton graphs. For each skeleton graph there are $(^{N-2}_{m-1})$ \
$TOD$'s.\\
\noindent {\bf Mapping}

The mapping from a recursive algorithm to a time-ordered theory is
straightforward for the flat loop. The order in which the poles $H_i$ are
removed in the reduction formula is the same as the time ordering. We refer to
this reduction as time ordered reduction.

A flat Feynman diagram gives the topology of all $TOD$'s contained in it and
all combinations between trivial events appear respecting the causal order.
Conservation of $p^+$-momentum determines whether a trivial vertex is an
absorptive or emissive event. However, this is not a new element, because we
have seen that a skeleton graph is determined by the external lines besides the
value of the plus-component of the loop momentum. Of course, early events
absorb external particles while late events emit them.

Because Feynman diagrams form the basis of our treatment of light-front field
theory, the basic elements we are concerned with are the single particle
propagators and the vertices derived from the underlying Lagrangian.
The interaction ${\cal H}_{int}$ is derived from ${\cal L}_{int}$ with an
additional phase factor { $\left(\sqrt{2^np^+_1p_2^+\cdots p^+_N}\
\right)^{-1}$} where the
longitudinal momenta of all the lines from a vertex are
included. A wave function must also be multiplied with the phase factor, as
compared to the covariant wave function.

For the success of our reduction algorithm the details of ${\cal L}_{int}$
are of minor importance. The algebra is connected to plus-momentum flow
in loops.
The only internal lines in the diagram of interest are those in the loop,
how momentum is extracted from and added to the loop is of less importance.

The most important feature of the reduction algorithm is the fact that it
always starts from an early state with positive longitudinal momentum
(there is always an early state as the result of momentum conservation),
so we exclude "vacuum"-type diagrams.

First, reduction is performed on the skeleton graph starting from the early
events, {\it i.e.}, removing lines directly connected to the early events. This
can be followed by removing poles corresponding to consecutive pieces of the
loop until a late vertex is reached. This is the point where $W_{m,n}/\Delta$
is
reduced to a sum of terms of the form $W_{j,1}/\Delta^{\prime}$ or
$W_{1,k}/\Delta^{\prime\prime}$. Secondly, the simultaneous-parts theorem is
applied to write all these terms as sums of terms containing true energy
denominators

Because we could have started the reduction from the late vertices in stead of
the early ones, we see that the algorithm can be written in different ways.
The final result, however, is the same. The same is true, {\it mutatis
mutandis},
for the application of the simultaneous-parts theorem.

So we peel off more and more of the Feynman diagram in a manner which is
 locally (for causally connected events) equal to time ordered reduction.

The relation of the results of the reduction process to l.c.t.-ordered diagrams
in the crossed loop case is more complicated than in the case of a flat loop.
The simple heuristic of
cutting lines representing constant l.c. time surfaces leads to a more
elaborate
bookkeeping in the crossed-loop case. This is so, because only the global
structure of the Feynman diagram determines which simultaneous parts are
joined by early or late vertices. The general strategy proposed here is to
start
at an early vertex, use the reduction algorithm locally until a late vertex is
attained. This procedure is to be repeated until all late vertices have been
processed. Next apply the simultaneous-parts theorem repeatedly.\\
\noindent {\bf Extension mapping}

Multi-loop diagrams can be reduced one loop after another. The loop momenta
that are not integrated over are kept fixed. The skeleton graph tells us again
 what is the general form of the time-ordered diagrams. Therefore, the skeleton
graphs, associated with different finite domains in $p^+$-space need to be
determined first.

 Upon  application of the reduction algorithm to the first loop, energy
 denominators occur that are combinations of two propagator poles. When the
 next loop is treated, some poles come from those energy denominators, while
 the others are due to propagator poles occurring in the part  of the original
 Feynman diagram unaltered by the reduction so far.
 These different types of poles play the same role in the integration of the
 next variable.  The pole is again
 the difference of the $p^-$-flow and the on-shell energies from the poles and
 their associated lines in the Feynman diagram. The question which lines must
be combined to generate energy denominators is related to the imaginary parts
 of the propagator poles and thus answered when the skeleton
 graph is determined. During the reduction process these combinations
 remain fixed.

The integration is invariant under coordinate transformations and reordering of
the integrations. The mapping from a recursive algorithm to a l.c.t.-ordered
approach is more complicated here than in the single-loop case, but in essence
the same. The most complicated task is the construction of the skeleton
graphs. After this job is done, the reduction algorithm is  applied to one
loop after another, and the interpretation of the result is the same as in the
case of one loop.\\
\noindent {\bf Theorem} Spectrum condition

The spectrum condition $p^+\geq 0$ holds for all particles in the internal
loops of the Feynman diagram.\\
\noindent {\it Proof}

The spectrum condition follows from two ingredients. First, at any vertex there
is conservation of four-momentum, in particular plus-momentum. Secondly, lines
with negative $p^+$, antiparticle lines, can be reinterpreted as particle lines
with positive  $p^+$ by reversing the direction of the four momentum on such
lines. This reversal is in agreement with four-momentum conservation and
l.c.t. ordering.
\subsection{Algorithm proofs}

In a number of cases we restricted ourselves to a recursive formula, which
could be applied to any structure \cite{Klop}. We gave a {\it rule} that could
be applied
successively and leads to l.c.t-ordered diagrams with the right structure.
For the flat loop case  we obtain all l.c.t.-ordered
diagrams, since the algorithm is unique, and directly related to a time-ordered
picture. In the case of the crossed loop diagrams
the algorithm is not unique; the rule could be applied to different
parts of the structure at the same time. We could start with the reduction
at one early vertex or another early vertex. Strictly speaking, this poses
a problem only when the rule can be applied to overlapping regions of the
structure, the different
orders in which the rule is applied might lead to different answers.
Our {\it conjecture} is that in our case the final answer is unique and does
not
depend on the order in which the rule  is applied. We base this conjecture
on the observation that each term obtained after applying the first rule
(with the simultaneous parts still present), can be associated with a
topological object, such that all terms give the complete set of objects with
specific properties. The topology  is invariant, thus unique and independent
of the way it is obtained.
We have checked this for a number of cases, and found it to hold in all these
cases.  Note that we used a similar argument for deriving an invariant
multi-loop integration.

We would rather have shown that the reduction algorithm would lead to
all time-ordered diagrams which satisfy the spectrum condition, which
would also be some kind of invariant, but this
turned out to be too complicated a problem in the general case. Therefore we
did not
write down a general formula, directly relating any Feynman diagram to the sum
of l.c.t.-ordered diagrams.

At the moment we satisfy at each step the spectrum condition and allow
all possible orderings. The global properties follows from the local
ones.

\section{Discussion}
\label{Chapter7}
We have established the degree of equivalence between light-cone-time ordered
perturbation theory and covariant perturbation theory for spin-0 and spin-1/2
particles. This effort might seem superfluous since the connection between
ordinary time-ordered perturbation theory and covariant perturbation
theory is well established \cite{Feyn49}. One might be tempted to believe that
the methods that work in the case of ordinary time-dependent perturbation
theory
apply to the l.c.t.-ordered theory too. This belief belongs to folklore. In
practice
the understanding of light-front field theory is growing only slowly.
Basic results in covariant field theory were often proven to obtain also in
light-front field theory along a path through much trial and error.
It took  years to obtain a proper light-front version of the
Schwinger-model \cite{SchwMod} and to prove spontaneous symmetry breaking in
the light-front version of $\phi^4$ \cite{phi4}.
The renormalization of light-front versions of known covariant, renormalizable
field theories is still an unsolved problem \cite{renorm}.
In general, approaches are followed that are specially tuned to the problems of
light-front field theory. Therefore it turns out to be difficult to relate the
solutions obtained to basic results in covariant field theory \cite{Tho79},
\cite{MPSW91}.

Conceptual problems were already present from the onset of light-front field
theory, when Weinberg showed \cite{W66} that only some physical processes, each
represented by an ordinary time-ordered diagram, contribute to the Feynman
diagram if this diagram was calculated in a frame that moves with the speed of
light \cite{KSuss}. This so called {\it infinite momentum frame}
(IMF) cannot
be connected to any other reference reference frame by a finite Lorentz
transformation. Thus, a limiting procedure is involved.
This limit has to compete with other limits present in field theory:
infinite space integration, regularization of singular expressions.
This is separate from the additional problems that might arise when we are
dealing with fermions \cite{Ahl92}.
That the IMF  is naturally described with light-front coordinates is only
apparent for coordinates and momenta \cite{LKS70}. That only some diagrams
survive \cite{CM69} is puzzling. However, we have shown that this, in general,
is indeed true.
Another possible approach to light-front field theory is the direct
quantization on a
light-front. There are many different ways to do this, which become more
elaborate if the theories are supposed to incorporate more features
\cite{KR94}. As a
classical theory, light-front field theory is ill-defined; the standard initial
values problem on the light-front is overdetermined. In addition, it leads to a
non-unique evolution in time \cite{Hormander}. The first problem can be solved
in principle through methods devised for the quantization of constrained
systems
\cite{constrained}. The second problem is more serious. One needs to introduce
degrees of freedoms associated with
different evolutions in l.c. time and then introduce constraints which can
restrict the space of solutions to the one considered physical on some grounds.
This, in essence, is what people are dealing with when they introduce
zero-modes, degrees of freedom of which the evolution is unknown (zero or
infinite?). This problem is often disguised in practical calculations
where a $(p^+)^{-1}$-singularity occurs \cite{pplus}, \cite{BL91}.

Another way to quantize a light-front theory is to use axiomatic commutation
relations on a complete set of free fields \cite{Jack72}, \cite{Ida77},
\cite{SS72}, an approach guided by the results of current algebra \cite{A+73}.
On the light front, different points with ($\Delta x_\perp=0$) are light-like
separated. The question arises what is  the equal-time commutator between
fields: a delta
function in $x^-$ which violates covariance, or a sign function in $x^-$ which
leads to non-integrable fields (except at the loss of covariance) \cite{Nak77}.
Whether such theories can describe physical processes has not been established.
This approach became less favorable in the late seventies.
The approach most favored nowadays is based on two methods: with the covariant
results in mind derive a "constrained Hamiltonian" \cite{KS70}, \cite{BRS73},
\cite{Mus90}.
Due to zero-modes it is hard to make a one-to-one correspondence between
normalized states on a space-like surface and a light-like surface.
As long as one deals only with tree graphs all these problems are rather
formal.
The presence of loops makes them acute. In loops one has to
"sum over all states". This rule forces one to think over states with
$p^+\to 0$.
The advantages of light-front field theory are paid for by the occurrence of
problems. In the approaches discussed here the problems arise in
different forms and are dealt with in different ways.  They might
be disguised as technical difficulties. Due to the fundamental nature of these
problems the final results depend strongly on the choices one has to make
when defining the (finite) theory, e.g.,
boundary conditions of fields quantized in a box \cite{NR71}, \cite{HL93},
or regularization of the $(p^+)^{-1}$ singularity \cite{pplus}.
We also had to make some choices. Wherever we had to do so, we emphasized the
relation with Feynman diagrams.
In a manifestly covariant approach there is no $(p^+)^{-1}$ singularity.
It is a distinct advantage of our approach that this singularity also
does not occur (see sect. \ref{Chapter5}). We seem to have cured one of
the diseases of light-front field theory.
However, presently we do not know whether our regularization procedure
leads to the same answer as the covariant approach.

We have shown (see sect. \ref{zeromodes}) that in some cases there are
$\delta (p^+)$ contributions to the S-matrix. Maybe these terms indicate
a coupling to the "vacuum"
or they  may represent contributions which relate one version of light-front
field theory to another by a finite renormalization.
However, it would be good practice to try to separate the mathematical question
"how to calculate?", from the physical one "how to interpret?".

We consider the present situation in light-front field theory to be confused.
We
give three reasons for this point of view:
\begin{itemize}
\item[(i)]{The paucity of  comparisons to standard covariant theories;}
\item[(ii)]{The mixing of mathematical problems with physical ones;}
\item[(iii)]{The lack of consensus on what are the established results (with
proper,
mathematically rigorous derivations).}
\end{itemize}
Still, there are a four good reasons to work on light-front field theory:
\begin{itemize}
\item[(a)]{It is the only theory  distinctly different from covariant field
theory
which allows for a comparison at intermediate levels. Such a comparison
increases the understanding in both theories;}
\item[(b)]{It is the most natural way to describe nucleons in terms of quarks;}
\item[(c)]{Our understanding increases with each answer to  questions that
light-front
field theory raises;}
\item[(d)]{It is a Hamiltonian formulation, sharing the intuitive picture
inherent in elementary quantum mechanics.}
\end{itemize}
\vspace{10mm}

\centerline{\large\bf Acknowledgement}

\vspace{10mm}

The work described in this paper is part of the research program of the
"Stichting voor Fundamenteel Onderzoek der Materie (FOM)", which is
financially supported by the "Nederlandse Organisatie voor Wetenschappelijk
Onderzoek (NWO)".

\end{document}